\documentclass[preprint,11pt,nofootinbib,amsmath,amssymb,floatfix,prd]{revtex4-1}
\usepackage{graphicx}
\usepackage{dcolumn}
\usepackage{bm}
\usepackage{hyperref}
\usepackage[mathlines]{lineno}
\usepackage{multirow}
\usepackage{url}
\usepackage{float}
\hypersetup{pdfstartview=FitV,colorlinks=true,linkcolor=blue,citecolor=red,filecolor=black,urlcolor=blue}

\def\issue(#1,#2,#3){{\bf #1}, #2 (#3)}

\catcode`\@=11
\def\lsim{\mathrel{\mathpalette\@versim<}}
\def\gsim{\mathrel{\mathpalette\@versim>}}
\def\@versim#1#2{\vcenter{\offinterlineskip
\ialign{$\m@th#1\hfil##\hfil$\crcr#2\crcr\sim\crcr } }}
\catcode`\@=12

\catcode`@=12

\newcommand{\met}{$\cancel E_T$}

\newcommand{\newc}{\newcommand}
\newc{\wt}{\widetilde}
\newc{\ra}{\rightarrow}
\def\beq {\begin{equation}}
\def\eeq {\end{equation}}
\def\bi {\begin{itemize}}
\def\ei {\end{itemize}}
\def\bea {\begin{eqnarray}}
\def\eea {\end{eqnarray}}

\def \met{\rm E{\!\!\!/}_T}

\newcommand{\br}{\begin{eqnarray}}
\newcommand{\er}{\end{eqnarray}}
\newcommand{\be}{\begin{equation}}
\newcommand{\ee}{\end{equation}}

\newcommand{\ch}{\widetilde \chi^\pm}

\newcommand{\gb}{\textcolor{blue}}

\def \ch2p {{\wt\chi_2^+}}

\def \ch2m {{\wt\chi_2^-}}

\def \chjpm{{\wt\chi_j}^{\pm}}
\def \chonepm{{\wt\chi_1}^{\pm}}

\newc{\dmchi}{\Delta m_{\wt\chi}}

\def \chtwopm{{\wt\chi_2}^{\pm}}

\def \lspi{\wt\chi_i^0}

\def \lspone{\wt\chi_1^0}

\def \lsptwo{\wt\chi_2^0}

\def \lspthree{\wt\chi_3^0}

\def \lspfour{\wt\chi_4^0}

\usepackage{array}
\newcolumntype{L}[1]{>{\raggedright\let\newline\\\arraybackslash\hspace{0pt}}m{#1}}
\newcolumntype{C}[1]{>{\centering\let\newline\\\arraybackslash\hspace{0pt}}m{#1}}
\newcolumntype{R}[1]{>{\raggedleft\let\newline\\\arraybackslash\hspace{0pt}}m{#1}}

\def\issue(#1,#2,#3){{\bf #1}, #2 (#3)}

\hypersetup{
   colorlinks=true,       
   linkcolor=blue,        
   citecolor=red,         
   filecolor=magenta,      
   }

\begin{document}

\title{Current bounds and future prospects of light neutralino dark matter in NMSSM }

\author{Rahool Kumar Barman}
\email{rahoolbarman@iisc.ac.in}
\affiliation{Indian Association for the cultivation of Science, 
Jadavpur, Kolkata 700032, India}
\affiliation{Centre for High Energy Physics, 
Indian Institute of Science, Bangalore 560012, India}

\author{Genevieve B\'elanger}
\email{belanger@lapth.cnrs.fr}
\affiliation{LAPTh, Universit\'e Savoie Mont Blanc, CNRS, B.P. 110,  
F-74941 Annecy Cedex, France}

\author{Biplob Bhattacherjee}
\email{biplob@iisc.ac.in}
\affiliation{Centre for High Energy Physics, 
Indian Institute of Science, Bangalore 560012, India}

\author{Rohini Godbole}
\email{rohini@iisc.ac.in}
\affiliation{Centre for High Energy Physics, 
Indian Institute of Science, Bangalore 560012, India}

\author{Dipan Sengupta}
\email{disengupta@physics.ucsd.edu}
\affiliation{Department of Physics and Astronomy, University of California, San Diego, 9500 Gilman Drive, La Jolla, USA}

\author{Xerxes Tata}
\email{tata@phys.hawaii.edu}
\affiliation{Department of Physics and Astronomy, University of Hawaii, Honolulu, HI 96822, USA}



\begin{abstract}

Unlike its minimal counterpart, the Next to Minimal supersymmetric Standard Model (NMSSM) allows the possibility that the lightest neutralino could have a mass as small as $\sim 1~{\rm GeV}$ while still providing  a significant component of relic dark matter (DM). Such a neutralino can provide an invisible decay mode to the Higgs as well. Further, the observed SM-like Higgs boson ($H_{125}$) could also have  an invisible branching fraction as high as $\sim 19\%$. Led by these facts,  we first delineate the region of parameter space of the NMSSM with a light
neutralino ($M_{\lspone} < 62.5~{\rm GeV}$) that yields a thermal neutralino
relic density smaller than the measured relic density of cold dark
matter, and is also compatible with constraints from collider
searches, searches for dark matter, and from flavour physics. We then
examine the prospects for probing the NMSSM with a light neutralino via
direct DM detection searches, via invisible Higgs boson width
experiments at future $e^+e^-$ colliders, via searches for a light
singlet Higgs boson in $2b2\mu$, $2b2\tau$ and $2\mu2\tau$ channels
and via pair production of winos or  doublet higgsinos at the high
luminosity LHC and its proposed energy upgrade. For this last-mentioned
electroweakino search, we perform a detailed analysis to map out the
projected reach in the $3l+\met$ channel, assuming that chargino
decays to $W \lspone$ and the neutralino(s) decay to $Z$ or
$H_{125}$ + $\lspone$. We find that the HL-LHC can discover SUSY in just part of the parameter space in each of these channels, which together can probe almost the entire parameter space. The HE-LHC probes essentially the entire region with higgsinos~(winos) lighter than 1~TeV~(2~TeV) independently of how the neutralinos decay, and leads to significantly larger signal rates.
\end{abstract}

\maketitle

\tableofcontents

\section{Introduction}
The LHC Run-II has ushered in a new era in terms of energy, luminosity and discovery potential. At the 
end of the Run-II with about 140 $\rm fb^{-1}$ of data collected at 13 TeV, the status of the Higgs as the last frontier to be conquered as concerns  
the Standard Model has been firmly established~\cite{Aad:2015zhl}. As yet, there is no unambiguous evidence for any physics beyond the Standard Model(BSM). In case of the primary candidate for  BSM physics $viz$ TeV scale supersymmetry (SUSY), large swaths of parameter space  have been ruled out.   
However the LHC 13 TeV still holds promise of further exploring the BSM landscape including  SUSY.  In particular, the electroweak sector of SUSY is still largely unconstrained. A better coverage of this sector is expected with larger integrated luminosities, as production of electroweakinos  not only suffers from weak production cross sections but limits from the current searches can also be evaded as an intricate combination of parameters can lead to compressed spectra, see for example~\cite{Kawagoe:2006sm,Giudice:2010wb,LeCompte:2011cn,
Murayama:2012jh,Dreiner:2012gx,Bhattacherjee:2012mz,
Rolbiecki:2012gn,Bhattacherjee:2013wna,Tobioka:2015vsv,
Dutta:2015exw,Delgado:2016gqn,Chowdhury:2016qnz,Nagata:2017gci,
Chakraborti:2017dpu,Dutta:2017jpe}. Compressed higgsinos are expected in models of natural SUSY~\cite{Chan:1997bi,PhysRevLett.109.161802}. Experiments at the LHC have also begun to probe such scenarios~\cite{PhysRevD.97.052010,Aad:2019qnd,Sirunyan:2018iwl}. Moreover, the electroweak sector  contains one of the most promising dark matter (DM) candidate, the lightest neutralino.  It is imperative that we continue to  probe the complementarity between LHC searches for SUSY and astrophysical observations as far as supersymmetric dark matter is concerned~\cite{Carena:2018nlf,Carena:2019pwq}. A study of complementarity between future direct detection experiments and future collider searches for the case of a light neutralino DM ($M_{\lspone} \leq 62.5~{\rm GeV}$), both thermal and non-thermal, in the MSSM framework was carried out in \cite{Barman:2017swy}. Results in \cite{Barman:2017swy} indicated that the MSSM parameter space with a light neutralino DM and with a correct or under-abundant relic density could be entirely probed via the future dark matter experiments.
 
This then naturally motivates us to examine the extent to which the considerations of \cite{Barman:2017swy} would be altered in the simplest extension of the MSSM, the Next-to-Minimal Supersymmetric Standard Model (NMSSM)~\cite{Ellwanger:2009dp,Djouadi:2008uj}.
As in \cite{Barman:2017swy} we focus on $M_{\lspone} \leq 62.5~{\rm GeV}$, that can potentially contribute to invisible decays of the Higgs boson discovered at
CERN~\cite{Djouadi:1996mj,Belanger:2001am,AlbornozVasquez:2011aa,Ananthanarayan:2013fga}, hereafter referred to as $H_{125}$.
In the MSSM with heavy sfermions, the current limit for light neutralinos after taking into account various cosmological, astrophysics and collider constraints lies around $30~{\rm GeV}$ ~\cite{Hooper:2002nq,Belanger:2003wb,Feldman:2010ke,Calibbi:2011ug,Boehm:2013gst,Belanger:2013pna,Hamaguchi:2015rxa,Barman:2017swy}. This arises mainly from  a combination of  the direct detection constraint which requires a weak coupling of the DM to the Higgs and of the relic density constraint which requires that the neutralino  annihilates efficiently through a mediator that is nearly on-shell. 
Within the NMSSM, much lighter neutralinos, as light as $1~{\rm GeV}$, 
can satisfy all current constraints~\cite{Vasquez:2010ru,Cao:2011re,AlbornozVasquez:2011js,Kozaczuk:2013spa,Ellwanger:2014dfa,Barducci:2015zna,Ellwanger:2018zxt,Cao:2013mqa}. This is because additional scalars present in the NMSSM may provide an efficient annihilation mechanism for light neutralinos in the early universe~\cite{Belanger:2005kh,Mahmoudi:2010xp,Cheung:2014lqa}. Morever, the singlet component of these new scalars entails that they can evade detection at the LHC even if their masses are below that of the standard model-like Higgs~\cite{Domingo:2008rr,Cerdeno:2013cz,Sirunyan:2018pzn,Sirunyan:2018mbx,Wang:2020dtb,Guchait:2015owa,Guchait:2016pes,Guchait:2020wqn}.
In this article, assuming a standard cosmological scenario, we perform 
a comprehensive exploration  of the light neutralino as a thermal cold dark matter candidate in the NMSSM  by imposing current low energy, collider and astrophysical constraints.
These include flavor physics constraints, LEP bounds, recent results from LHC measurements of the Higgs sector including searches for light Higgses~\cite{Khachatryan:2015nba,Khachatryan:2017mnf,Khachatryan:2015wka,Sirunyan:2018pzn} and measurements of the Higgs signal strengths,  direct chargino/neutralino searches at the LHC~\cite{Ellwanger:2013rsa,Domingo:2018ykx,Sirunyan:2018ubx} as well as limits from direct detection experiments in spin-independent neutralino-nucleon scattering (Xenon-1T~\cite{Aprile:2018dbl}) and from  indirect detection of DM in the photon channel (Fermi-LAT~\cite{Fermi-LAT:2016uux}).
We concentrate on the electroweakino sector and assume that squarks and sleptons are heavy and do not play an important role in DM  or collider observables. After having established the region in parameter space compatible with current experimental results, we assess the impact of future searches. We reemphasize that for the (thermally produced) neutralino LSP (lightest SUSY particle) to not produce too much dark matter in the early universe, we require a light spin-zero particle close to twice the LSP mass. Correspondingly, we choose parameter space points which also feature $M_{A_{1}}$ and $M_{H_{1}}$ below $ 122~{\rm GeV}$.  We focus on three specific directions that we find more promising for discovery:  
\begin{enumerate}
\item multi-ton direct detection experiments, \item the measurement of the invisible decay width of the Higgs at the LHC or a future collider like FCC, CEPC and ILC, \item direct searches for light Higgses and  electroweakinos at the LHC and its future upgrades $viz$ the proposed high luminosity LHC (HL-LHC: $\sqrt{s}=14~{\rm TeV}$, $3000~{\rm fb^{-1}}$) and the high energy LHC (HE-LHC: $\sqrt{s}=27~{\rm TeV}$, $15~{\rm ab^{-1}}$).
\end{enumerate}
Specifically, 
we assess the reach of direct electroweakino searches in the $WZ$ mediated and $WH_{125}$ mediated $3l+\met$ search channel, 
 at HL-LHC as well as HE-LHC. The impact of the projected search limits from direct light Higgs searches in the $2b2\tau$, $2b2\mu$, $2\mu2\tau$ and $4\mu$ channels at the future upgrades of LHC are also examined~\cite{Cepeda:2019klc}. Finally we emphasize the complementarity of future  measurements of an invisible branching ratio for the Higgs,  searches for Higgs and electroweakino  at colliders, and 
direct and indirect detection experiments for discovering or probing the light neutralino in the NMSSM.

The rest of the paper is organized as follows. In Sec.~\ref{sec:model_framework}, we present  the NMSSM framework and discuss the  parameter space of interest in Sec.~\ref{sec:parameter_space}. The relevant constraints and their implications are described in Sec.~\ref{sec:constraints}. The characteristic features of the allowed parameter space region obtained after imposing the current constraints are discussed in Sec.~\ref{Sec:NMSSM:thermal_neutralino}. In Sec.~\ref{Sec:NMSSM:future_1}, we investigate the future reach of Xenon-nT (through Spin-independent WIMP-nucleon interactions), the future prospects for ILC and CEPC (through Higgs invisible width measurements) and the scope of direct light Higgs searches at future runs of LHC, and study their implications on the allowed parameter space. In Sec.~\ref{Sec:NMSSM:future_2}, we  explore the reach via searches for direct production of electroweakinos in the $3l+\met$ final state at the HL-LHC and the HE-LHC. We conclude in Sec.~\ref{Sec:NMSSM:Conclusion}.

\section{The NMSSM framework}
\label{sec:model_framework}

The NMSSM Higgs sector consists of two doublet Higgs superfields, $\hat{H_{u}}$ and $\hat{H_{d}}$, and a Higgs singlet superfield $\hat{S}$. The scale invariant NMSSM superpotential has the form~\cite{ELLWANGER20101}

\begin{equation}
W = W_{MSSM} (\mu = 0) + \lambda \hat{S}\hat{H_{u}}\cdot\hat{H_{d}} + \frac{\kappa}{3}\hat{S}^{3}
\label{eqn:superpot}
\end{equation} 
where $W_{MSSM}$ ($\mu= 0$) refers to the MSSM superpotential without the $\mu$-term, while $\lambda$ and $\kappa$ are dimensionless parameters, and $\hat{H_{u}}\cdot \hat{H_{d}} = \hat{H_{u}^{+}}\hat{H_{d}^{-}} - \hat{H_{u}^{0}}\hat{H_{d}^{0}}$. The soft SUSY breaking terms which contain the Higgs scalar fields are 

\begin{equation}
-V_{soft} = m_{H_{d}}^{2} H_{d}^{\dagger}H_{d} + m_{H_{u}}^{2} H_{u}^{\dagger}H_{u} + m_{S}^{2} S^{\dagger}S + \left\lbrace\lambda A_{\lambda} H_{u}\cdot H_{d} S + \frac{1}{3}\kappa A_{\kappa}S^{3} + h.c.\right\rbrace \quad  
\end{equation}
where $A_{\kappa}$ and $A_{\lambda}$ are the trilinear soft-breaking parameters, while $m_{H_{d}}$, $m_{H_{u}}$ and $m_{S}$ are the soft breaking Higgs masses. An effective $\mu$-term with $\mu = \lambda v_{s}$ is generated when $S$ develops a vacuum expectation value. 
The F- and D-terms also contribute to the Higgs scalar potential, and are given by

\begin{equation}
V_{F} = |\lambda H_{u}\cdot H_{d} + \kappa S^{2}|^{2} + \lambda^{2}S^{\dagger}S\left(H_{u}^{\dagger}H_{u} + H_{d}^{\dagger}H_{d}\right)
\end{equation}
\begin{equation}
V_{D} = \frac{g_{1}^{2} + g_{2}^{2}}{8} {\left(H_{u}^{\dagger}H_{u} - H_{d}^{\dagger}H_{d}\right)}^{2} + \frac{g_{2}^{2}}{2} |H_{d}^{\dagger}H_{u}|^{2}
\end{equation} 
where, $g_{1}$ and $g_{2}$ are the $U(1)_{Y}$ and $SU(2)_{L}$ gauge couplings of the SM. 
Expanding the scalar potential, $V_{soft}+V_{F}+V_{D}$, around the real neutral vacuum expectation values ($vevs$), $v_{u}$, $v_{d}$ and $v_{s}$ of $H_{u}$, $H_{d}$ and $S$, respectively, gives the physical Higgs states. Following the notation of \cite{PhysRevD.95.115036},

\begin{equation}
H_{u}^{0} = \frac{v_{u}+H_{u}^{R}+i H_{u}^{I}}{\sqrt{2}},\quad H_{d}^{0} = \frac{v_{d}+H_{d}^{R}+ i H_{d}^{I}}{\sqrt{2}},\quad S = \frac{s+H^{S}+ i A^{s}}{\sqrt{2}}
\end{equation}
where, $H_{u}^{R}$, $H_{d}^{R}$, $H^{s}$, are the real components while $H_{u}^{I}$, $H_{d}^{I}$, $A^{s}$ are the imaginary components. Three scalar neutral Higgs bosons are obtained from $H_{u}^{R}$, $H_{d}^{R}$ and $H^{S}$, while a pseudoscalar Higgs boson is obtained from the combination of $H_{u}^{I}$ and $H_{d}^{I}$, and a second pseudoscalar Higgs is obtained from $A^{S}$. At tree-level, the symmetric squared mass matrix ($M_{s}^{2}$) of the neutral scalar Higgs bosons in the Higgs interaction basis $\lbrace H^{NSM},H^{SM},H^{S}\rbrace$\footnote{$H^{SM}$ refers to the eigenstate which has SM Higgs boson like couplings with the SM particles, $H^{NSM}$ refers to the eigenstate which has couplings similar to the additional scalar Higgs boson in MSSM and $H^{S}$ refers to the singlet-like scalar eigenstate.} is given by~\cite{ELLWANGER20101,PhysRevD.95.115036,PhysRevD.93.035013}

\begin{scriptsize}
\begin{equation}
\begin{pmatrix}
\left(m_{Z}^{2} - \frac{1}{2}\lambda^{2}\right) {\sin2\beta}^{2} + \frac{\mu}{\sin\beta \cos\beta}\left(A_{\lambda} + \frac{\kappa \mu}{\lambda}\right) & \left(\frac{1}{2} \lambda^{2}v^{2} - m_{Z}^{2}\right)\sin2\beta \cos2\beta & -\frac{1}{\sqrt{2}}\lambda v \cos2\beta\left(\frac{2\kappa\mu}{\lambda} + A_{\lambda}\right) \\
\left(\frac{1}{2} \lambda^{2}v^{2} - m_{Z}^{2}\right)\sin2\beta \cos2\beta & m_{Z}^{2} {\cos2\beta}^{2} + \frac{1}{2}\lambda^{2}v^{2}{\sin2\beta}^{2} & \sqrt{2}\lambda v \mu \left( 1 - \frac{A_{\lambda}}{2\mu}\sin2\beta - \frac{\kappa}{\lambda}\sin2\beta\right) \\
-\frac{1}{\sqrt{2}}\lambda v \cos2\beta\left(\frac{2\kappa\mu}{\lambda} + A_{\lambda}\right) & \sqrt{2}\lambda v \mu \left( 1 - \frac{A_{\lambda}}{2\mu}\sin2\beta - \frac{\kappa}{\lambda}\sin2\beta\right) & \frac{1}{4} \lambda^{2} v^{2} \sin2\beta \left(\frac{A_{\lambda}}{\mu}\right) + \frac{\kappa\mu}{\lambda} \left(A_{\kappa} + \frac{4\kappa\mu}{\lambda}\right) \\
\end{pmatrix}
\label{eqn:S_11_mass}
\end{equation}
\end{scriptsize}

Here, $\beta$ is defined as $\tan^{-1}\frac{v_{u}}{v_{d}}$, $m_{Z}$ represents the $Z$ boson mass and $v =\sqrt{v_{u}^{2} + v_{d}^{2}}$. Defining \begin{equation}M_{A}^{2} = \frac{\mu}{\sin\beta \cos\beta}\left( A_{\lambda} + \frac{\kappa\mu}{\lambda}\right),
\end{equation}
the elements of the symmetric $2\times 2$ pseudoscalar Higgs squared mass matrix ($M_{p}^{2}$) in the interaction basis $\lbrace A^{NSM},A^{S}\rbrace$\footnote{Here, $A^{NSM}$ refers to the pseudoscalar eigenstate with couplings similar to the MSSM pseudoscalar Higgs boson and $A^{S}$ refers to the singlet-like pseudoscalar eigenstate.} can be written as follows~\cite{ELLWANGER20101,PhysRevD.95.115036,PhysRevD.93.035013}:

\begin{small}
\begin{equation}
\begin{pmatrix}
M_{A}^{2} & -\frac{1}{\sqrt{2}}\lambda v \left(\frac{3\kappa\mu}{\lambda} - \frac{M_{A}^{2}}{2\mu}\sin2\beta\right) \\
-\frac{1}{\sqrt{2}}\lambda v \left(\frac{3\kappa\mu}{\lambda} - \frac{M_{A}^{2}}{2\mu}\sin2\beta\right) & \frac{1}{2}\lambda^{2} v^{2}\sin2\beta\left(\frac{M_{A}^{2}}{4\mu^{2}}\sin2\beta + \frac{3\kappa}{2\lambda}\right) - \frac{3\kappa A_{\kappa}\mu}{\lambda} \\
\end{pmatrix}
\label{eqn:A_22_mass}
\end{equation}
\end{small}

The Higgs mass eigenstates are defined in terms of the the Higgs interaction basis
\begin{equation}
H_{i} = s_{i1}H^{NSM} + s_{i2}H^{SM} + s_{i3}H^{S},~ i = 1,2,3
\end{equation}
\begin{equation}
A_{j} = p_{j1}A^{NSM} + p_{j2}A^{S}, ~ j = 1,2
\end{equation}
where, $s_{ik}$ and $p_{jk}$ are obtained by diagonalizing the corresponding mass squared matrices. 

In addition to the $3$ CP-even neutral Higgs states, $H_{1},H_{2},H_{3}$, one of which is identified with $H_{125}$ ($H_{1}$ being the lightest and $H_{3}$ being the heaviest), and, $2$ CP-odd neutral Higgs bosons, $A_{1},A_{2}$ ($A_{1}$ being the lighter one), the NMSSM Higgs spectrum also contains two charged Higgs boson. The tree-level mass of the charged Higgs bosons is given by
\begin{equation}
M_{H^{\pm}}^{2} = M_{A}^{2} + M_{W}^{2} - \frac{1}{2}\lambda^{2}v^{2}
\label{eqn:charged_higgs_mass}
\end{equation} 
where, $M_{W}$ represents the mass of $W$ boson. 

It can be observed from Eq.~(\ref{eqn:S_11_mass}-\ref{eqn:charged_higgs_mass}) that the tree-level Higgs sector of the NMSSM can be parametrized by $6$ input parameters:
\begin{equation}
\lambda,~\kappa,~A_{\lambda},~A_{\kappa},~\tan{\beta},~\mu
\label{Eqn:higgs_input_par}
\end{equation}  

Compared with the MSSM, the electroweakino sector of the NMSSM is phenomenologically richer and contains $5$ neutralinos and $2$ charginos. The neutralino mass matrix, in the basis of $\lbrace \tilde{B}, \tilde{W_{3}}, \tilde{H_{d}^{0}}, \tilde{H_{u}^{0}}, \tilde{S} \rbrace$ ($\tilde{B}$: bino, $\tilde{W^{3}}$: neutral wino, $\tilde{H_{d}^{0}}$ and $\tilde{H_{u}^{0}}$: neutral Higgsinos, $\tilde{S}$: singlino) is given by (following the notation of \cite{PhysRevD.95.115036})
\begin{small}
\begin{equation}
M_{\lspi}=
  \begin{pmatrix}
    M_{1} & 0 & -m_{Z}\sin\theta_{W}\cos\beta & m_{Z}\sin\theta_{W}\sin\beta & 0 \\
    0 & M_{2} & m_{Z}\cos\theta_{W}\cos\beta & -m_{Z}\cos\theta_{W}\sin\beta & 0 \\
    -m_{Z}\sin\theta_{W}\cos\beta & m_{Z}\cos\theta_{W}\cos\beta & 0 & -\mu & -\lambda v\sin\beta \\
    m_{Z}\sin\theta_{W}\sin\beta & -m_{Z}\cos\theta_{W}\sin\beta & -\mu & 0 & -\lambda v\cos\beta \\
    0 & 0 & -\lambda v\sin\beta & -\lambda v\cos\beta & 2\kappa v_{s} \\
    \label{eqn:neut_mass_matrix}
  \end{pmatrix}
\end{equation}
\end{small}
where, $M_{1}$ is the bino mass parameter, $M_{2}$ is the wino mass parameter, $m_{Z}$ is the mass of the $Z$ boson and $\theta_{W}$ is the Weinberg  angle.  

The neutralino mass eigenstates are given by
\begin{equation}
\lspi = N_{i1}\tilde{B} + N_{i2}\tilde{W^{3}} + N_{i3}\tilde{H_{d}^{0}} + N_{i4}\tilde{H_{u}^{0}} + N_{i5} \tilde{S}
\end{equation}
where, the $N_{ij}$'s are obtained by diagonalizing the neutralino mass matrix in Eq.~(\ref{eqn:neut_mass_matrix}). It follows from Eq.~(\ref{eqn:neut_mass_matrix}) that the neutralino sector at tree level is governed by the following input parameters:
\begin{equation}
M_{1},~M_{2},~\mu,~\tan\beta,~\lambda,~\kappa
\label{Eqn:electroweak_input_par}
\end{equation} 

It must be noted that the phenomenology of the electroweakino and Higgs sectors is modified from that in the MSSM due to additional parameters $\lambda, \kappa, A_{\lambda}, A_{\kappa}$.

Standard Big Bang cosmology with thermally produced neutralinos with $M_{\lspone} \lesssim 62.5~{\rm GeV}$ as the cold dark matter results in too large a density of cold dark matter, unless resonance enhancements of annihilation are operative in the early universe. 
The $Z$ boson and the SM-like Higgs boson enables the LSP neutralino to generate a correct or under-abundant relic density in the $M_{\lspone} \sim M_{Z}/2$ and $M_{\lspone} \sim M_{H_{125}}/2$ regions. At  lower LSP masses ($\lesssim M_{Z}/2$), efficient annihilation can be achieved via exchange of a light scalar Higgs ($H_{1}$) or a pseudoscalar Higgs ($A_{1}$), with mass, $M_{A_{1}/H_{1}} \sim 2 M_{\lspone}$. 

\section{Scanning the NMSSM parameter space}
\label{sec:parameter_space}

The first step of our analysis is to delineate the NMSSM parameter space region which is compatible with current constraints from collider and dark matter searches. A detailed discussion of constraints is found in Sec.~\ref{sec:constraints}. Note that a light $\lspone$ is completely consistent with all the current data in the NMSSM unlike in the MSSM and we intend to focus on the case of a light neutralino LSP with mass $M_{\lspone} < 62.5~{\rm GeV}$. We find that there is a significant region with $M_{\lspone} < 62.5$~GeV where the invisible decay of $H_{125}$ into a $\lspone$ pair is kinematically allowed. We zero in on this region and also explore prospects for invisible Higgs boson searches at future facilities.

The main focus of this study involves the exploration of the Higgs and electroweakino sectors of the NMSSM parameter space. The relevant input parameters which captures the physics of these two sectors are: $\lambda,~\kappa$, $A_{\lambda}$, $A_{\kappa}$, $\tan\beta$, $\mu$, $M_{1},~M_{2}$ (from Eq.~(\ref{Eqn:higgs_input_par}) and (\ref{Eqn:electroweak_input_par})), $M_{3}$: the gluino mass parameter, $A_{t}$,~$A_{t}$,~$A_{\tau}$: the stop, sbottom and stau trilinear coupling, and $M_{U^{3}_{R}}$,~$M_{D^{3}_{R}}$,~$M_{Q^{3}_{L}}$: the mass of the third generation squarks. The first and second generation squark masses, and the slepton masses are fixed at $3~{\rm TeV}$. 

The scenario with $\lambda,\kappa \approx 0$ is referred to as the effective MSSM~\cite{Ellwanger:2009dp}. From Eq.~(\ref{eqn:superpot}), it becomes clear that the singlet superfield $\hat{S}$ does not interact with the MSSM Higgs superfields, $\hat{H}_{u}$ and $\hat{H_{d}}$, when $\lambda \to 0$. In the limit of effective MSSM, the singlet scalar, the singlet pseudoscalar and the singlino-like neutralino does not couple with the MSSM sector. Thus, it is not possible to distinguish between the $\lambda,\kappa \to 0$ limit~(with
$\lambda\langle s\rangle$ fixed) of the NMSSM and the MSSM scenario. It is however possible for the singlino-like neutralino to be the LSP with mass $\sim 2 \kappa \langle s \rangle$. In order to allow the interaction between the singlet sector and the MSSM sector, it is essential to have a non-zero $\lambda$. For the case of a singlino-like LSP and $\lambda << 1$, the MSSM-like NLSPs would undergo cascade decay into singlino-like LSP + SM particles with a long lifetime which could also lead to potential LLP (long-lived particles) signatures, the study of which is beyond the scope of this work. Therefore, in this article, we focus on the regions of parameter space where $\lambda$ and $\kappa$ are not $\approx 0$ or $<<~1$.

The \texttt{NMSSMTools-5.3.1}~\cite{Djouadi:1997yw,Ellwanger:2004xm,
Belanger:2005kh,Ellwanger:2005dv,
Domingo:2007dx,Degrassi:2009yq} package is used to generate the particle spectrum, and to compute the couplings and branching fraction of the Higgses and the branching ratios of the SUSY particles. The presence of a large number of input parameters makes it difficult to find parameter space points in the $M_{\lspone} \lesssim 62.5~{\rm GeV}$ region and it becomes essential to choose an optimized scan range for the input parameters. Initially, a random scan of the parameter space is performed using the \texttt{NMSSMTools-5.3.1} package, for a wide range of input parameters. In the next step, the parameter space points, thus generated, are checked against the theoretical, collider and astrophysical constraints, implemented within \texttt{NMSSMTools-5.3.1}. Parameter space points are then chosen from the previous step to be used as seeds for the Markov Chain Monte Carlo (mcmc) scanning technique implemented in the \texttt{NMSSMTools-5.3.1} package. The parameter space points generated from the mcmc scan are distributed over the following range of input parameters: 

\begin{eqnarray}
0.01 < \lambda < 0.7,~10^{-5} < \kappa < 0.05,~3<\tan\beta < 40   \nonumber \\
100~{\rm GeV}< \mu < 1~{\rm TeV}, ~1.5~{\rm TeV}< M_{3} < 10~{\rm TeV} \nonumber \\
2~{\rm TeV} < A_{\lambda}<10.5~{\rm TeV},~-150~{\rm GeV} <A_{\kappa} <100~{\rm GeV} \nonumber \\
M_{1} = 2~{\rm TeV},~70~{\rm GeV} < M_{2} <2~{\rm TeV}\nonumber \\
 A_{t} = 2~{\rm TeV},~A_{b,\tilde{\tau}} = 0 ,~M_{U^{3}_{R}},M_{D^{3}_{R}},M_{Q^{3}_{L}} = 2~{\rm TeV},~M_{e^{3}_{L}},M_{e^{3}_{R}} = 3~{\rm TeV} 
\label{Eqn:scan_range}
\end{eqnarray}

The collider and astrophysical observables that constrain the points generated from this scan are detailed in the next section.

\section{Constraints}
\label{sec:constraints}
 
As we have just mentioned, our demand that the singlino is lighter than 62.5~GeV, along with the chosen range of input parameters, leads to $H_1$ and $A_1$ lighter than 125~GeV, so that $H_2$ then plays the role of the observed SM-like Higgs boson. The SM-like Higgs boson is required to lie within the mass range allowed by the measurements at the LHC and its couplings must be compatible with those measured at the LHC. The NMSSM parameter space considered in our study is further constrained by the low-energy flavor physics limits, LEP limits, 
searches of directly produced light Higgs bosons and measurements of the Higgs boson signal strengths  at the LHC, gluino searches, direct searches of electroweakinos in the $3l+\met$ and the $l^{+}l^{-}+\met$ final state as well as from direct and indirect DM searches.  These constraints are discussed in more detail below.

\begin{itemize}
\item{\textbf{Mass of SM-like Higgs boson:}}  
The combined measurement by the ATLAS and CMS collaborations has determined the Higgs boson mass to lie within $124.4-125.8~{\rm GeV}$ ($3\sigma$)~\cite{Aad:2015zhl}. Adopting a conservative approach, we require $H_{2}$ (we will henceforth also refer to this as $H_{125}$) to be within the interval $122-128~{\rm GeV}$, to allow for theoretical uncertainties in the Higgs boson mass computation~\cite{Allanach:2004rh,Heinemeyer:2007aq,Borowka:2015ura}. 

\item{\textbf{Limits from LEP:}} Measurements at LEP have excluded a chargino with mass below $M_{\chonepm} \lesssim 103.5~{\rm GeV}$~\cite{Abbiendi:2003sc}. 
This constraint, together with our choice $M_{1} = 2~{\rm TeV}$ (see Eq.~(\ref{Eqn:scan_range})) requires that the LSP below $62.5~{\rm GeV}$ is dominantly singlino, with only small higgsino, wino and bino admixture. We also impose the upper limit on the production cross section of $\lspone\lsptwo$ pair at $95\%$ C.L.($\sigma_{\lspone\lsptwo} \lesssim 0.1~{\rm pb}$~\cite{Abbiendi:2003sc}) for $|M_{\lsptwo} - M_{\lspone}| > 5~{\rm GeV}$ as well as upper limits on $e^{+}e^{-} \to ZH_{j}$ and $e^{+}e^{-} \to A_{i}H_{j}$ processes in various final states. The \texttt{NMSSMTools-5.3.1} framework was used to implement these constraints.

\item{\textbf{Upper bound on relic density:}} Results from the PLANCK Collaboration have put the DM relic density at $\Omega_{DM}^{obs.}h^{2} = 0.120\pm 0.001$~\cite{Aghanim:2018eyx}, and assuming a $2\sigma$ window, the relic density can fall within an interval of $0.118 - 0.122$. \texttt{MicrOMEGAs}~\cite{Belanger:2014vza,Belanger:2001fz} is used to compute the relic density of the LSP neutralino in the standard cosmological scenario, and adopting a conservative approach, we only exclude over-abundant DM, that is we require  $\Omega_{DM(\lspone)}^{obs.}h^{2} \leq 0.122$. This constraint would not apply if DM production entails non-standard cosmology~\cite{Moroi:1999zb,Gelmini:2006pw,Baer:2014eja}. In this work, the scenario of non-standard cosmology has not been considered. 

\item{\textbf{Flavor physics constraints:}} Constraints from the measurement of flavor physics observables offer sensitive probes of new physics scenarios. We impose the flavor physics constraints through bounds on the branching fraction of the rare decay modes $B \to X_{s}\gamma$, $B_{s} \to \mu^{+}\mu^{-}$ and $B^{+}\to \tau^{+}\nu_{\tau}$. Recent measurements obtain  $Br(B\to X_{s}\gamma) = (3.32\pm 0.16)\times 10^{-4}$~\cite{Amhis:2016xyh},  $Br(B_{s} \to \mu^{+}\mu^{-})=(3.0\pm 0.6^{+0.3}_{-0.2})\times 10^{-9}$~\cite{PhysRevLett.118.191801} and $Br(B^{+}\to \tau^{+}\nu_{\tau})=(1.06\pm 0.19)\times 10^{-4}$~\cite{Amhis:2016xyh}. In the current study, we use \texttt{micrOMEGAs-5.0.6}~\cite{Belanger:2018mqt,Belanger:2014vza,
Belanger:2004yn} to compute these branching fractions and we allow $2\sigma$ experimental uncertainty. Note that we do not use the \texttt{NMSSMTools-5.3.1} framework to implement the aforementioned $B$-physics constraints, rather, we impose the more recent bounds on the flavor physics observables discussed in this section. Additionally, the constraints on $\Upsilon(1s) \to H/A 
\gamma$, $\Delta M_{s}$ and $\Delta M_{d}$ are also imposed using the \texttt{NMSSMTools-5.3.1} package.  

\item{\textbf{Higgs signal strength measurements:}} 
The signal strength constraints in the $b\bar{b}$, $\tau^{+}\tau^{-}$, $ZZ$, $W^{+}W^{-}$ and $\gamma\gamma$ final states, derived from LHC Run-II data, are imposed using the \texttt{NMSSMTools-5.3.1} package. 

\item{\textbf{Invisible decay width of the Higgs boson:}} The CMS Collaboration has derived an upper limit on the total decay width of the observed $125~{\rm GeV}$ Higgs boson, using the dataset collected at $5.1~{\rm fb^{-1}}$ and $19.7~{\rm fb^{-1}}$, for $\sqrt{s}=7~{\rm TeV}$ and $\sqrt{s}=8~{\rm TeV}$, respectively. At $95\%$ C.L., the upper limit stands at $\Gamma_{H_{125}} < 22~{\rm MeV}$~\cite{Khachatryan:2014iha}. Correspondingly, in the current analysis, we require the total decay width of $H_{125}$ to lie below $22~{\rm MeV}$.

The total invisible branching fraction of the SM like Higgs can also be probed by directly searching for the invisibly decaying
Higgs boson through its production in association with a vector boson~\cite{Godbole:2003it,Ghosh:2012ep} and $jets$~\cite{Djouadi:2011aa} or vector boson fusion~\cite{Eboli:2000ze}. We have imposed upper limits 
obtained from such studies. The ATLAS Collaboration has also used LHC Run-II data ($\mathcal{L} \sim 140~{\rm fb^{-1}}$) to probe the invisibly decaying Higgs, produced via 
$VBF$ mode, and have set an upper limit of $13\%$~\cite{ATLAS-CONF-2020-008}. We must note that Ref.~\cite{ATLAS-CONF-2020-008} is a preprint and the published result from the ATLAS collaboration used LHC Run-I data, considered Higgs production via $WH_{125}$, $ZH_{125}$ and $VBF$ modes, and have set an upper limit of $25\%$~\cite{Aad:2015pla}. Similar search by the CMS Collaboration, using the entire Run-I data and $35.9~{\rm fb^{-1}}$ of the Run-II data, have derived an upper limit at $19\%$~\cite{Sirunyan:2018owy}. The prospects of probing the invisibly decaying Higgs boson at the future LHC has also been studied, see for example Refs.~\cite{Dawson:2013bba,Cepeda:2019klc}. In the context of our analysis, the invisible decay modes of  the SM-like Higgs ($H_{125}$) are: $H_{125} \to \lspone\lspone$, $H_{125} \to H_{1} H_{1} \to 4\lspone$ and $H_{125} \to A_{1}A_{1} \to 4\lspone$. We impose an upper limit of $13\%$ on the sum of these branching fractions. We note that the Higgs signal strength constraints also impose an indirect upper limit on the invisible branching fraction of the Higgs boson. The indirect limit can be comparable or, at times, even stronger than the direct upper limit, $viz.$ see Ref.~\cite{Barman:2017swy}. In our case, we find the indirect upper limit to be nearly comparable with the latest direct upper limit from ATLAS~\cite{ATLAS-CONF-2020-008} and our results do not change upon increasing the direct upper limit on the invisible branching fraction of the Higgs boson from $13\%$ to the published limit of $19\%$. 

\item{\textbf{Gluino searches at the LHC:}} Searches by the ATLAS and CMS collaborations using the LHC Run-II data collected at $\sim 36~{\rm fb^{-1}}$ and $\sim 139~{\rm fb^{-1}}$ of integrated luminosity has excluded gluinos up to $ 2~{\rm TeV}$~\cite{Aaboud:2018mna} and $ 2.2~{\rm TeV}$~\cite{Sirunyan:2019ctn}, respectively, at $95\%$ C.L. for a bino-like LSP with mass up to $\sim 600~{\rm GeV}$. Correspondingly, we impose a lower limit of $2.2~{\rm TeV}$ on the gluino mass.

\item{\textbf{Direct search of light Higgs bosons at the LHC:}} The ATLAS Collaboration has searched for a Higgs boson decaying into a pair of light spin-zero particles ($A_{1}$ or $H_{1}$), one of which further decay into $b\bar{b}$ and the other decays into a pair of muons~\cite{Aaboud:2018esj}. This search probed the mass range of $20~{\rm GeV} < M_{H_{1},A_{1}} < 60~{\rm GeV}$ using the LHC dataset collected at $\sqrt{s} = 13~{\rm TeV}$ corresponding to an integrated luminosity of $36.1~{\rm fb^{-1}}$ and derived upper limits on the production cross-section of the SM-like Higgs ($H_{125}$) normalized with its SM value ($\sigma_{H_{125}}/\sigma_{H_{SM}}$) times the branching fraction of $H_{125} \to A_{1}A_{1} \to 2b2\mu$. We have computed the value of $\sigma_{H_{125}}/\sigma_{H_{SM}} \times Br(H_{125} \to A_{1}A_{1} \to 2b2\mu)$ at $\sqrt{s}=13~{\rm TeV}$ for each point in the parameter scape and exclude points which exceed the upper limit.

The CMS collaboration has also searched for the exotic decay of a Higgs boson into two pseudoscalar Higgses in the  $2b2\tau$ final state~\cite{Sirunyan:2018pzn} and $2\mu2\tau$ final state~\cite{Sirunyan:2018mbx}. From the  $\sqrt{s} = 13~{\rm TeV}$ dataset collected with  ${\cal L}\sim 36~fb^{-1}$, upper limits were derived on $\sigma_{H_{125}}/\sigma_{H_{SM}} \times BR(H_{125} \to A_{1}A_{1} \to b\bar{b}\tau^{+}\tau^{-})$ and $\sigma_{H_{125}}/\sigma_{H_{SM}} \times BR(H_{125} \to A_{1}A_{1} \to \mu^{+}\mu^{-}\tau^{+}\tau^{-})$ over the mass range of $15~{\rm GeV} < M_{A_{1}} < 60~{\rm GeV}$.
We have excluded points which exceed these upper limits as well.

\item{\textbf{Direct electroweakino searches at the LHC:}} Numerous searches have been performed by both ATLAS and CMS, to probe the neutralinos and charginos, and limits have been derived on the mass of electroweakinos within simplified model scenarios~\cite{Alves:2011wf}. The most stringent limits are offered by the search channel $pp \to (\lspi \to Z/H_{125} \lspone)(\chonepm \to W^{\pm} \lspone)$ resulting in a $3l+\met$ final state. A recent study by CMS which looked into the $3l+\met$ final state, originating from the cascade decay of directly produced mass degenerate wino-like $\lsptwo\chonepm$ pair ($\lsptwo\to Z \lspone,~\chonepm \to W^{\pm} \lspone$), has excluded wino-like $\lsptwo,\chonepm$ up to $\sim 600~{\rm GeV}$ for $M_{\lspone} \lesssim 60~{\rm GeV}$ (Fig.~7  in \cite{Sirunyan:2018ubx}). It must be noted that this search assumes a $100\%$ branching ratio for $Br(\lsptwo\to Z \lspone)$ and $Br(\chonepm \to W^{\pm} \lspone)$. In regards to the parameter space considered in this study, $\lsptwo$ has additional decay modes namely $\lsptwo \to H_{1} \lspone$, $\lsptwo \to A_{1} \lspone$ and $\lsptwo \to H_{2} \lspone$, and, therefore, the assumption of $Br(\lsptwo \to \lspone Z) \sim 100\%$ does not always hold true. As a result, the wino exclusion limit derived in \cite{Sirunyan:2018ubx} 
 cannot be directly applied to the parameter space of interest. The limits derived in \cite{Sirunyan:2018ubx} have been translated to the case of higgsino-like NLSP's ($M_{2} > \mu$) in \cite{Ellwanger:2018zxt}, where the direct production of mass-degenerate higgsino-like $\lsptwo\chonepm$ and $\lspthree\chonepm$, is considered. This translation procedure enables recasting of the electroweakino search limits derived within a simplified model framework to any generic parameter space point. This translation scheme is implemented in the \texttt{NMSSMTools-5.3.1} package, and has been used to evaluate the impact of direct electroweakino search limits in the $3l+\met$ final state, performed using LHC $\sqrt{s}=13~{\rm TeV}$ dataset collected at $\sim 36~fb^{-1}$~\cite{Sirunyan:2018ubx}, on the parameter space under study.  

The ATLAS Collaboration has also probed the electroweakino sector via direct chargino pair-production in the $WW$-mediated opposite-sign di-lepton + $\met$ final state within a simplified model framework with wino-like $\chonepm$~\cite{ATLAS-CONF-2019-008}. This search was performed using the LHC Run-II data collected with ${\cal L} \sim 139~{\rm fb^{-1}}$. The limits obtained from this search exclude a wino-like chargino up to $M_{\chonepm} \sim 400~{\rm GeV}$ for a bino-like $\lspone$ with mass up to $M_{\lspone} \sim 90~{\rm GeV}$ 
assuming $100\%$ branching ratio in the $\chonepm \to W^{\pm} \lspone$ channel. This condition  holds true for the parameter space considered in this study since the only possible two body decay mode for $\chonepm$ is into a $W^{\pm}\lspone$ pair. Therefore these search limits  can  directly be translated in the current analysis by excluding dominantly wino-like charginos (wino-content in $\chonepm \gtrsim 90\%$) with mass $M_{\chonepm} \lesssim 400~{\rm GeV}$.

\item{\textbf{Direct detection constraints:}} The spin-independent (SI) and spin-dependent (SD) DM-nucleon scattering cross sections form the basis of DM direct detection experiments. 
\begin{figure}[!htpb]
\begin{center}
\includegraphics[scale=0.20]{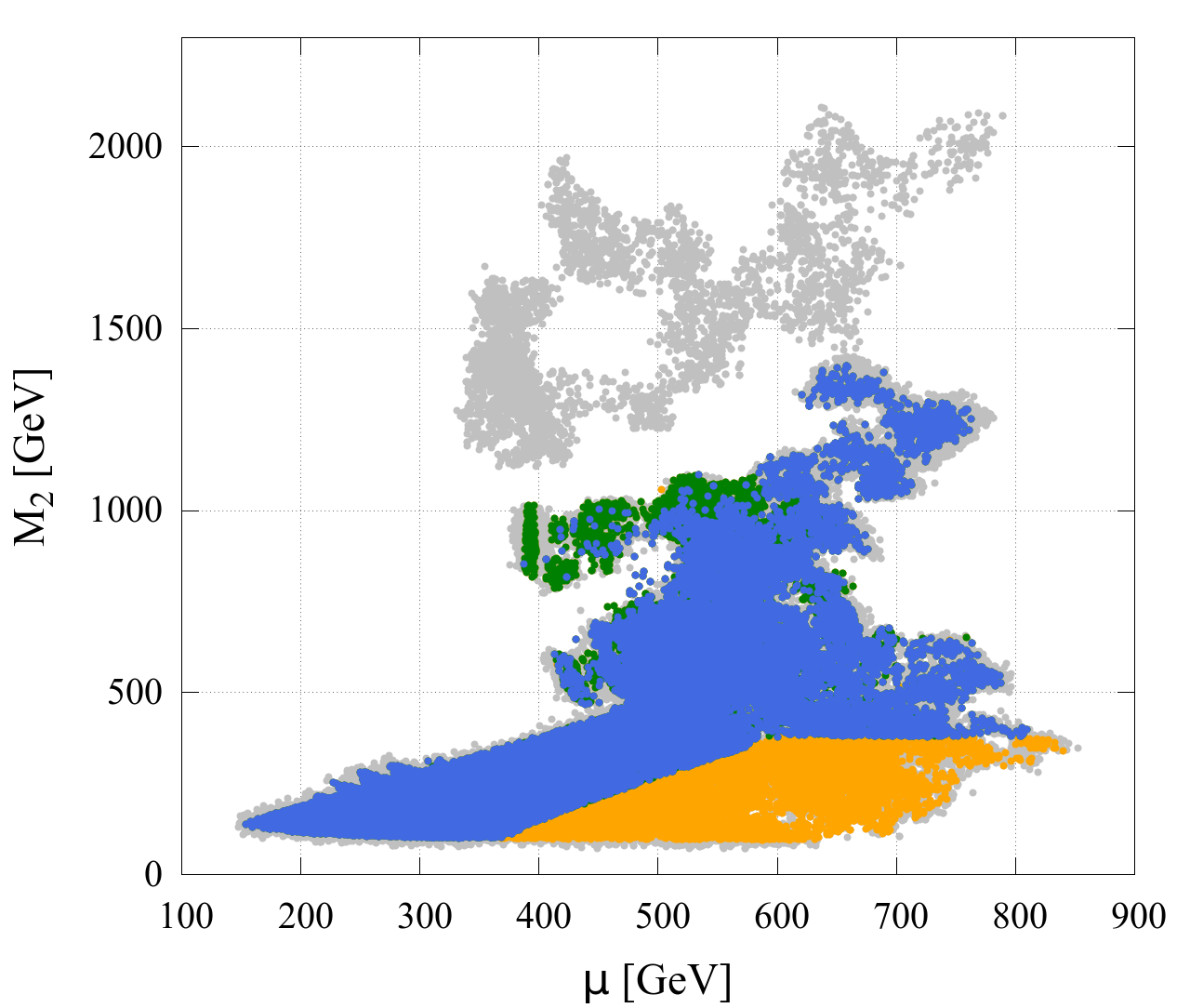}
\caption{Parameter space points are shown in the $\mu$-$M_{2}$ plane. The grey colored points are excluded by the LEP limits, LHC signal strength constraints, $B$ physics constraints, direct light Higgs searches at the LHC and sparticle searches at the LHC. The green colored points are excluded by the latest limits on $\sigma_{SI}$~(\cite{Aprile:2018dbl}), $\sigma_{\rm SD-neutron}$~(\cite{Aprile:2019dbj}) and $\sigma_{\rm SD-proton}$~(\cite{Amole:2019fdf}) at $90\%$ C.L. while the orange colored points are excluded by the current $95\%$ C.L. limits from direct electroweakino searches in $3l+\met$ and $2l+\met$ channels. The blue colored points represent the currently allowed parameter space.}
\label{fig:m2_mu_current}
\end{center}
\end{figure}
We impose the latest upper limits on spin-independent WIMP-nucleon cross-section~($\sigma_{SI}$; Xenon-1T~\cite{Aprile:2018dbl}), the spin-dependent WIMP-neutron cross-section~($\sigma_{\rm SD-neutron}$; Xenon-1T~\cite{Aprile:2019dbj}), and the spin-dependent WIMP-proton cross-section~($\sigma_{\rm SD-proton}$; PICO-60~\cite{Amole:2019fdf}) derived at $90\%$ C.L.. Among the points allowed by the latest upper limits on $\sigma_{SI}$, a small fraction~($\lesssim 4\%$) of points are excluded by the current upper limits on $\sigma_{\rm SD-proton}$ and $\sigma_{\rm SD-neutron}$. These points which are exclusively excluded by the current upper limits on the spin-dependent WIMP-neutron/proton cross-sections feature a large $\lambda~(\gtrsim 0.1)$ and a relatively large higgsino component is $\lspone$.
 

We illustrate in Fig.~\ref{fig:m2_mu_current} the impact of current constraints in the $M_{2}$-$\mu$ plane. The grey points are excluded by either the LEP limits,   LHC signal strength constraints, B physics constraints or direct light Higgs searches and sparticle searches at the LHC. In particular, the points generated in the large $M_2$ region are excluded by constraints on the Higgs sector including the requirement of having a Higgs around 125 GeV as well as the constraints on the Higgs signal strengths and from the LEP searches for light Higgses. Indeed in this region,  the one-loop corrections to the Higgs masses from charginos can be large ~\cite{ELLWANGER20101}, thus impacting both the mass spectrum of the scalars and their singlet/doublet content. Note also that the Higgs sector  depends strongly on the third generation squark sector, hence the impact of the Higgs constraints could be relaxed in a more general framework where parameters of the squark sector are allowed to vary, and not kept fixed as we have in Eq.~(\ref{Eqn:scan_range}).
We also indicate those points (green)  that are excluded by the latest upper limits 
from the direct detection experiments and those (orange) that are excluded by the current direct electroweakino searches limits discussed in Sec.~\ref{sec:constraints}. The blue colored points represent the set of currently allowed parameter space points with light neutralinos that we will use to study the impact of current indirect detection constraints and the reach of future experiments.

\item{\textbf{Indirect detection constraints:}} The indirect detection of  DM aims at identifying the visible signatures originating from interactions between the dark matter already present in the universe.
\begin{figure}[!htb]
\begin{center}
\includegraphics[scale=0.18]{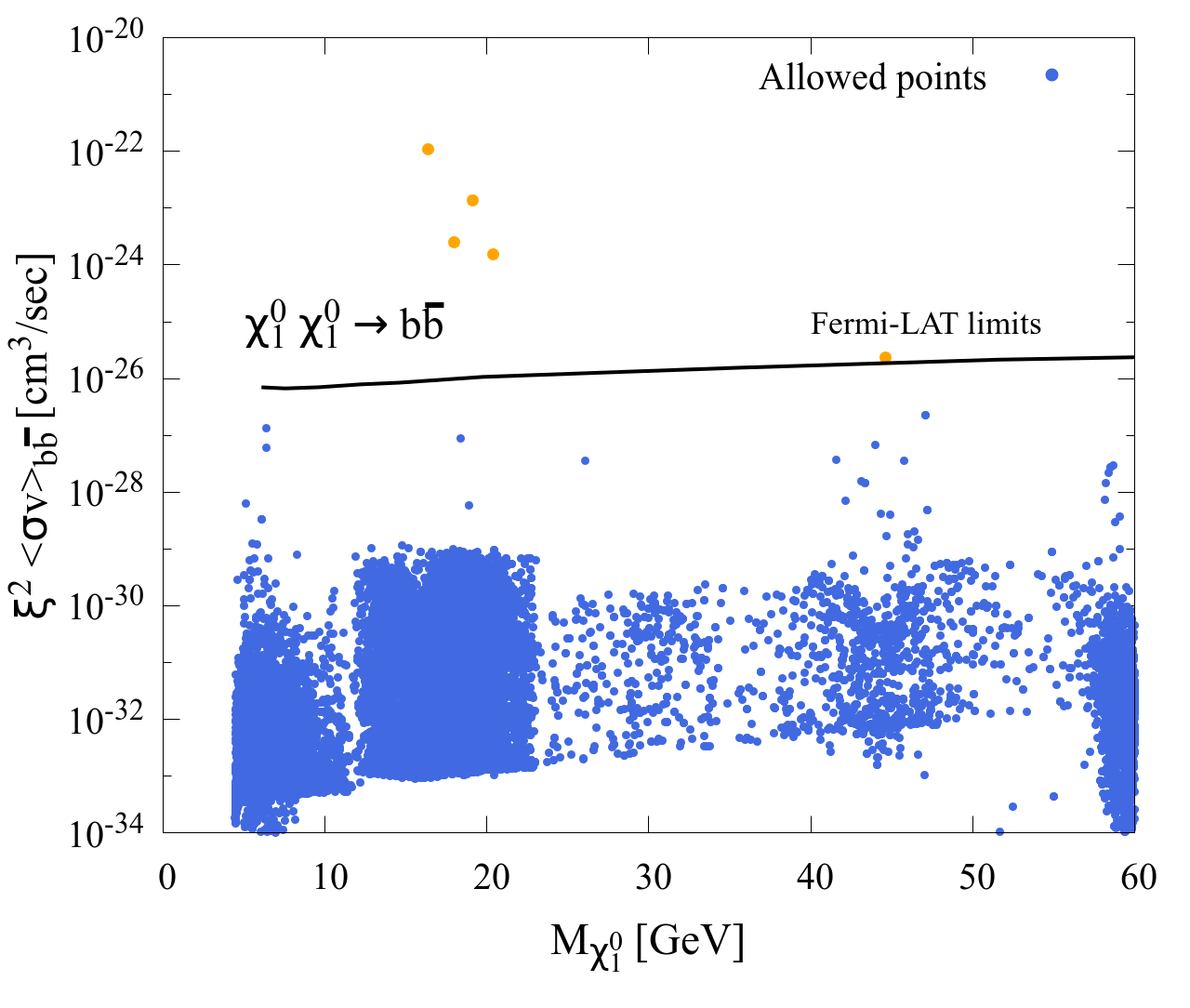}
\includegraphics[scale=0.18]{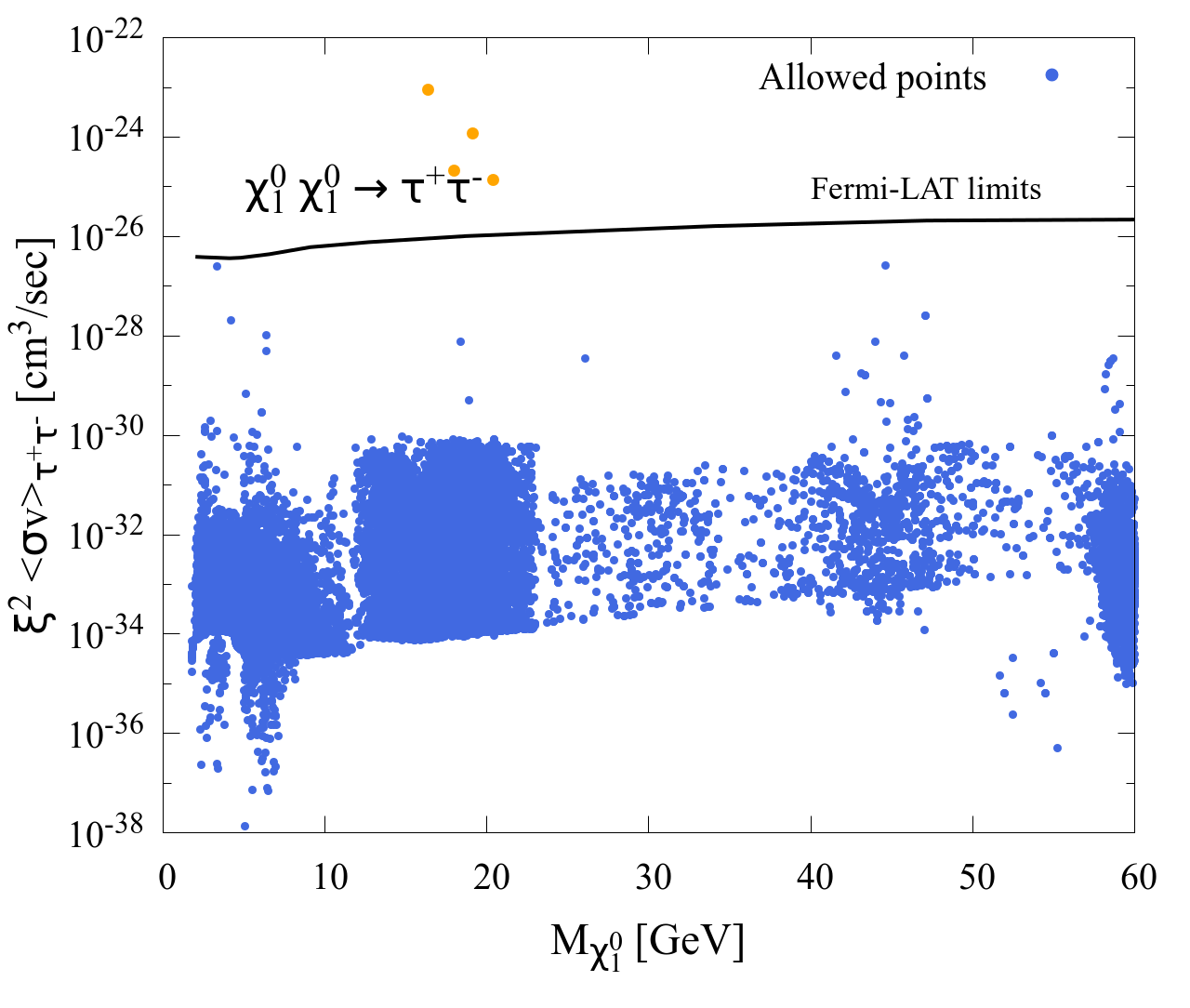}
\caption{The $\xi^{2}$ scaled thermally averaged DM annihilation cross section times velocity ($<\sigma v>$) in the $\lspone\lspone \to b\bar{b}$ (left) and $\lspone\lspone \to \tau^{+}\tau^{-}$ (right) channels has been shown in the y-axis against $M_{\lspone}$ in the x-axis. The parameter space points shown here are allowed by all the current constraints discussed till now in the present section. The black dashed line represents the $95\%$ C.L. upper limits on the $<\sigma v>$ from search for excess $\gamma$-ray emission in dwarf galaxies by the FERMI-LAT Collaboration~\cite{Fermi-LAT:2016uux}. The orange colored points are excluded by the current indirect limits, while the blue colored points are still allowed.}
\label{fig:indi_det_current}
\end{center}
\end{figure} 
The FERMI-LAT Collaboration has derived constraints on the thermally averaged DM annihilation cross section times the relative velocity between the DM candidates ($\left\langle\sigma v\right\rangle$) in $\lspone\lspone \to b\bar{b},~\tau^{+}\tau^{-}$ channels~\cite{Fermi-LAT:2016uux}. In the current study, we use \texttt{micrOMEGAs-5.0.6} to compute $\left\langle\sigma v\right\rangle$. Recall that the annihilation rate scales as the square of the local dark matter density. Assuming thermal DM, this then is scaled by  a factor $\xi = \Omega^{\rm thermal}_{\lspone}/0.12$. This then means that the detection rate will scale by $\xi^{2}$. After this scaling, we find that only a few points are constrained by current FERMI-LAT limits, see Fig.~\ref{fig:indi_det_current}. Moreover values of $\xi^2 \langle \sigma v \rangle$ can lie several orders of magnitude below current upper limits as well as below the typical value expected, $3 \times 10^{-26} {\rm cm}^3/{\rm sec}$,  even for points with $\xi$ close to unity.

The very small values of $\xi^2 \langle \sigma v \rangle$ that we found in our scenarios can be explained as follows.   As already noted, the relic density  constraint requires that DM annihilation be enhanced by $s$-channel  exchange of a boson ($Z$ or scalar/pseudoscalar) near a resonance. For low dark matter masses below $\sim 10$~GeV where one may also have to  worry about CMB constraints (from DM annihilation injecting electromagnetic energy during the era of matter-radiation decoupling), the enhancement is due to {\em very narrow}  $A_1/H_1$ resonances which have very small couplings and tiny values of $\Gamma/M_{A_1/H_1}$, $\sim 10^{-9}-10^{-7}$, resulting in a very strong velocity dependence of the annihilation cross section. Typically the thermal energy of the DM in the early universe will allow for the required enhancement of the cross-section while for the lower DM galactic velocities, there is no such enhancement. For the same reason, CMB constraints which involve even smaller velocities, will also not apply~\cite{Natarajan:2012ry}, even though the thermal relic density may be close to its observed value. The only exception is when the DM mass is very near $M_{A_1/H_1}/2$ or even slightly above, then the full resonant enhancement can occur at small velocities $v \sim 10^{-3} c$ while for the higher velocities in the early universe $v\sim 0.3c$ only the tail of the resonance will contribute. The fact that DM annihilation cross-section in the galaxy can be enhanced as compared with the one in the early universe -- the so-called Breit-Wigner enhancement -- was discussed in \cite{Ibe:2008ye,AlbornozVasquez:2011js}. It requires a very narrow resonance, as is  the case here for the light scalar/pseudoscalar here, as well as a fine-tuned mass difference $M_{\lspone}-M_{A_{1}/H_{1}}/2$. In our scan we find a few points which feature such a large cross-section and are thus excluded by FermiLAT. Following the same line of argument, we also expect only a few fine-tuned points to be subject to CMB constraints.

\end{itemize}

The parameter space points from Fig.~\ref{fig:indi_det_current} which are allowed by all constraints discussed above
will be referred to as the allowed parameter space points in the remainder of this paper.

\section{The thermal neutralino}
\label{Sec:NMSSM:thermal_neutralino}
Having discussed the relevant experimental constraints, we will move on to discuss the characteristic features of the NMSSM parameter space under study. At this point, we re-iterate that we remain confined to the region where the decay of SM-like Higgs boson into a pair of LSP neutralinos is kinematically feasible, and, correspondingly, we impose an upper bound on the mass of $\lspone$, $M_{\lspone} \leq 62.5~{\rm GeV}$. In addition, the parameter space is also subjected to the collider and astrophysical constraints discussed in Sec.~\ref{sec:constraints}. 

In the MSSM parameter space considered in our previous work ~\cite{Barman:2017swy}, it was observed that the imposition of the relic density constraint resulted in a lower bound ($M_{\lspone} \gtrsim 34~{\rm GeV}$) on the mass of the LSP neutralino. The lower bound on $M_{\lspone}$ was implied by the presence of only the Z boson and the Higgs boson as mediators for efficient $\lspone \lspone$ annihilation. Correspondingly, the MSSM parameter space points were confined to the $Z$ funnel ($\sim M_{Z}/2$) and Higgs-funnel ($M_{H_{125}}/2$) regions only. The NMSSM framework, on the other hand, features additional light Higgses ($H_{1}$ and $A_{1}$) with mass below $M_{Z}$, which can potentially mediate the annihilation of lighter LSP neutralinos with mass below the $Z$ funnel region. Correspondingly, we obtain allowed parameter space points with $M_{\lspone}$ as low as $\sim 1 ~{\rm GeV}$, which are consistent with the upper bound on relic abundance. 

\begin{figure}[!htb]
\begin{center}
\includegraphics[scale=0.19]{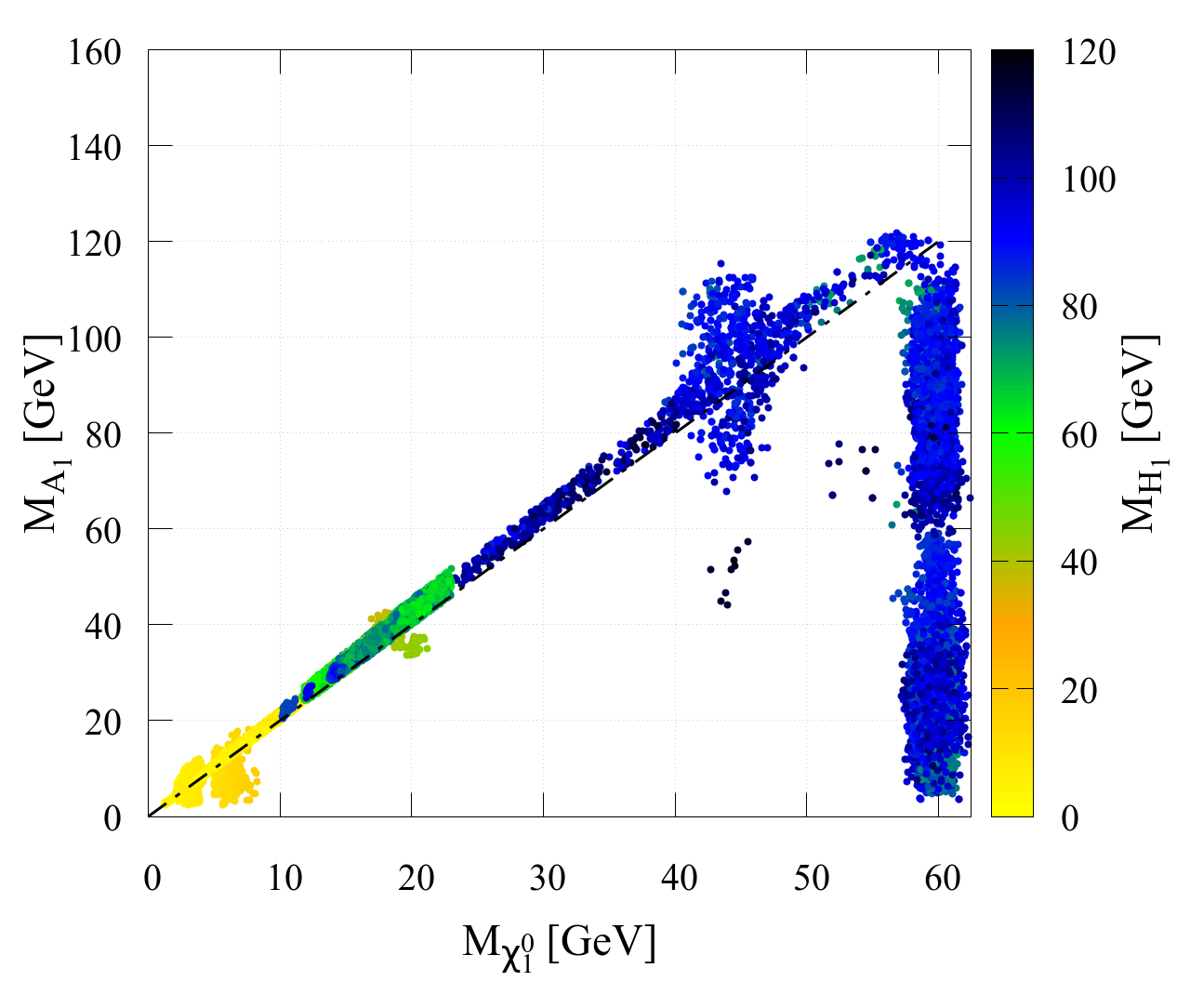}
\caption{ {\gb Correlation between the mass of the LSP neutralino ($M_{\lspone}$) and the mass of the light pseudoscalar Higgs ($M_{A_{1}}$) in the  $M_{\lspone}$ and $M_{A_{1}}$ plane.} The mass of lightest scalar Higgs ($M_{H_{1}}$) is represented along the color axis. The black dashed line corresponds to  $M_{A_{1}} = 2 M_{\lspone}$. The parameter space points shown here are allowed by all the constraints listed in Sec.~\ref{sec:constraints}.} 
\label{fig:ma1_mDM_masscorr}
\end{center}
\end{figure}

\begin{figure}
\begin{center}
\includegraphics[scale=0.20]{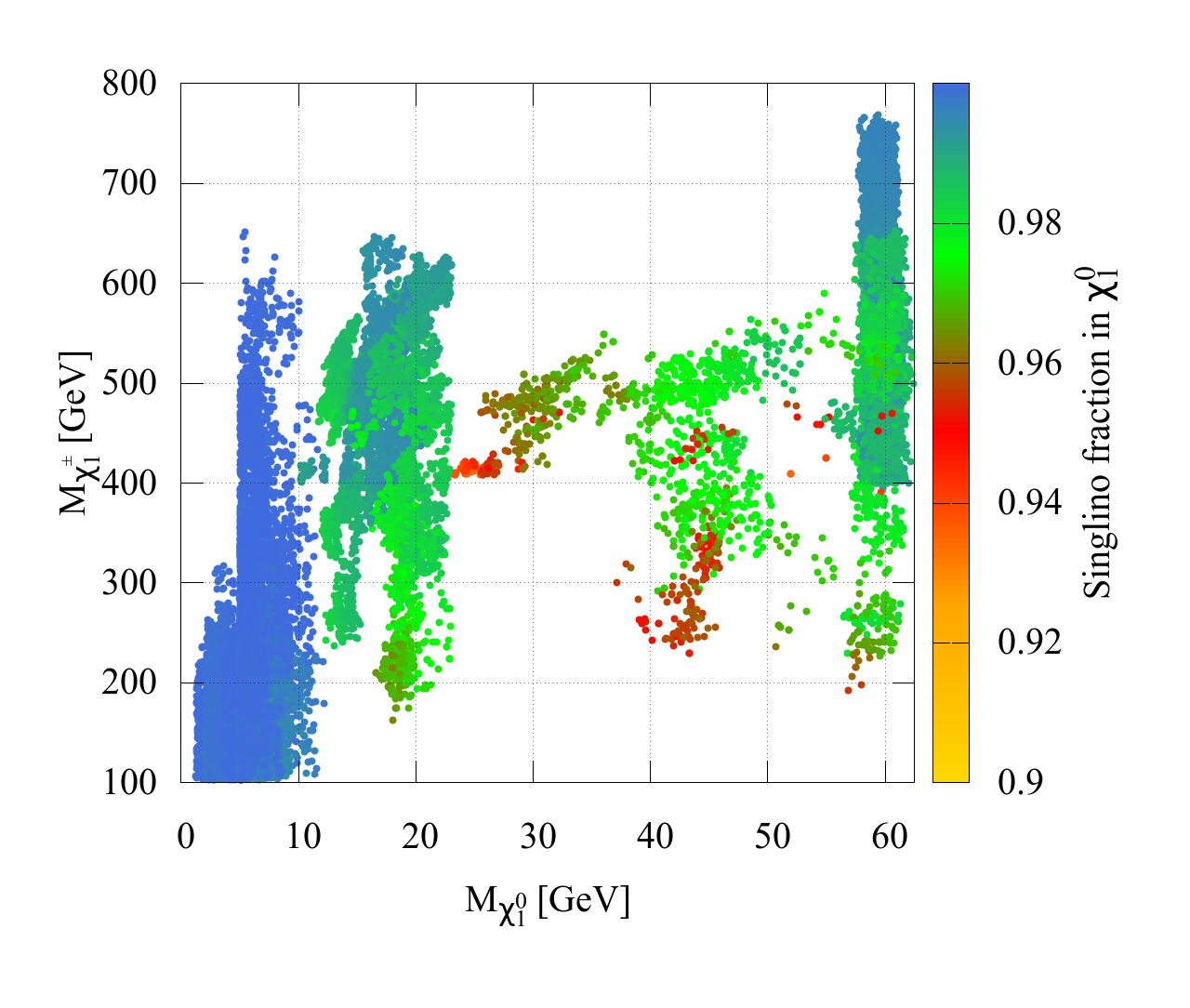}
\caption{The allowed parameter space points are shown in the $M_{\lspone}-M_{\chonepm}$ plane. The color palette represents the fraction of singlino content in the LSP $\lspone$.}
\label{fig:mchi1_mlsp}
\end{center}
\end{figure}

In order to emphasize upon the mass correlations between the LSP neutralino and light Higgs boson states, we plot the allowed parameter space points from Fig.~\ref{fig:indi_det_current} in the $M_{\lspone}$-$M_{A_{1}}$ plane shown in Fig.~\ref{fig:ma1_mDM_masscorr}. The color palette in Fig.~\ref{fig:ma1_mDM_masscorr} represents the mass of lightest scalar Higgs boson, $M_{H_{1}}$.  It is evident from Fig.~\ref{fig:ma1_mDM_masscorr} that the allowed  points below the $Z$ funnel region are mostly populated along the line  $M_{A_{1}} \sim 2 M_{\lspone}$, those that lie below this line actually  satisfy the condition $M_{H_{1}} \sim 2 M_{\lspone}$. Accord with the relic density constraint is then attained by efficient annihilation via the $A_{1}/H_{1}$ resonance.
We also exhibit the currently allowed parameter space points in the $M_{\lspone}-M_{\chonepm}$ plane in Fig.~\ref{fig:mchi1_mlsp}. The color palette in Fig.~\ref{fig:mchi1_mlsp} corresponds to the singlino content of the LSP.

The LEP limit on the chargino mass ($M_{\chonepm} > 103.5~{\rm GeV}$) implies a lower limit on the doublet higgsino and wino mass parameters and restricts them to values roughly above $\sim 100~{\rm GeV}$. Since our region of interest concerns $M_{\lspone} \leq M_{H_{SM}}/2$, the $\lspone$ has to be either bino-like or singlino-like. However, a bino or singlino dominated $\lspone$ can only satisfy the relic density upper bound if it undergoes  co-annihilation or pair annihilates via a resonance at roughly twice its mass. Within the NMSSM parameter space considered in the present analysis, co-annihilation is not feasible since there is a large mass splitting between the LSP and all other  sparticles. Thus $\lspone\lspone$ annihilation through an intermediate resonance remains the only feasible option with either a  light scalar and pseudoscalar Higgs states at mass $\sim 2M_{\lspone}$.
In the present analysis, $M_{1}$ has been fixed at $\sim 2~{\rm TeV}$. Therefore, the possibility of obtaining a bino-dominated $\lspone$ is eliminated. Moreover, a bino-dominated $\lspone$ requires a non-zero doublet higgsino admixture in order to couple with a Higgs state and since the chargino mass limits constrains the amount of higgsino fraction in $\lspone$, the relic 
density limit disfavors a bino-like LSP neutralino below the $Z$ funnel region\footnote{The possibility of Z funnel annihilation of a bino-like LSP also
occurs in the MSSM and has been examined in detail in Ref.~\cite{Barman:2017swy}. Since
our goal was to study new possibilities in the MSSM, we have taken the
bino to be very heavy in our study.}.
The singlino-like neutralino, on the other hand, can couple with either a singlet scalar or pseudoscalar Higgs with the coupling being proportional to $\kappa N_{15}^{2}$. Therefore the only possibility to obtain a LSP neutralino below $\lesssim 34~{\rm GeV}$ that satisifes  the relic density constraint, is to have a singlino-like $\lspone$ and at least one singlet  Higgs states at roughly twice its mass.

A study of the composition of light pseudoscalar and scalar Higgses within the allowed parameter space points show that $A_{1}$ and $H_{1}$ are dominantly singlet in nature\footnote{Similar features have also been reported in a recent work (see Ref.~\cite{Wang:2020dtb}).}, with singlet fraction $ \gtrsim 90\%$. Similarly, in accordance with the previous discussion, the LSP neutralinos are also found to be dominantly singlino in nature, as illustrated in Fig.~\ref{fig:mchi1_mlsp}.

\section{Prospects at future experiments}
\label{Sec:NMSSM:future_1}
This section is subdivided into two parts. In the first part, we study the projected reach of Xenon-nT (via bounds on $\sigma_{\rm SD-neutron}$), PICO-250~(via bounds on $\sigma_{\rm SD-proton}$) and at the FCC, ILC and CEPC (through the measurement of the invisible width of $H_{125}$). The second part investigates the scope of direct light Higgs boson searches in the $2b2\mu$, $2b2\tau$ and $2\mu2\tau$ channels at the future upgrades of LHC. This is, of course, in addition to continuing searches for signals from direct production of superpartners at the LHC, some of which are discussed in Sec.~\ref{Sec:NMSSM:future_2}.

\subsection{Reach of Xenon-nT, FCC-hh and the future electron-positron colliders}

\begin{figure}[!htb]
\begin{center}
\includegraphics[scale=0.19]{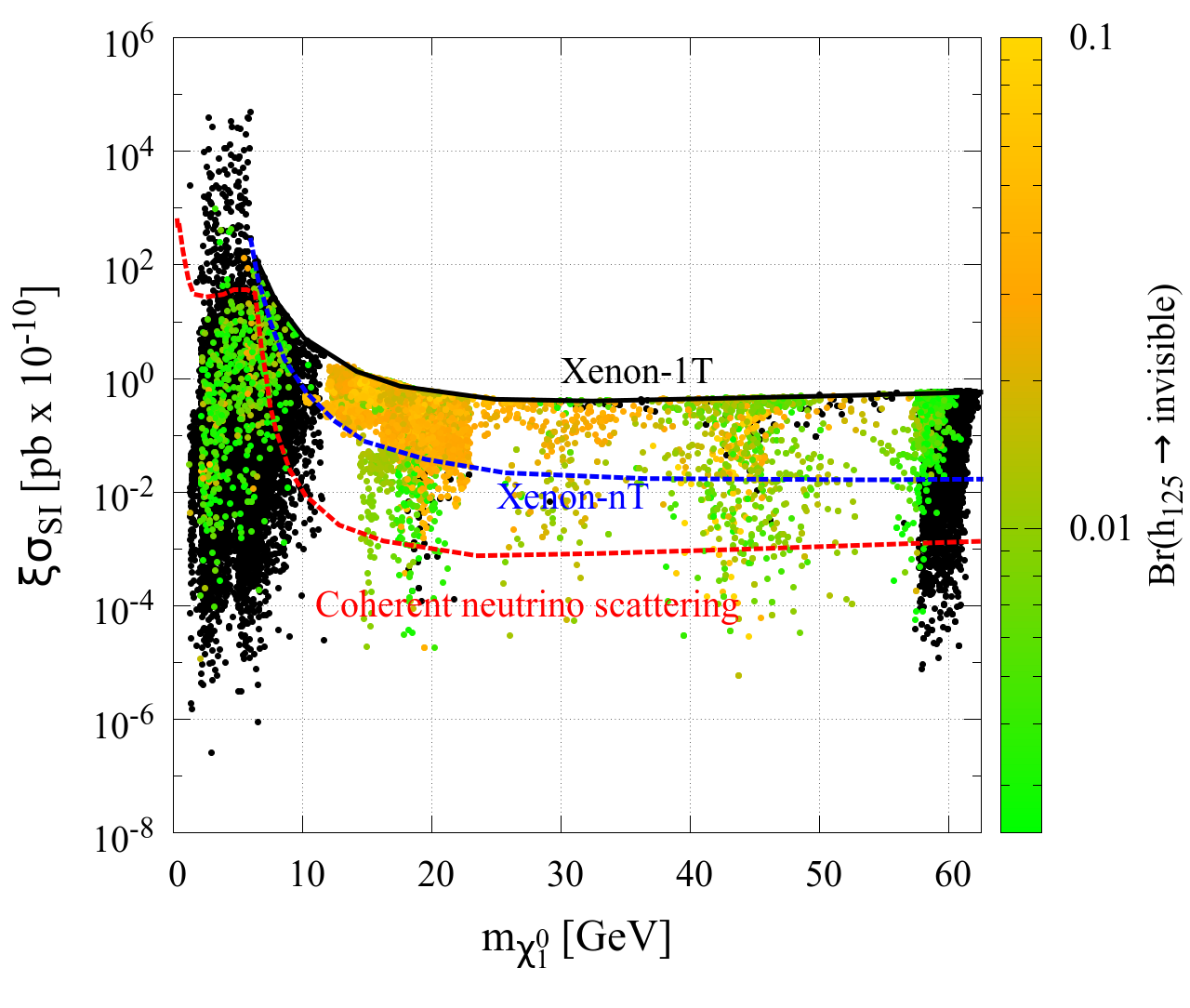}\includegraphics[scale=0.19]{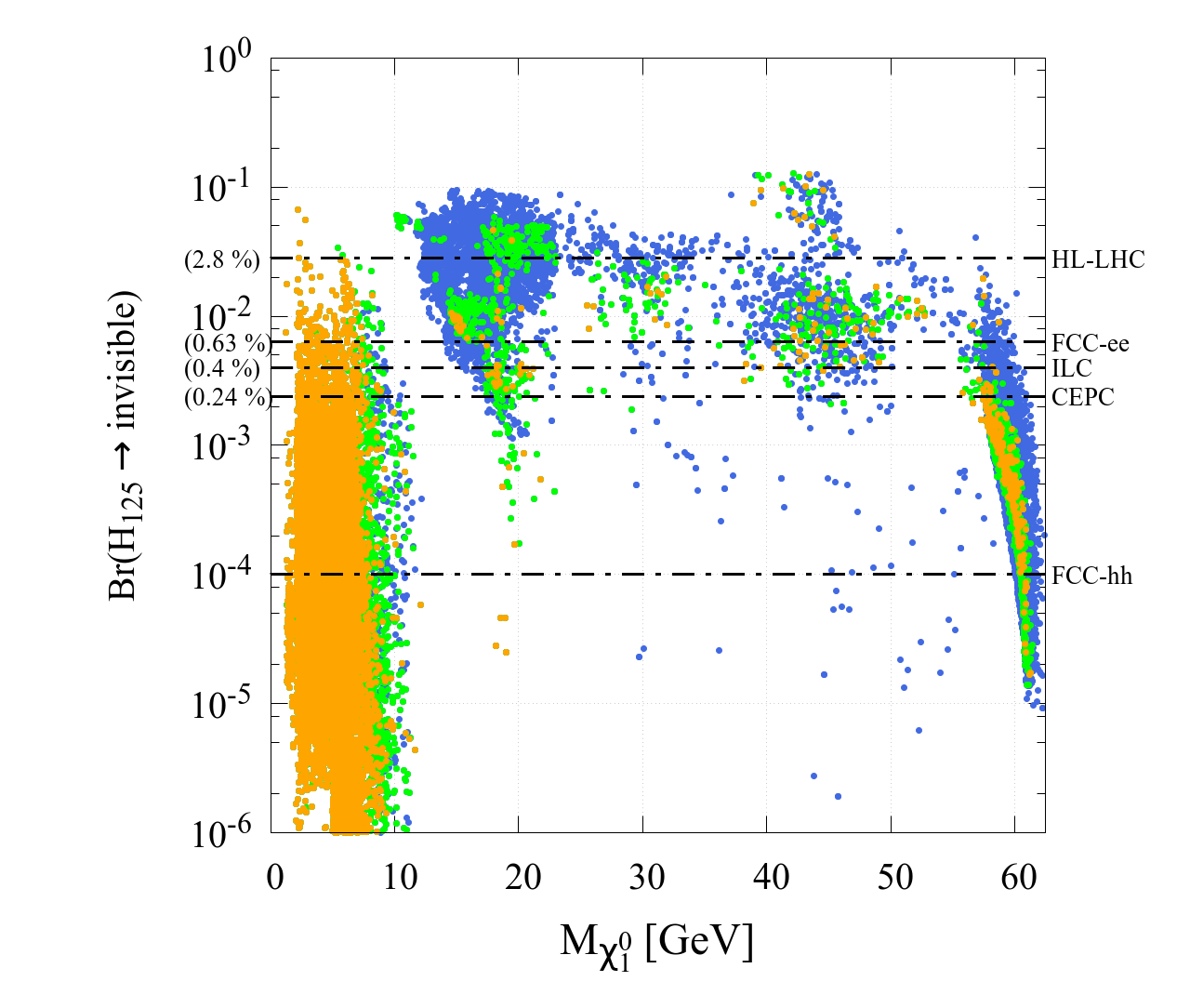}
\caption{Left: The currently allowed parameter space points from Fig.~\ref{fig:indi_det_current} are shown in the $\xi \sigma_{SI}-M_{\lspone}$ plane. The color palette represents the value of $Br( H_{125} \to invisible)$ computed by summing over all invisible decay modes of $H_{125}$. The black colored points have $Br(H_{125} \to invisible) \leq 0.24\%$, and therefore, are outside the projected reach of Higgs invisible width measurement capability of CEPC. The solid black, dashed blue and red dashed lines represent the current upper limits on $\sigma_{SI}$ from Xenon-1T, the projected upper limits on $\sigma_{SI}$ from Xenon-nT and the coherent neutrino scattering floor, respectively. Right: Allowed parameter space points are shown in the $Br(H_{125} \to invisible)-M_{\lspone}$ plane. The black dashed lines represent the projected capability of measuring the Higgs invisible branching fraction at(from top to bottom): HL-LHC~\cite{CMS-PAS-FTR-16-002}, FCC-ee~\cite{Cerri:2016bew}, ILC~\cite{Asner:2013psa}, CEPC~\cite{CEPCStudyGroup:2018ghi} and FCC-hh~\cite{LBorgonovi:2642471}. The orange colored points fall below the coherent neutrino scattering floor, while the green colored points are outside Xenon-nT's projected reach and above the coherent neutrino scattering floor. The blue colored points are within the projected reach of Xenon-nT.}
\label{Fig:ILC_CEPC_Higgs_SI_2}
\end{center}
\end{figure}

The projected reach of Xenon-nT~\cite{Aprile:2015uzo} in probing SI WIMP-nucleon cross-sections extends upto a factor of $\sim 50$ beyond the current limits from Xenon-1T in the DM mass range of $\sim 15~{\rm GeV}$ to $62.5~{\rm GeV}$. Correspondingly, a considerable region of the currently allowed parameter space is expected to lie within the future reach of Xenon-nT. We show the projected reach of Xenon-nT~(represented as blue dashed line) in the $\xi\sigma_{SI} - M_{\lspone}$ plane in Fig.~\ref{Fig:ILC_CEPC_Higgs_SI_2}~(left). The points shown in Fig.~\ref{Fig:ILC_CEPC_Higgs_SI_2}~(left) correspond to the currently allowed parameter space points. A significant fraction of those points fall within the projected reach of Xenon-nT while another large fraction remains out of reach. Moreover, an important fraction of these points, especially at low masses, lie below the coherent neutrino scattering floor and thus will remain out of reach of even larger detectors. In the same figure, the black  points represent those parameter space points for which  the invisible branching fraction of the SM-like Higgs boson, $Br(H_{125} \to invisible)< 0.24\%$, and thereby, are outside the projected reach of even the CEPC invisible Higgs boson branching fraction measurements~\cite{CEPCStudyGroup:2018ghi}. The color palette represents the value of  $Br(H_{125} \to invisible)$ for those allowed parameter space points which have $Br(H_{125} \to invisible)> 0.24\%$ and hence fall within the CEPC's projected capability of Higgs invisible branching fraction measurements. The invisible branching fraction has been computed by adding the contributions from the following decay modes:

\begin{itemize}
\item $H_{125} \to \lspone \lspone$
\item $H_{125} \to A_{1}A_{1} \to 4\lspone$
\item $H_{125} \to H_{1}H_{1} \to 4\lspone$
\item $H_{125} \to \lsptwo \lspone \to (\lsptwo \to H_{1} \lspone) \lspone \to (H_{1} \to \lspone \lspone) 2\lspone \to 4\lspone$ 
\item $H_{125} \to \lsptwo \lspone \to (\lsptwo \to A_{1} \lspone) \lspone \to (A_{1} \to \lspone \lspone) 2\lspone \to 4\lspone$ 
\end{itemize}

An important observation to be made from Fig.~\ref{Fig:ILC_CEPC_Higgs_SI_2} is that CEPC will be able to probe a small fraction of parameter space points in the $M_{\lspone} \lesssim 10~{\rm GeV}$ region which may be forever outside the reach of DM detectors unless directional detection technologies become available.

The complementarity between direct detection experiments and Higgs boson invisible width measurements at the future electron-positron colliders is further explored in Fig.~\ref{Fig:ILC_CEPC_Higgs_SI_2}~(right), where the allowed parameter space points are displayed in the $Br(H_{125} \to invisible)$ - $M_{\lspone}$ plane together  with  the future reach of HL-LHC ($\gtrsim 2.8\%$)~\cite{CMS-PAS-FTR-16-002}, FCC-ee ($\gtrsim 0.63\%$)~\cite{Cerri:2016bew}, ILC ($\gtrsim 0.4\%$)~\cite{Asner:2013psa}, CEPC ($\gtrsim 0.24\%$)~\cite{CEPCStudyGroup:2018ghi} and FCC-hh ($\gtrsim 0.01\%$)~\cite{LBorgonovi:2642471}. The color code illustrates whether points are within (blue) or outside (orange and green) the projected reach of Xenon-nT~(via $\sigma_{SI}$ measurements).  The orange colored points also lie below the coherent neutrino scattering floor. It can be observed from Fig.~\ref{Fig:ILC_CEPC_Higgs_SI_2} that a  majority of the  points in the $M_{\lspone} \gtrsim 10~{\rm GeV}$ region would be accessible via invisible Higgs boson branching fraction measurements at future $e^{+}e^{-}$ colliders, except when the invisible decay of $H_{125}$ is kinematically suppressed.

In the NMSSM, the spin-independent $\lspone$-nucleon interaction is mainly mediated through the CP-even Higgs bosons and the squarks. Since we have set the squark mass to a large value~($M_{U^{3}_{R}},M_{D^{3}_{R}}$,$M_{Q^{3}_{L}} =$~2~TeV), the main contribution to $\sigma_{SI}$ comes from the $t$-channel exchange of $H_{1}$ and $H_{125}$ between the $\lspone$ and the light flavored quarks. The respective amplitudes from the exchange of $H_{1}$ and $H_{125}$ can also undergo destructive interference for some specific choice of input parameters with substantial fine-tuning~\cite{Ellis:2000jd,Ellis:2000ds,Baer-2007,Huang:2014xua,Huang:2017kdh}. Such scenarios, also referred to as blind spots, can escape the current constraints on $\sigma_{SI}$ and also be compatible with the relic density limits. In NMSSM with singlino-like $\lspone$ LSP, these blind spots can arise when $M_{\lspone}/\left(\mu\sin\beta\right) \sim 1$~\cite{Baum:2017enm}.  In this work, although we do not probe the blind spot scenarios exclusively, however, we do obtain parameter points with small $\sigma_{SI}$ values~(which fall below the projected reach of Xenon-nT), many among which even fall below the neutrino scattering floor~(orange colored points in Fig.~\ref{Fig:ILC_CEPC_Higgs_SI_2}~(right)). We obtain such points over the entire range of $M_{\lspone} \lesssim M_{H_{125}}/2$  region as illustrated in Fig.~\ref{Fig:ILC_CEPC_Higgs_SI_2}. Spin-dependent measurements at the future experiments can be potential probes for these points in the $M_{\lspone} \gtrsim 15~{\rm GeV}$ region.

\begin{figure}[!htb]
\begin{center}
\includegraphics[scale=0.17]{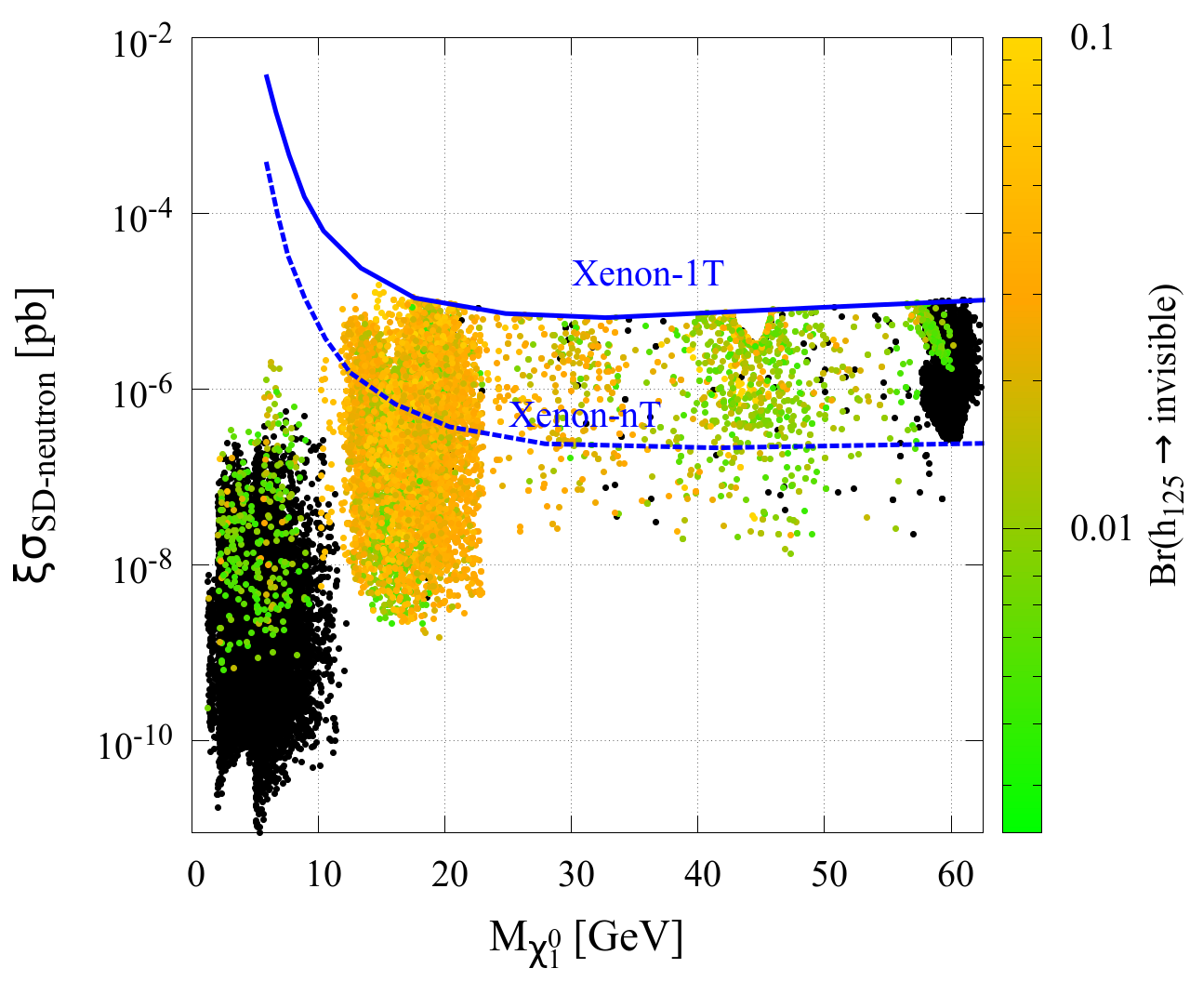}\includegraphics[scale=0.17]{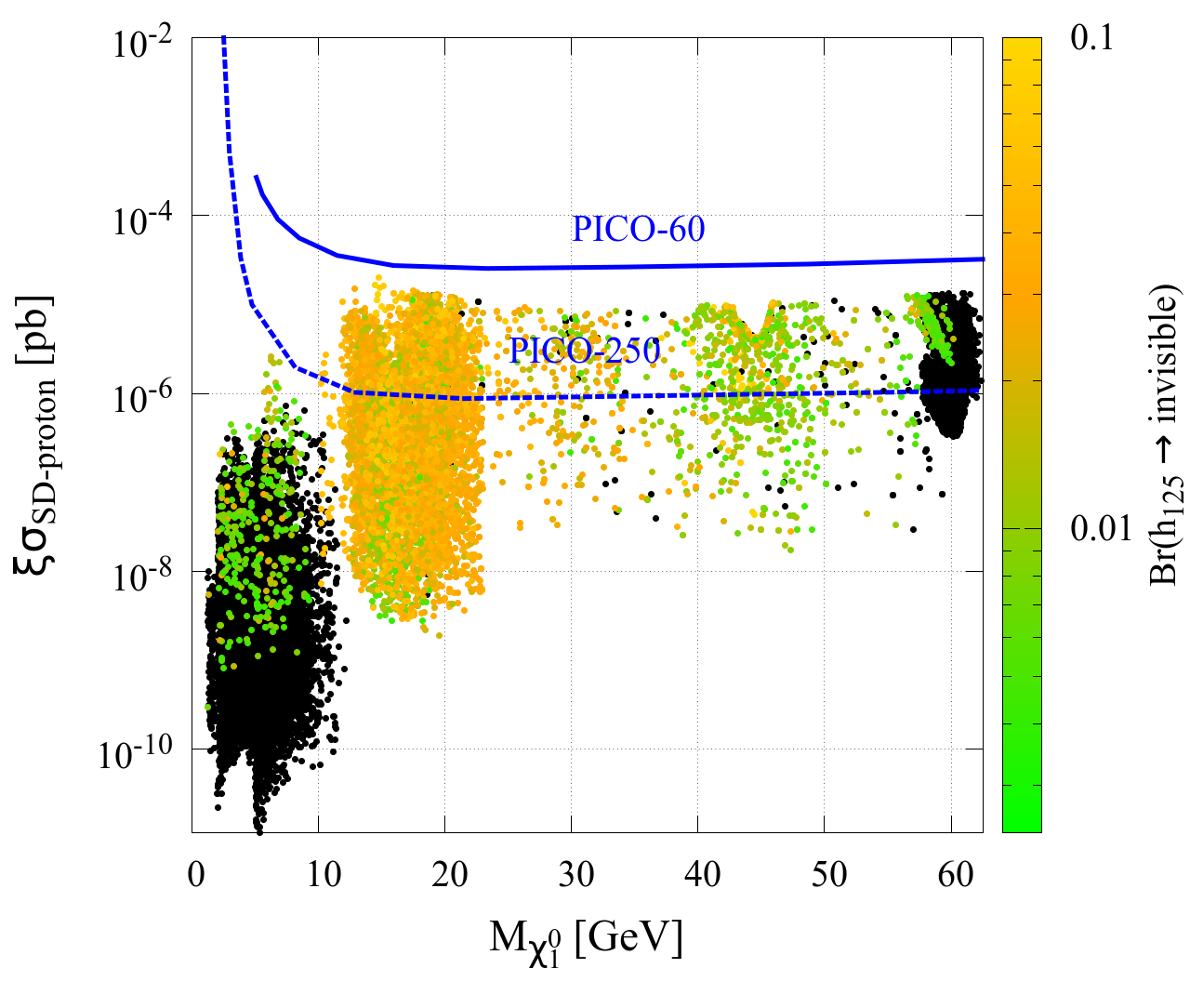}
\caption{Left: The currently allowed parameter space points are shown in the $\xi \sigma_{\rm SD-neutron}-M_{\lspone}$ plane. The solid blue and dashed blue lines represent the current upper limits on $\sigma_{\rm SD-neutron}$ from Xenon-1T~\cite{Aprile:2019dbj} and the projected upper limits on $\sigma_{\rm SD-proton}$ from Xenon-nT~\cite{Aprile:2020vtw}, respectively. Right: Allowed parameter space points are shown in the $\xi \sigma_{\rm SD-neutron}-M_{\lspone}$ plane. The solid blue and dashed blue lines represent the current upper limits on $\sigma_{\rm SD-proton}$ from PICO-60~\cite{Amole:2019fdf} and the projected upper limits on $\sigma_{\rm SD-proton}$ from PICO-250~\cite{Cushman:2013zza}, respectively. The color code is similar to that followed in Fig.~\ref{Fig:ILC_CEPC_Higgs_SI_2}~(left). }
\label{Fig:ILC_CEPC_Higgs_SI}
\end{center}
\end{figure}

\begin{figure}[!htb]
\begin{center}
\includegraphics[scale=0.2]{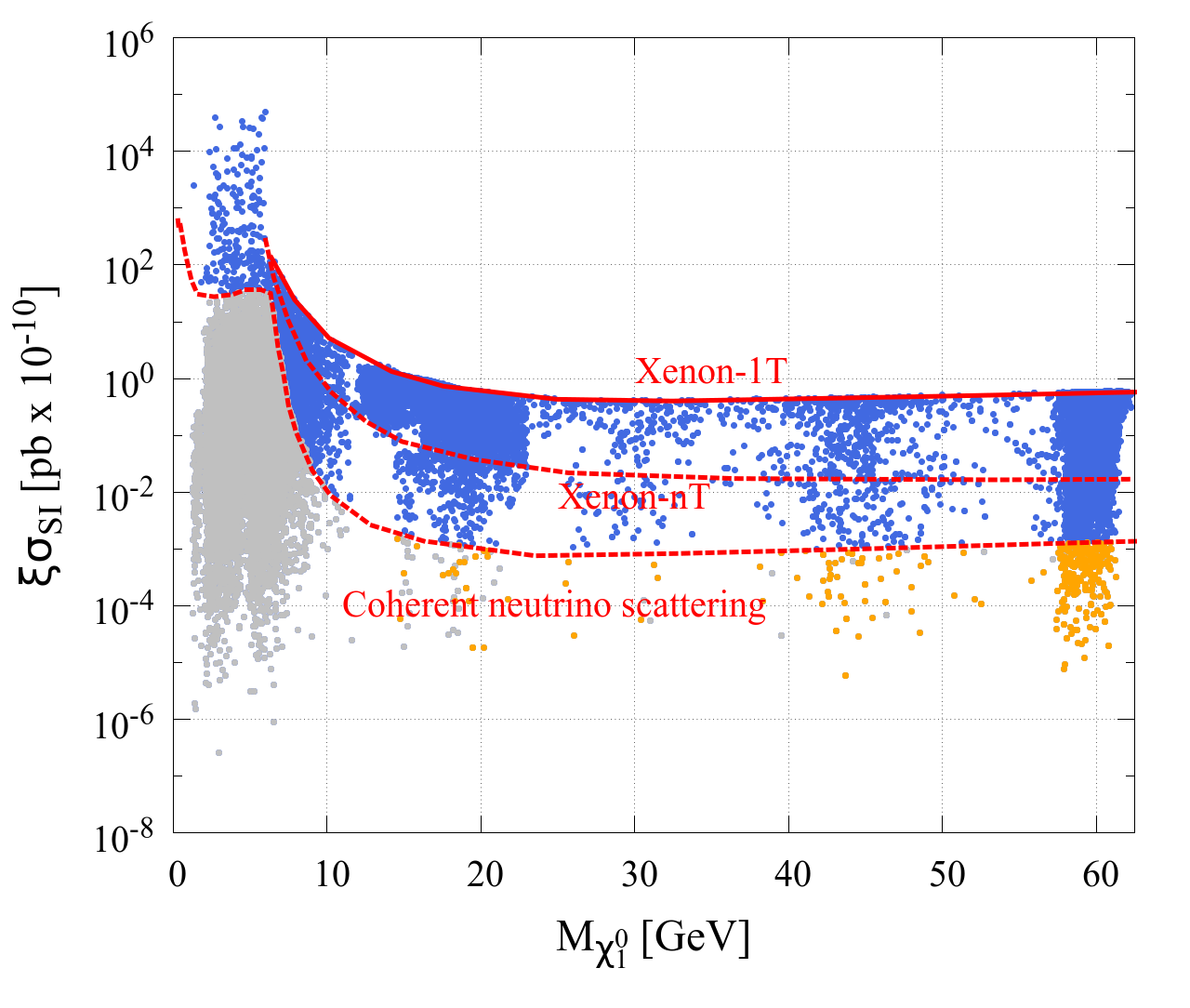}
\caption{The currently allowed parameter space points are shown in the $\xi\sigma_{SI}-M_{\lspone}$ plane. The blue points fall above the coherent neutrino scattering floor while the grey and orange points fall below. The orange points also fall within the projected reach of Xenon-nT~(via $\sigma_{\rm SD-neutron}$ measurements) or PICO-250~(via $\sigma_{\rm SD-neutron}$ measurements).}  
\label{Fig:SD_SI_corr}
\end{center}
\end{figure}

The projected reach of PICO-250~\cite{Cushman:2013zza} and Xenon-nT~\cite{Aprile:2020vtw} in probing SD WIMP-proton and WIMP-neutron cross-sections extend up to a factor of $\sim 75$ and $\sim 7$, respectively, beyond the current limits from PICO-60~\cite{Amole:2019fdf} and Xenon-1T~\cite{Aprile:2019dbj} in the DM mass range of $\sim 15~{\rm GeV}$ to $62.5~{\rm GeV}$. We illustrate the future reach of PICO-250 and Xenon-nT in probing the currently allowed parameter space points in the $\xi\sigma_{\rm SD-proton}$-$M_{\lspone}$ and $\xi\sigma_{\rm SD-neutron}$-$M_{\lspone}$ plane in Fig.~\ref{Fig:ILC_CEPC_Higgs_SI}~(left) and (right), respectively. Here as well, a significant region of currently allowed parameter space points fall within the projected reach of Xenon-nT as well as PICO-250 while another considerable fraction remains out of reach. In Fig.~\ref{Fig:SD_SI_corr} we again illustrate the currently allowed parameter space points in the $\xi\sigma_{SI}-M_{\lspone}$ plane. The grey and orange points fall below the coherent neutrino scattering floor~(they correspond to the orange colored points in Fig.~\ref{Fig:ILC_CEPC_Higgs_SI_2}~(right)) and will remain outside the projected coverage of any future $\sigma_{SI}$ measurement experiment. The orange points, however, fall within the projected reach of Xenon-nT~\cite{Aprile:2020vtw} and PICO-250~\cite{Cushman:2013zza} via $\sigma_{\rm SD-neutron}$ and $\sigma_{\rm SD-proton}$ measurements, respectively. Thus, it can be observed from Fig.~\ref{Fig:SD_SI_corr} that  spin-dependent measurements can provide coverage of parameter space points which lie even below the coherent neutrino scattering floor in the $M_{\lspone} \gtrsim 15~{\rm GeV}$ region.



\subsection{Direct searches for the light Higgs bosons at LHC luminosity and energy upgrades}

\begin{figure}[!htb]
\begin{center}
\includegraphics[scale=0.19]{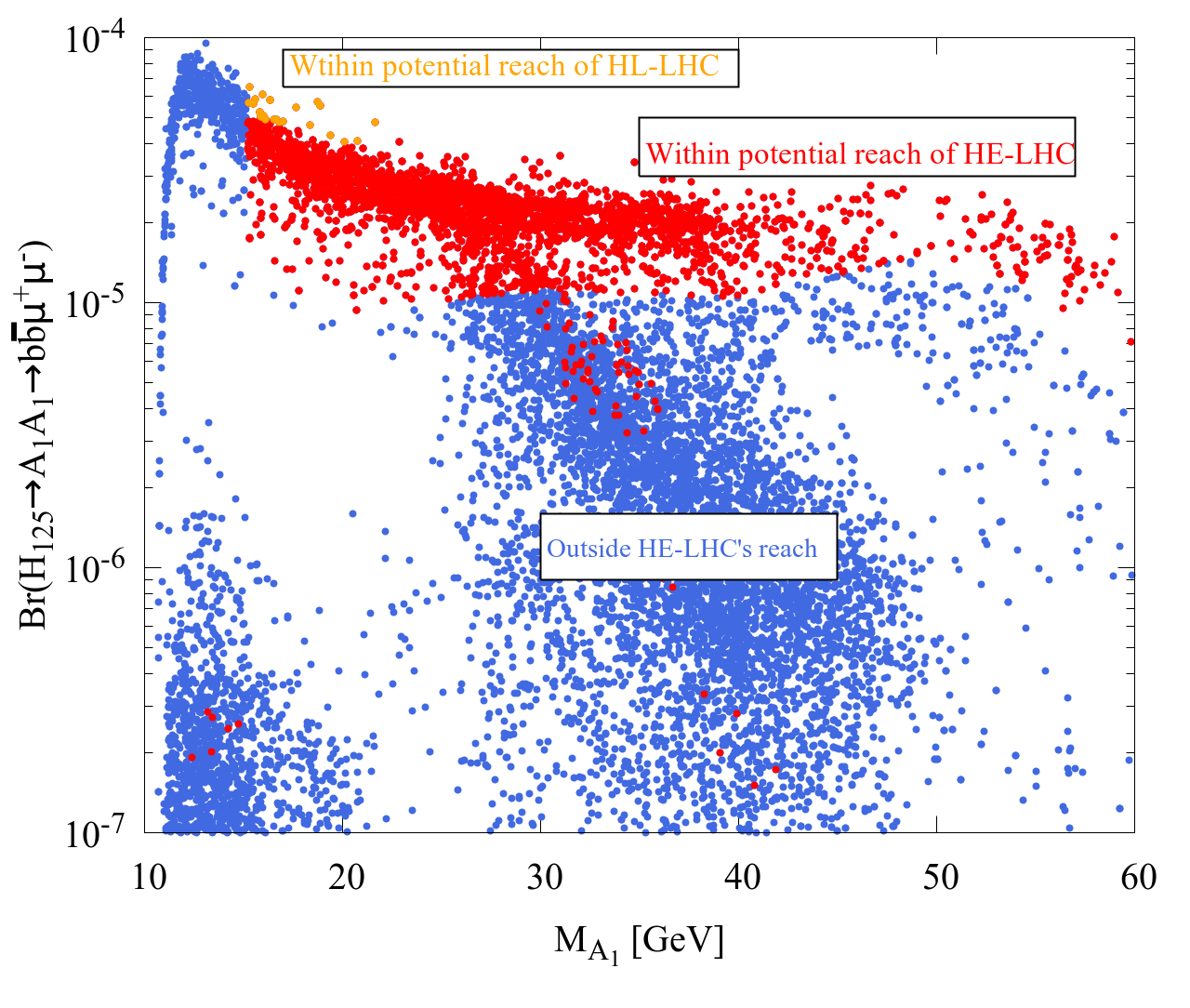}
\caption{ The allowed parameter space points (allowed by all constraints listed in Sec.~\ref{sec:constraints}) are shown in the $Br(H_{125} \to A_{1}A_{1} \to 2b2\mu)$ - $M_{A_{1}}$ plane. The orange and red colored points fall within the potential reach of direct light Higgs searches in the $H_{125} \to A_{1}A_1/H_1H_1 \to 2b2\mu$ channel, at HL-LHC and HE-LHC~\cite{Cepeda:2019klc}, respectively. The blue colored points fall outside the projected reach of light Higgs searches in $2b2\mu$ channel at HE-LHC.}
\label{fig:future_2b2mu}
\end{center}
\end{figure}

The projected sensitivity of HL-LHC and HE-LHC in probing light $A_{1}$ via direct searches in the $H_{125} \to A_{1}A_{1}/H_{1}H_{1} \to 2b2\mu$ channel has been discussed in \cite{Cepeda:2019klc}, where the  corresponding $95\%$ C.L. projection contours on $\sigma_{H_{125}}/\sigma_{H_{SM}} \times Br(H_{125} \to A_{1}A_{1}/H_{1}H_{1} \to 2b2\mu)$ have also been derived, however, under the assumption that $\sigma_{H_{125}}/\sigma_{H_{SM}} = 1$. We have translated these projection limits on to the currently allowed NMSSM parameter space of our interest and the projected reach has been shown in Fig.~\ref{fig:future_2b2mu}, where, the horizontal and vertical axes represent $M_{A_{1}}$ and $Br(H_{125} \to A_{1}A_{1} \to 2b2\mu)$, respectively. The orange colored points in Fig.~\ref{fig:future_2b2mu} are within the projected reach of HL-LHC at $95\%$ C.L. while the red colored points fall within the projected reach of HE-LHC at $95\%$ C.L. The blue colored parameter space points would be undetectable at HE-LHC via direct light Higgs searches in the $2b2\mu$ final state. These projections show that the discovery potential of light Higgs bosons produced via direct decays of $H_{125}$ is not very strong. Indeed no points are observable with $5\sigma$ sensitivity. We note though that we have made no attempt to optimize the analysis (which was based on an ATLAS analysis for $\sim 36~{\rm fb^{-1}}$) for either the increased luminosity or increased energy of the HE-LHC. Our conclusion should be viewed with caution.

Projected limits from direct light Higgs searches in the $2b2\mu$ channel, at the HL-LHC and the HE-LHC, has the potential to probe the $M_{A_{1}/H_{1}} \gtrsim 15~{\rm GeV}$ region (as shown in Fig.~\ref{fig:future_2b2mu}), however, it could be observed that a major region of the currently allowed parameter space lies outside its reach. It is to be noted that, along with light scalars and pseudoscalars, the particle spectrum of the parameter space under study also features light electroweakinos. Therefore, the question that arises next is, whether it would be possible to probe the parameter space through direct electroweakino searches at the future colliders. The next section intends to provide a conservative answer to this question.

\section{Future reach of direct electroweakino searches at HL- and HE-LHC}
\label{Sec:NMSSM:future_2}

The ATLAS Collaboration has made MSSM projections for the wino reach of the HL-LHC via trilepton searches $pp \to \lsptwo\chonepm$, assuming that $\chonepm \to \lspone + (W^\pm \to l^{\prime\pm}\nu)$ ($l^{\prime} = e,\mu,\tau$ and $\nu = \nu_{e},\nu_{\mu},\nu_{\tau}$) and $\lsptwo \to \lspone + (Z \to l^{\prime +}l^{\prime -})$ or $\lsptwo \to \lspone + (H_{125} \to WW^{*} \to l^{\prime +}l^{\prime -} + \nu\nu, H_{125} \to  \tau^{+}\tau^{-})$\footnote{Here both taus decay leptonically.} \cite{ATL-PHYS-PUB-2014-010,ATL-PHYS-PUB-2018-048}. Hereafter, we will refer to these as the $WZ$- and $WH_{125}$-mediated channels, respectively. In Sec.~\ref{subsec:ewino_hllhc}, we perform our own analysis of the wino reach in the $WZ$-mediated channel, and compare our results with those in \cite{ATL-PHYS-PUB-2018-048}. Our purpose in doing so is to assess how our ``theorist's computation" does, {\it vis a vis} the ATLAS analysis. We then project the reach for {\em doublet higgsino production} via the $WZ$- as well as the $WH_{125}$-channels, assuming that these higgsinos directly decay to a lighter LSP.\footnote{For direct higgsino searches via the $WH_{125}$-mediated trilepton channel, we closely follow the analysis strategy of \cite{ATL-PHYS-PUB-2014-010}. It may be possible to further optimize the analysis for the higher luminosity, and also the higher energy that may be available in the future.} This is, of course, not possible in the MSSM with our assumption of a heavy bino. We perform this analysis as a prelude to our goal of mapping out how well experiments at the HL-LHC will be able to probe the NMSSM parameter space with a light LSP even though this analysis covers a much larger range of masses for the LSP. In Sec.~\ref{sec:case1_ewino_hllhc}, we translate the discovery projections (and exclusion reach) derived in Sec.~\ref{subsec:ewino_hllhc} to the currently allowed NMSSM parameter space of interest. In Sec.~\ref{subsec:ewino_helhc}, we perform another collider study to obtain the discovery reach (and exclusion regions) of higgsinos at the HE-LHC. These are then translated in Sec.~\ref{sec:ewino_helhc} to the corresponding reach in the parameter space of the NMSSM.

A simplified model with degenerate higgsino-like $\lsptwo$, $\lspthree$, $\chonepm$, and bino-like $\lspone$ is considered while evaluating the HL-LHC and HE-LHC projections. Within the simplified framework, the relevant branching fractions: $Br\left(\lspthree/\lsptwo \to \lspone Z \right)$ or $Br\left(\lspthree/\lsptwo \to \lspone H_{125} \right)$, and, $Br\left(\chonepm \to \lspone W^{\pm}\right)$, are assumed to be $100\%$. Note that in the NMSSM, this assumption remains valid for the lighter chargino but not for the heavier chargino or for neutralinos. Indeed the $\lsptwo$, $\lspthree$ and $\lspfour$ are an admixture of the winos and higgsinos while the LSP is a singlino resulting in variedly different values of the relevant branching fractions. Moreover, additional decay channels into light Higgses are possible. 
In order to translate the projected discovery/exclusion regions on to the NMSSM parameter space, we first map out the efficiency grid for the corresponding search strategy in the doublet-higgsino LSP mass plane. Since this grid is determined largely by the kinematics, the details of the composition of the parent higgsinos or the daughter LSP are completely irrelevant: the efficiency grid is determined by the kinematics and the branching fractions of the parent higgsinos to decay to the LSP and the associated boson ($W$, $Z$ or $H_{125}$).

\subsection{Direct electroweakino searches at HL-LHC}
\label{subsec:ewino_hllhc}

As stated previously, the aim of the present subsection is twofold: to derive the projected discovery reach and exclusion range for direct higgsino searches ($pp \to \lsptwo\chonepm + \lspthree \chonepm$) in the $WZ$ mediated and $WH_{125}$ mediated $3l+\met$ final state at the HL-LHC, and, map out the efficiency grid for the $WZ$ and $WH_{125}$-mediated channels that we will use to delineate the region of NMSSM parameter space that can be probed at the luminosity and energy upgrades of the LHC.

\subsubsection{$WZ$-mediated $3l~+~\met$ channel at HL-LHC} \label{sec:wz_hllhc}

We begin by considering the trilepton signal produced via $pp \to \lsptwo\chonepm + \lspthree \chonepm \to (\lspthree/\lsptwo \to \lspone Z)(\chonepm \to W^{\pm} \lspone)$. We assume that the $W$ and $Z$ bosons are close to their mass shell and focus on the trilepton +$\met$ final state from their leptonic decays to electrons, muons and taus.
We use \texttt{Pythia-6}~\cite{Sjostrand:2001yu,Sjostrand:2014zea} to simulate the signal events for different values of $M_{\lspone}$ and $M_{\lsptwo,\lspthree,\chonepm}$ ($M_{\lspone}$ is varied from 0 - 800 GeV with a step size of $30$ GeV while $M_{\lsptwo}$($=M_{\lspthree} =M_{\chonepm}$) is varied from $100$ GeV to $1300$ GeV with a step size of $30$ GeV, with an additional condition: $M_{\lsptwo,\lspthree} - M_{\lspone} > M_{Z}$, in order to ensure on-shell production of $W$ and $Z$ bosons.). The direct higgsino production cross section ($\sigma_{pp \to \lsptwo\chonepm+\lspthree\chonepm}$) has been computed at next to leading order (NLO) using \texttt{Prospino}~\cite{Beenakker:1994an,Beenakker:1995fp,
Beenakker:1996ed}, while the branching fraction of $\lsptwo/\lspthree \to Z\lspone$ and $\chonepm \to W^{\pm}\lspone$ has been considered to be $100\%$.

\begin{figure}
\begin{center}
\includegraphics[scale=0.19]{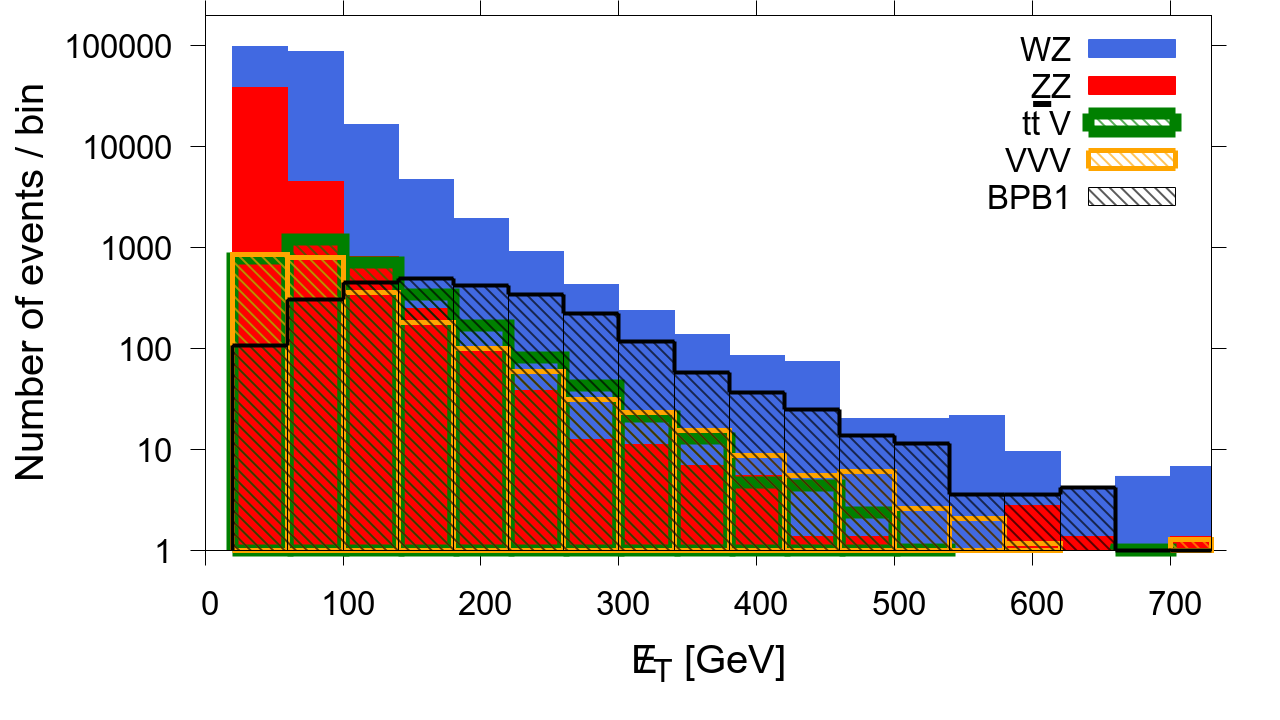}\includegraphics[scale=0.19]{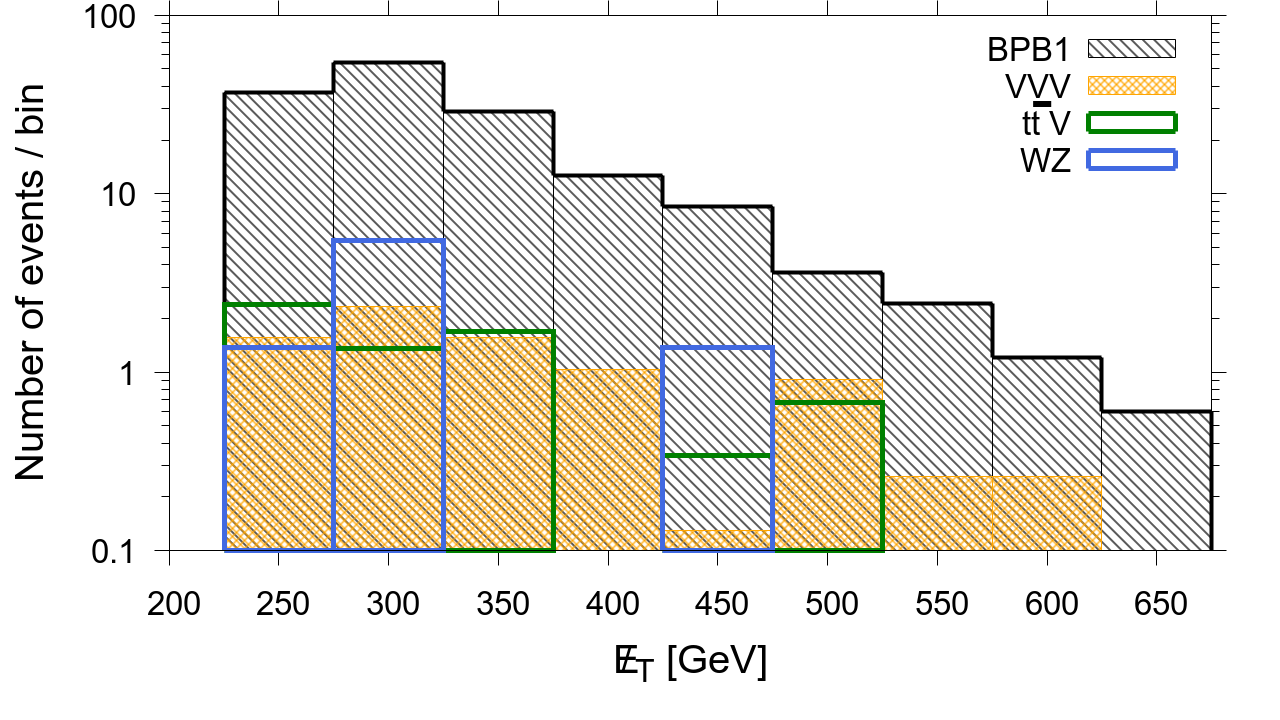}\\
\includegraphics[scale=0.19]{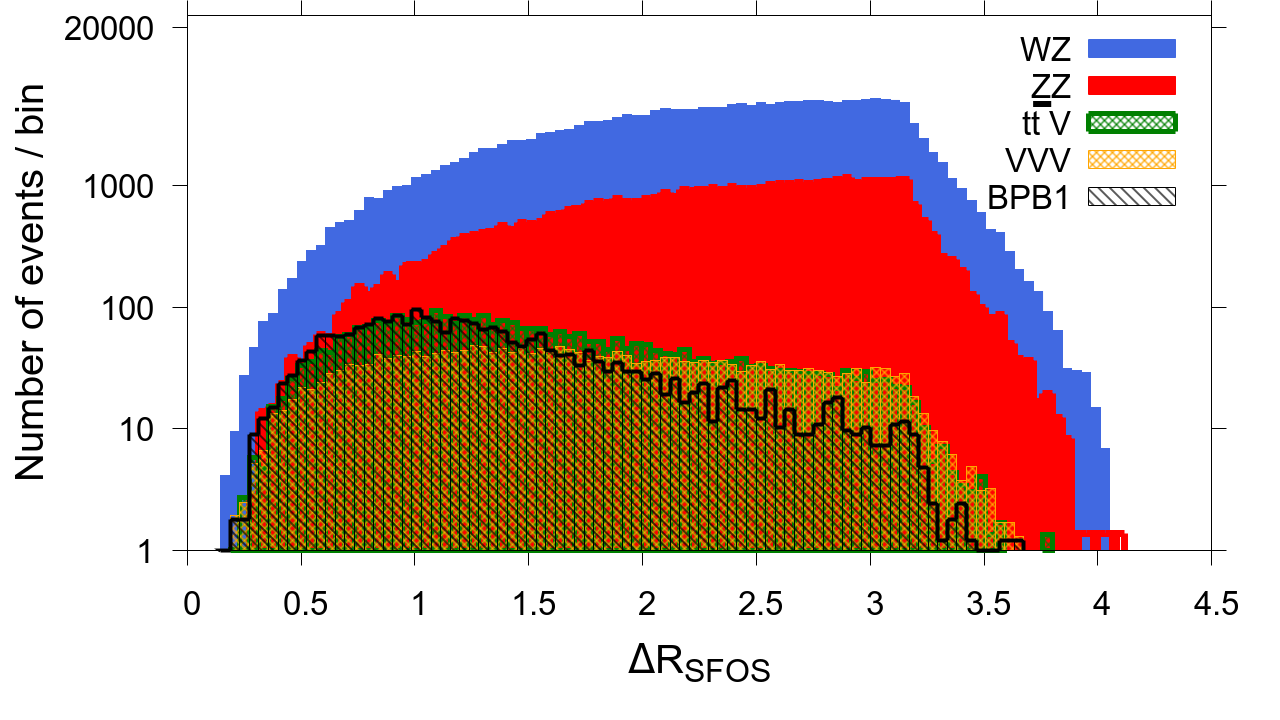}\includegraphics[scale=0.19]{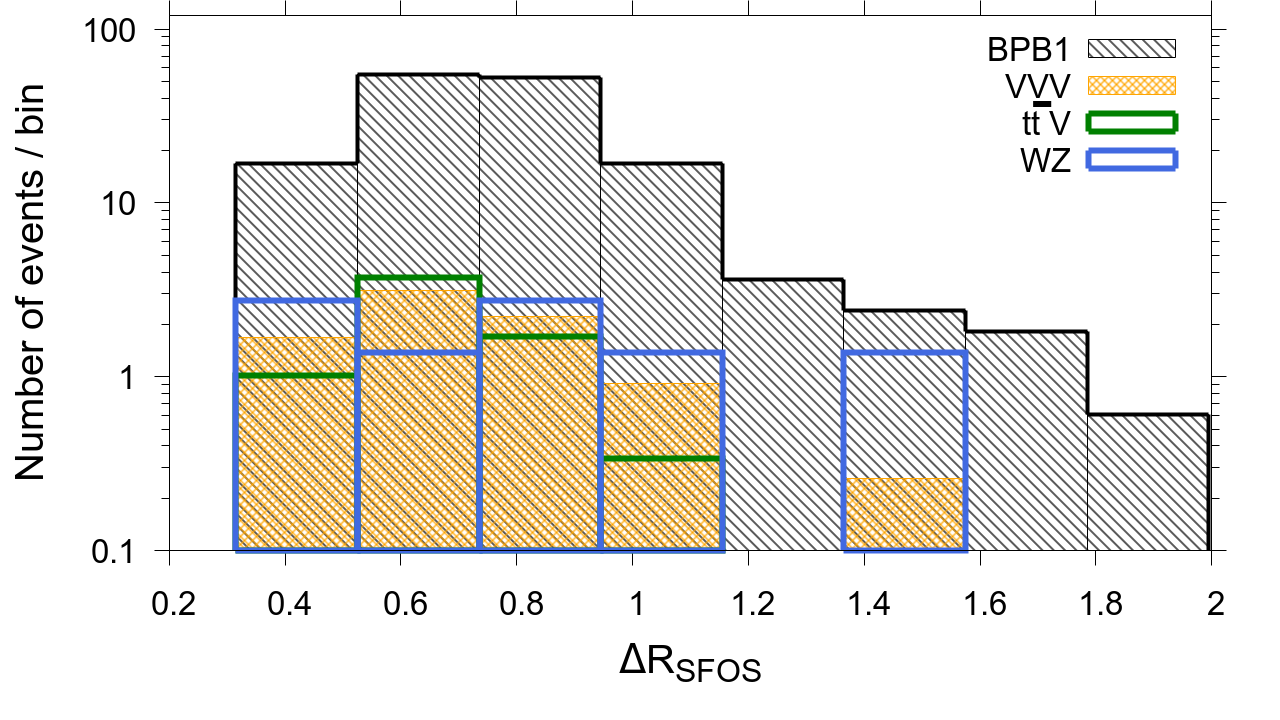}\\
\caption{The event distribution of $\met$ (top) and $\Delta R_{SFOS}$ (bottom) at the HL-LHC, for the signal benchmark point (BPB1: $M_{\lspthree,\lsptwo,\chonepm}=310~{\rm GeV}$, $M_{\lspone}=0~{\rm GeV}$) and the relevant backgrounds are shown. The signal events have been illustrated in black color while on- and off-shell $WZ$, $VVV$, $t\bar{t}V$ and $ZZ$ backgrounds have been shown in blue, orange, green and red colors, respectively. The left (right) panel shows the event distribution before (after) the application of SRB1 cuts.}
\label{fig:WZ_kin_dist_B1}
\end{center}
\end{figure}

A signal event is required to have three isolated leptons ($l = e,\mu$) in the final state. The lepton isolation criteria requires the ratio: $p_{T}^{charged}/p_{T}^{l}$ to be $\leq 0.15$ (for both electrons and muons), where, $p_{T}^{charged}$ is the scalar sum of transverse momentum ($p_{T}$) of all charged particles with $p_{T} \geq 1.0~{\rm GeV}$ within a cone of radius $\Delta{R} = 0.3$ around the lepton momentum direction while $p_{T}^{l}$ represents the transverse momentum of the lepton. Here, $\Delta R$ is defined as $\Delta R = \sqrt{\Delta \eta^{2} + \Delta \phi^{2}}$ with $\Delta \eta$ and $\Delta \phi$ representing the difference in pseudorapidity and azimuthal angle, respectively,  between the charged particle and the lepton under consideration. Candidate leptons are required to have $p_{T} > 20~{\rm GeV}$ and must lie within a pseudo-rapidity range of $|\eta| < 2.5$. Two of the three final state leptons are required to form a same flavor opposite sign (SFOS) lepton pair with invariant mass ($M_{ll}^{SFOS}$) in the range $M_{Z} \pm 10~{\rm GeV}$, where $M_{Z}$ is the mass of the $Z$ boson. In the presence of more than one pair of SFOS leptons with invariant mass in the above range, the SFOS pair which minimizes the transverse mass of the non-SFOS lepton and $\met$ is chosen to have its origin in the decay of the $Z$ boson, with the non-SFOS lepton coming from the decay of the $W$ boson. 

Standard model backgrounds to the $WZ$-mediated $3l+\met$ final state mainly come from: on-shell and off-shell $WZ$, $t\bar{t}V$ ($V = W^{\pm},Z$), triple vector boson ($VVV$) production and $ZZ$ processes. The $t\bar{t}V$ and the $ZZ$ samples have been produced by decaying $V (= W^{\pm},Z)$ and the $Z$ bosons, respectively, via leptonic decay modes. In the remaining sections of this work, the $t\bar{t}V$ and the $ZZ$ backgrounds have always been generated in a similar way with the $V$ and the $Z$ bosons, respectively, decaying leptonically. The leading order (LO) \texttt{MadGraph5$\_$aMC@NLO} cross-section values for $t\bar{t}V$ and the $ZZ$ process considered in this work accounts for the aforementioned decay channels. Events containing a $b$-tagged $jet$ with $p_{T} > 30~{\rm GeV}$ and $|\eta| < 2.5$ are also rejected in order to suppress the $tt\bar{V}$ background\footnote{A flat $b$-tagging efficiency of $70\%$ has been used. The $c \to b$ and $u,d,s \to b$ mistag rates have been fixed at $30\%$ and $1\%$, respectively~\cite{Collaboration:2017mtb}.}. These background events have been generated using the \texttt{MadGraph5$\_$aMC@NLO}~\cite{Alwall:2014hca,Sjostrand:2014zea} framework and for both, the signal as well as the background, detector effects are simulated using \texttt{Delphes-3.4.2}~\cite{deFavereau:2013fsa}. 

A cut based analysis is performed by considering $8$ signal regions with different set of cuts on the following kinematic variables: $p_{T}^{l_{1},l_{2},l_{3}}$ (transverse momenta of the three final state leptons with $l_{1}$ being the leading $p_{T}$ lepton and $l_{3}$ being the lepton with the smallest $p_{T}$), $\met$ (missing transverse energy), $M_{T}^{l_{W}}$ (transverse mass of the non-SFOS lepton ($l_{W}$) and $\met$ system), $M_{CT}^{W}$ (contransverse mass~\cite{Tovey:2008ui} of the $l_{W}-\met$ system), $\Delta \phi_{l_{W}\met}$ (difference between azimuthal angles of $l_{W}$ and $\met$), $\Delta \phi_{SFOS-\met}$ (difference between the azimuthal angles of the SFOS pair of leptons and $\met$) and $\Delta R_{SFOS}$ ($\Delta R = \sqrt{\Delta \eta^{2} + \Delta \phi^{2}}$, where $\Delta \eta$ and $\Delta \phi$ are the differences in pseudorapidity and azimuthal angle, respectively, of the two leptons which constitute the SFOS pair). The $8$ signal regions: SRA1, SRB1, SRC1, SRD1, SRE1, SRF1, SRG1 and SRH1, have been chosen by optimizing the signal significance for the following benchmark signal events $\left[M_{\lsptwo,\lspthree,\chonepm},~M_{\lspone}\right]$ (in GeV) : BPA1 [130,0], BPB1 [310,0], BPC1 [310,210], BPD1 [610,0], BPE1 [610,300], BPF1 [610,510], BPG1 [1000,0], BPH1 [1000,420], respectively. Here, we have chosen benchmark points with small, intermediate and large mass difference between the NLSP and the LSP, for several choices of the NLSP mass.

\begin{table}[!htb]
\begin{center}
\begin{tabular}{|| C{2.2cm} | C{1.55cm} C{1.5cm} C{1.55cm} C{1.9cm} C{1.55cm} C{1.55cm}  C{1.9cm} C{1.55cm} ||}
\hline \hline
 & \multicolumn{8}{c||}{Benchmark points} \\  \cline{2-9}
 & BPA1 & BPB1 & BPC1 & BPD1 & BPE1 & BPF1 & BPG1 & BPH1 \\ \hline
$M_{\lsptwo,\lspthree,\chonepm}$ [GeV] & 130 & 310 & 310 & 610 & 610 & 610 & 1000 & 1000 \\
$M_{\lspone}$ [GeV] & 30 & 0 & 210 & 0 & 300 & 510 & 0 & 420 \\ \hline \hline
Kinematic & \multicolumn{8}{c||}{Signal regions} \\ \cline{2-9}
variables & SRA1 & SRB1 & SRC1 & SRD1 & SRE1 & SRF1 & SRG1 & SRH1\\\hline
$\Delta \Phi_{l_{W} \met}$ & $ \leq 0.2$ & - & $\leq 1.5$ & - & - & - & - & - \\
$\Delta \Phi_{SFOS-\met}$ & - & $[2.7:\pi]$ & $[1.8:\pi]$ & $[1.5:\pi]$ & $[1.8:\pi]$ & - & $[1.6:\pi]$ & $[1.5:\pi]$ \\
$\Delta R_{SFOS}$ & $[1.4:3.8]$ & $[0.3:2.1]$ & - & $[0.1:1.3]$ & $[0.1:1.3]$ & $[1.6:4.0]$ & $[0.1:1.0]$ & $[0.1:1.3]$ \\ 
$\met$ [GeV]& $[50:290]$ & $\geq 220$ & $[100:380]$ & $\geq 200$  & $\geq 250$ & - & $\geq 200$ & $\geq 200$ \\ 
$M_{T}^{l_{W}}$ [GeV]& - & $\geq 100$ & $[100:225]$ & $\geq 300$ & $\geq 150$ & $[150:350]$ & $\geq 150$ & $\geq 200$\\ 
$M_{CT}^{l_{W}}$ [GeV]& - & $\geq 100$ & - & $\geq 100$ & $\geq 150$ & $[100:400]$ & $\geq 200$ & $\geq 200$\\ 
$p_{T}^{l_{1}}$ [GeV]& $[50:150]$ & $\geq 120$ & $[60:110]$ & $\geq 150$ & $\geq 150$  & $[60:150]$ & $\geq 210$ & $\geq 200$\\ 
$p_{T}^{l_{2}}$ [GeV]& $[50:110]$ & $\geq 60$ & $\geq 30$ & $\geq 100$ & $\geq 100$  & $[50:80]$ & $\geq 150$ & $\geq 100$\\
$p_{T}^{l_{3}}$ [GeV]& $\geq 30$ & $\geq 30$ & $\geq 30$ & $\geq 50$ & $\geq 50$  & $[30:60]$ & $\geq 50$ & $\geq 50$\\ \hline
\end{tabular}
\caption{ List of selection cuts corresponding to the signal regions for $WZ$ mediated $3l+\met$ final state. The signal regions have been optimized to yield maximum signal significance for the signal samples corresponding to the respective benchmark points.}
\label{tab:wz_hllhc_sr}
\end{center}
\end{table}

As an illustration, in Fig.~\ref{fig:WZ_kin_dist_B1}, we show the kinematic distribution of $\met$ and $\Delta R_{SFOS}$ for the signal benchmark point BPB1 (shown in black color) and the relevant backgrounds: on-shell and off-shell $WZ$ (blue color), $VVV$ (orange color), $t\bar{t}V$ (green color) and $ZZ$ (red color), at the HL-LHC. The distributions shown in the left panel of Fig.~\ref{fig:WZ_kin_dist_B1} have been obtained by imposing the following conditions on the generated event samples: presence of only three isolated leptons in the final state, presence of at least one SFOS pair with invariant mass in the range of $M_{Z} \pm 10~{\rm GeV}$, and $b$ $jet$ veto.  
The $\met$ distribution of BPB1 peaks roughly at $150~{\rm GeV}$ while the $\met$ distribution for the $WZ$, $VVV$, $t\bar{t}V$ and $ZZ$ backgrounds peaks roughly below $\lesssim 50~{\rm GeV}$. Due to the large mass gap between $\lspthree/\lsptwo$ and $\lspone$ in BPB1, the $Z$ boson produced from the cascade decay of $\lspthree/\lsptwo$ carries a relatively larger boost causing the $\Delta R_{SFOS}$ distribution of BPB1 to peak at smaller values. The SFOS lepton pairs for the background events are however, produced with relatively larger angular separation. The same is reflected in the $\Delta R_{SFOS}$ distribution of BPB1 and the background events shown in the lower left panel of Fig.~\ref{fig:WZ_kin_dist_B1}.

We optimize the signal significance of the $8$ benchmark points by applying various combination of selection cuts. The signal significance is computed as:
\begin{equation}
S_{\sigma} = \frac{S}{\sqrt{B+\left(sys\_un * B \right)^{2}}}
\label{Eqn:significance}
\end{equation}
where, $S$ and $B$ represents the signal and background yields, respectively, while $sys\_un$ corresponds to the systematic uncertainty taken to be $5\%$. The selection cuts corresponding to the $8$ signal regions, obtained from the cut-based analysis, have been listed in Table~\ref{tab:wz_hllhc_sr}. The figures on the right panel of Fig.~\ref{fig:WZ_kin_dist_B1} have been obtained on further imposing the selection cuts labeled SRB1. 

\begin{figure}[!htb]
\begin{center}
\includegraphics[scale=0.152]{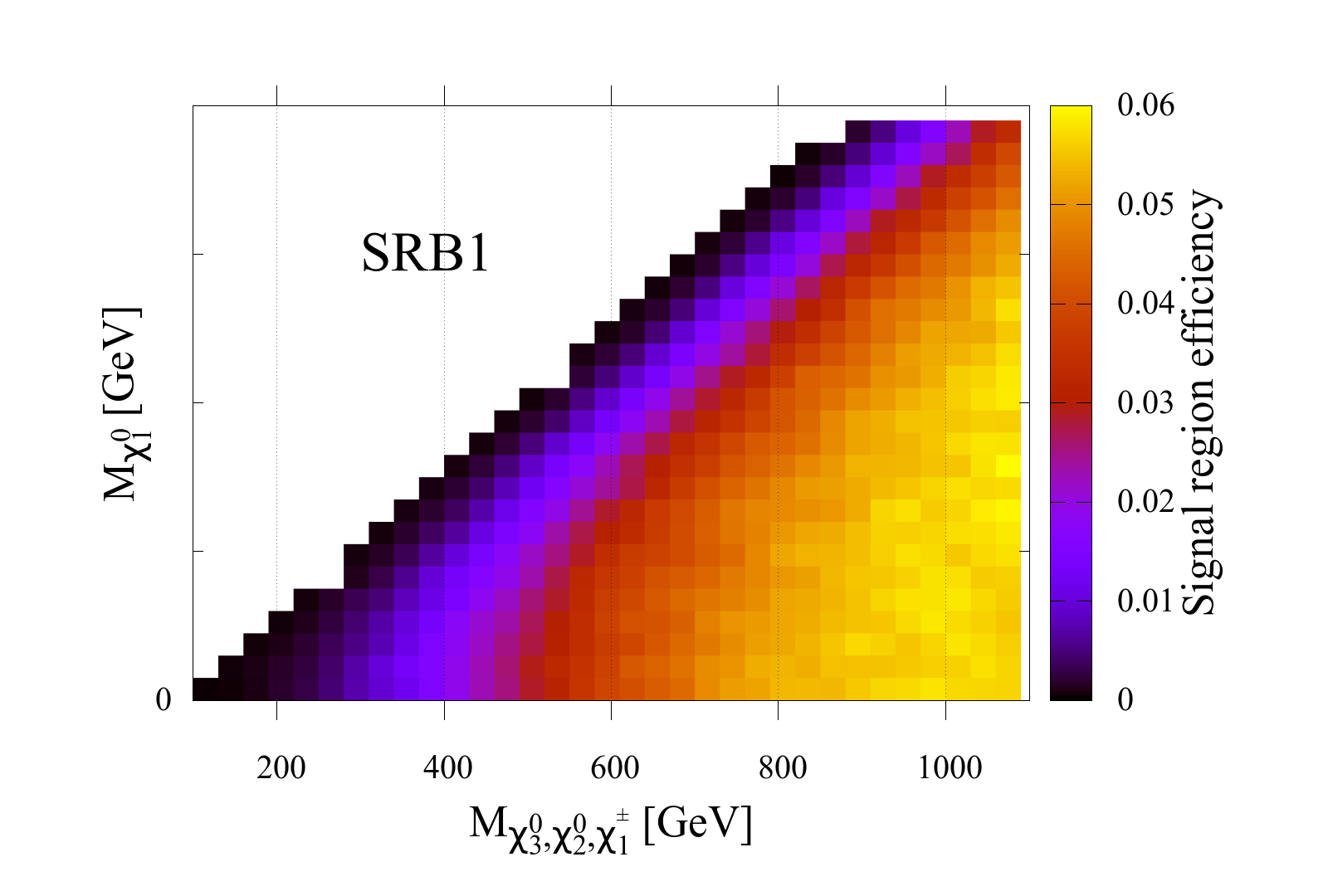}\includegraphics[scale=0.152]{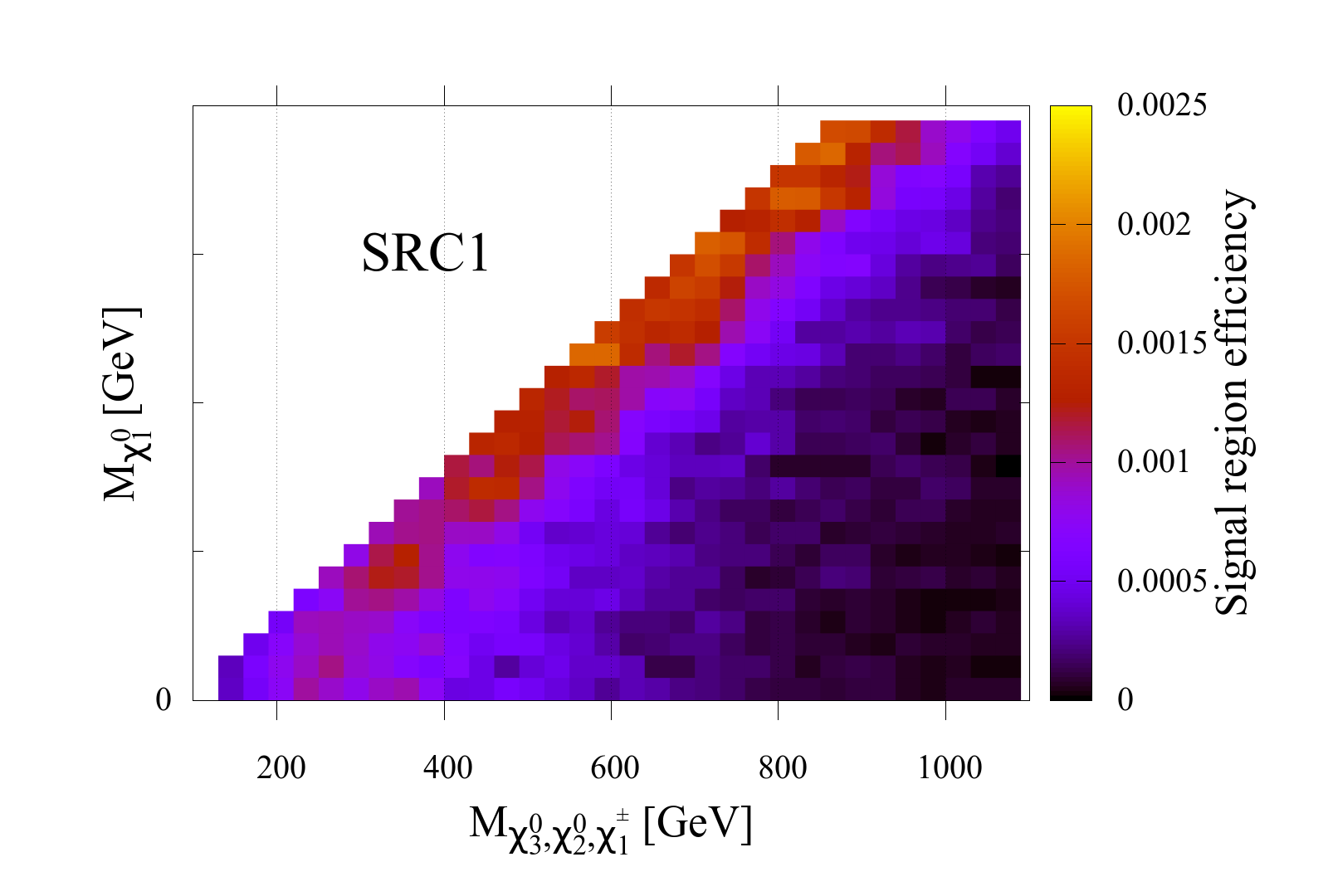}
\includegraphics[scale=0.152]{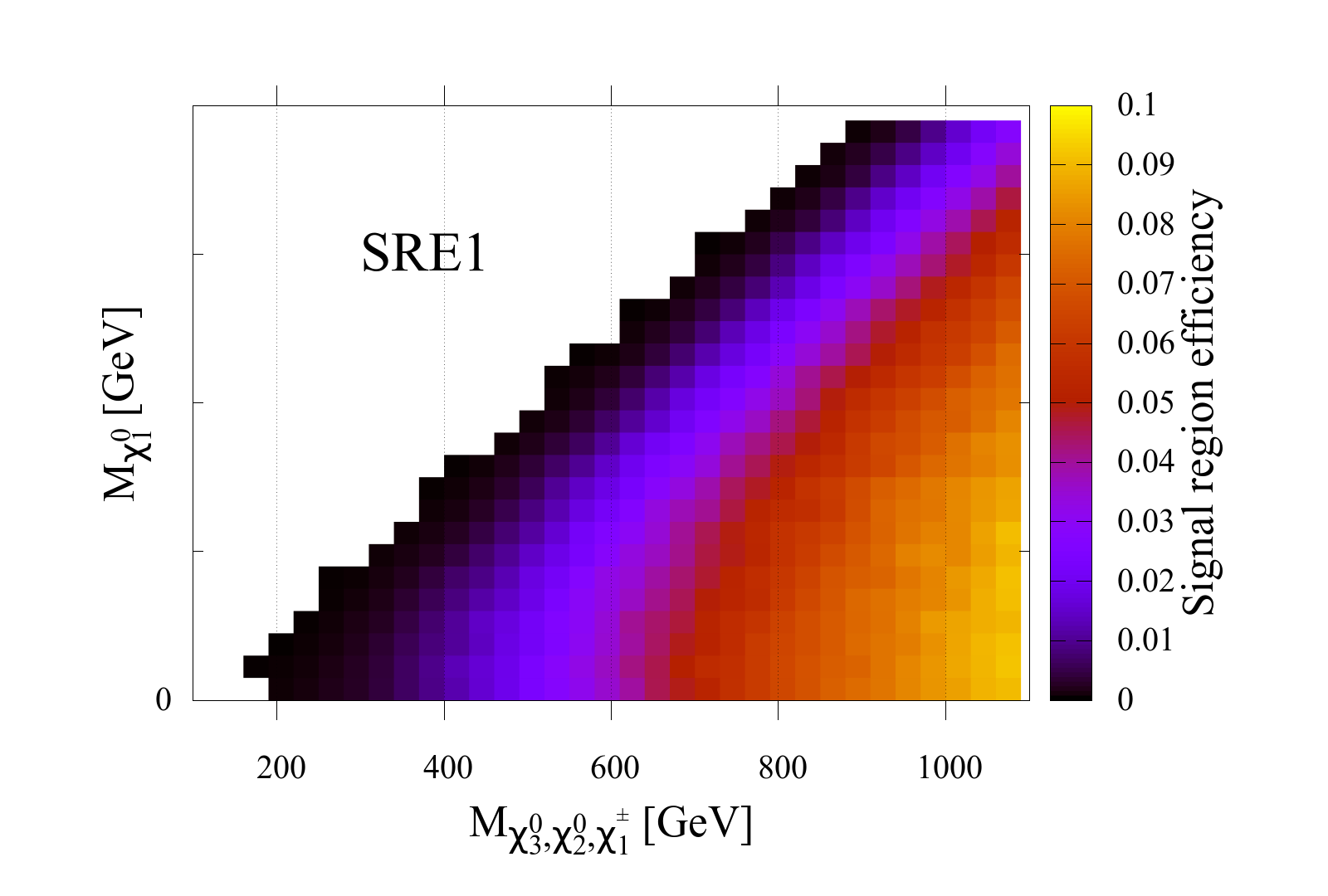}\includegraphics[scale=0.152]{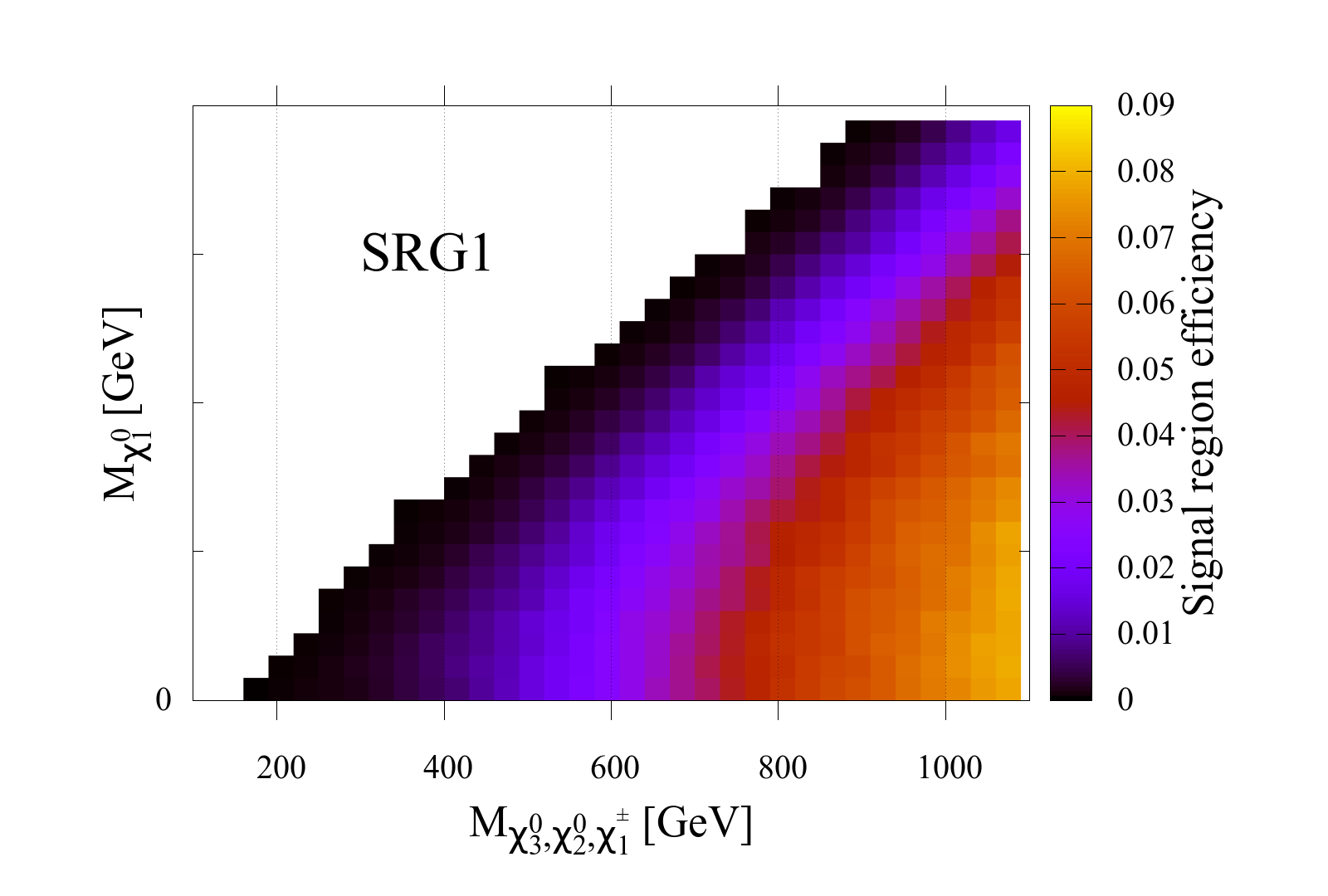}
\caption{ Efficiency map for the $WZ$-mediated $3l+\met$ signal from higgsino pair production at the HL-LHC with the SRB1 cuts~(top left), SRC1 cuts~(top right), SRE1 cuts~(bottom left) and SRG1 cuts~(bottom right) in the $M_{\lsptwo,\lspthree,\chonepm}$-$M_{\lspone}$ plane. Note that the z-axis scale for SRC1 is different from the other three frames.}
\label{fig:wz_hllhc_effmap}
\end{center}
\end{figure}

The efficiency grids for SRB1, SRC1, SRE1 and SRG1 are illustrated in the four panels of Fig.~\ref{fig:wz_hllhc_effmap} where the efficiency values are  shown as a color palette in the $M_{\lsptwo,\lspthree,\chonepm}$-$M_{\lspone}$ plane. The efficiency (${\rm Eff.}$) of a signal region for a particular signal point is computed by taking the ratio of the number of signal events which pass through the signal region selection cuts to the total number of generated signal events. We note that SRB1, SRD1 and SRG1 cuts were particularly optimized to maximize the signal significance of signal processes with larger mass difference between the directly produced doublet higgsinos and the LSP (BPB1, BPD1 and BPG1, respectively), while SRA1, SRC1 and SRF1 were optimized to maximize the signal significance of such signal events where the mass difference between the directly produced higgsinos and the LSP was just above the $Z$ mass threshold (BPA1, BPC1 and BPF1, respectively). On the other hand, SRE1 and SRH1 were optimized for signal process with intermediate mass difference between the NLSP higgsinos and the LSP (BPE1 and BPH1, respectively).

\begin{table}[!htb]
\begin{center}
\begin{tabular}{||C{3.2cm} || C{1.5cm} | C{1.2cm} C{1.2cm} C{1.2cm} C{1.2cm} C{1.2cm} C{1.2cm} C{1.2cm} C{1.2cm}||}
\hline \hline
Background & LO cross & \multicolumn{8}{c||}{Background yield ($14~{\rm TeV}$, $3~ab^{-1}$)} \\ \cline{3-10}
process & section & SRA1 & SRB1 & SRC1 & SRD1 & SRE1 & SRF1 & SRG1 & SRH1\\ \hline\hline 
\textbf{$WZ$} & 686 fb & 407 &9.60 & 12.3 & 5.49 & 4.12 & 9.60 & 5.49 & 6.86 \\ \hline 
\textbf{$t\bar{t}V$} & \multirow{3}{*}{343 fb} & \multirow{3}{*}{18.2} & \multirow{3}{*}{6.86} & \multirow{3}{*}{5.48} & \multirow{3}{*}{3.09} & \multirow{3}{*}{3.43} & \multirow{3}{*}{0.69} & \multirow{3}{*}{3.43} & \multirow{3}{*}{2.34} \\ 
($V=W,Z$) & & & & & & & & & \\ 
($W \to l\nu$, $Z \to ll$) & & & & & & & & & \\ \hline 
\textbf{$VVV$} ($V = W,Z$) & 261 fb & 13.6 & 8.22 & 2.22 & 2.48 & 3.00 & 1.57 & 2.48 & 2.22 \\ \hline 
\textbf{$ZZ$}  & \multirow{2}{*}{926 fb} & \multirow{2}{*}{62.5} & \multirow{2}{*}{0.0} & \multirow{2}{*}{1.4} & \multirow{2}{*}{0.0} & \multirow{2}{*}{0.0} & \multirow{2}{*}{1.4} & \multirow{2}{*}{0.0} & \multirow{2}{*}{0.0} \\  
(leptonic) & & & & & & & & & \\ \hline 
\multicolumn{2}{||c|}{\textbf{Total Background}} & 502 & 24.7 & 21.4 & 11.0 & 10.5 & 13.2 & 11.4 & 11.5 \\\hline\hline 
\end{tabular}
\caption{ The background yields for $\sqrt{s}=14~{\rm TeV}$ LHC corresponding to $3~ab^{-1}$ of integrated luminosity, for the $8$ different signal regions considered for the cut-based analysis, are tabulated. The leading order (LO) cross sections generated by \texttt{MadGraph5$\_$aMC@NLO} are also shown. The $t\bar{t}V$ background has been generated with $V$ decaying leptonically while the generic sample of triple vector boson production has been used in the analysis.}
\label{tab:bkg_hllhc_wz}
\end{center}
\end{table}

\begin{figure}[!htb]
\begin{center}
\includegraphics[scale=0.3]{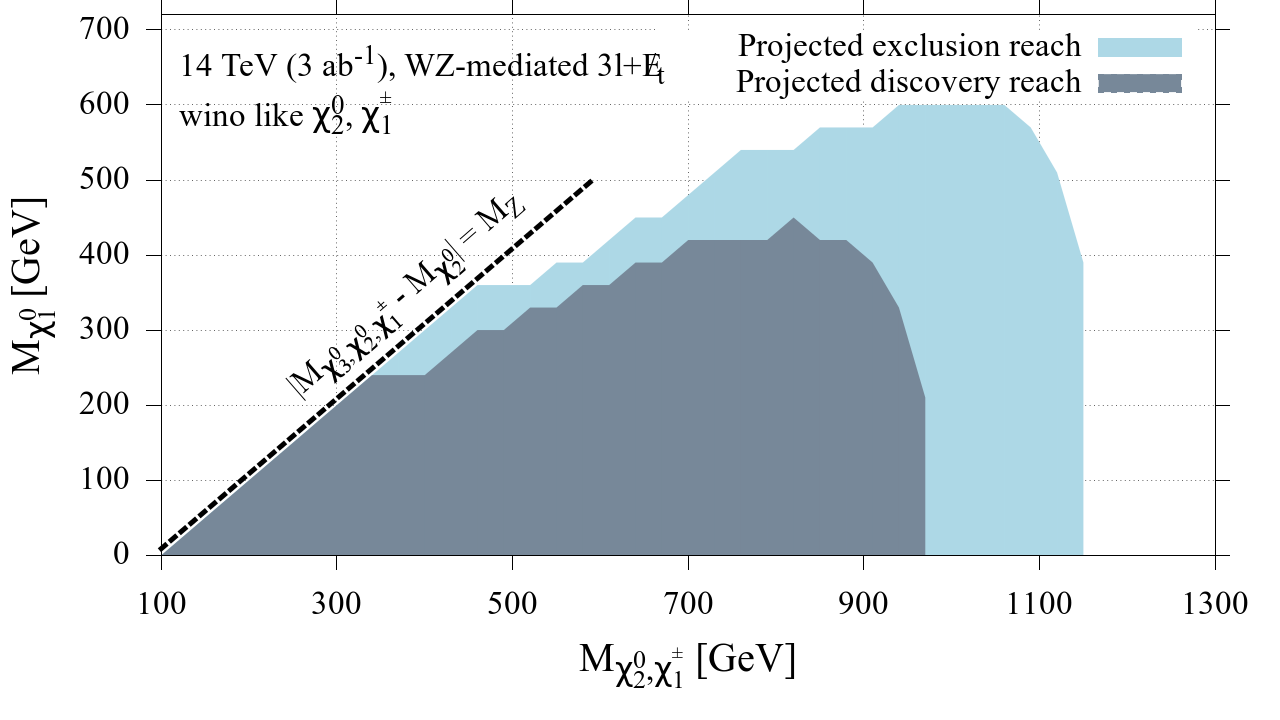}
\caption{The projected exclusion ($> 2\sigma$, light blue) and discovery ($> 5\sigma$, grey) contours in the $M_{\lsptwo,\chonepm}$ - $M_{\lspone}$ plane, derived from direct wino searches ($pp \to \lsptwo\chonepm$) in the $WZ$ mediated $3l+\met$ search channel, in the context of future HL-LHC. The projection contours shown here are comparable with the future projections from ATLAS~\citep{ATL-PHYS-PUB-2018-048}.}
\label{fig:wz_hllhc_excl_wino}
\end{center}
\end{figure}

\begin{figure}[!htb]
\begin{center}
\includegraphics[scale=0.3]{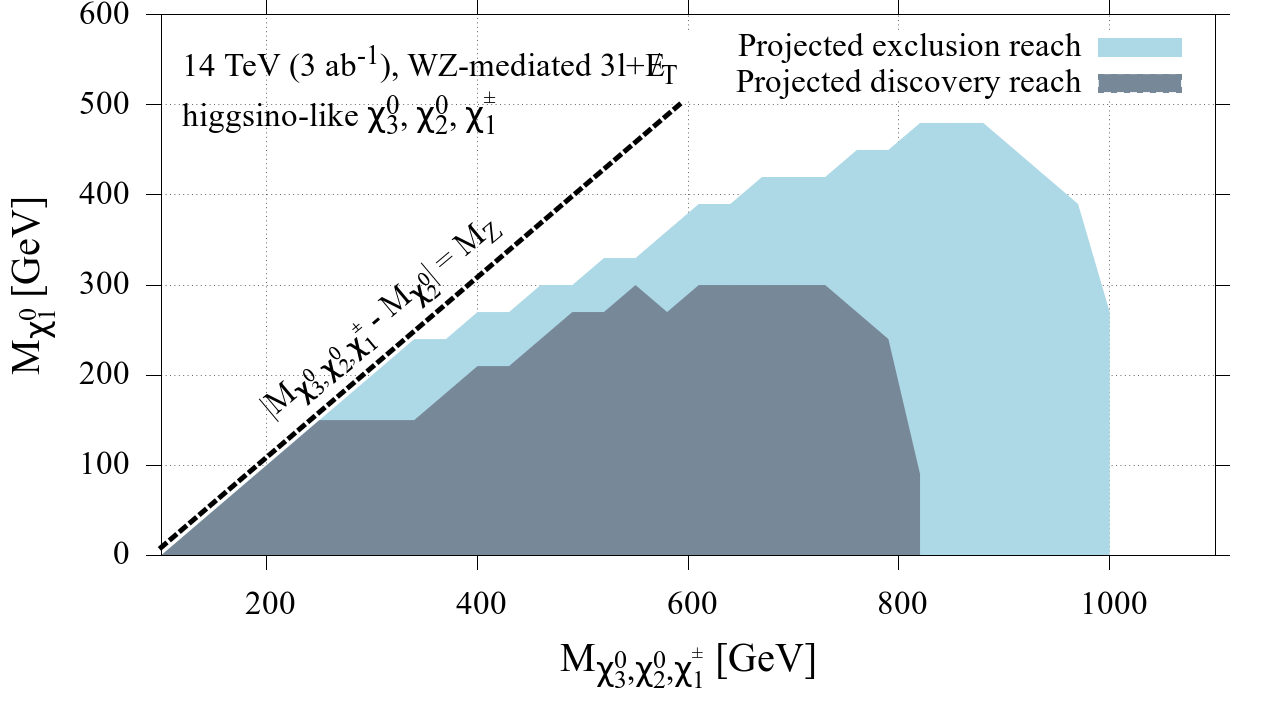}
\caption{The projected exclusion ($> 2\sigma$, light blue color) and discovery ($> 5\sigma$, grey color) contours in the $M_{\lspthree,\lsptwo,\chonepm}$ - $M_{\lspone}$ plane, derived from direct higgsino searches ($pp \to \lsptwo\chonepm + \lspthree\chonepm$) in the $WZ$ mediated $3l+\met$ search channel, in the context of future HL-LHC.}
\label{fig:wz_hllhc_excl}
\end{center}
\end{figure}

The signal yield is computed as $S = \sigma_{\lsptwo\chonepm + \lspthree\chonepm} \times Br\left(\lsptwo,\lspthree \to Z \lspone\right) \times Br\left(\chonepm \to W^{\pm}\lspone\right) \times Br\left(Z \to l^{\prime}l^{\prime}\right)\times Br\left(W \to l^{\prime}\nu \right) \times {\rm Eff.} \times \mathcal{L}$, where $\mathcal{L}$ corresponds to the integrated luminosity ($3000~{fb^{-1}}$ for HL-LHC) and $l^{\prime} = e,\mu,\tau$. The values of $Br\left(\lsptwo,\lspthree \to Z \lspone\right)$ and $Br\left(\chonepm \to W^{\pm}\lspone\right)$ are assumed to be $100\%$, while  SM values for $Br\left(Z \to l^{\prime}l^{\prime}\right)$ and $Br\left(W \to l^{\prime}\nu \right)$ are considered. Similarly, in the calculation of background yields, the leading order (LO) cross-section computed by \texttt{MadGraph5$\_$aMC@NLO} are considered. The background yields for $WZ$, $t\bar{t}V$, $VVV$ and $ZZ$ processes, corresponding to the $8$ different signal regions, are listed in Table~\ref{tab:bkg_hllhc_wz}. The leading order cross-section labeled $WZ$ refers to the tree-level cross-section for $l\nu ll$ production from \texttt{MadGraph5$\_$aMC@NLO} with generator-level cuts $p_{T}>5~{\rm GeV}$ and $|\eta| < 3.5$ on the leptons. These include contributions from {\em off-shell} $W,Z$ and $\gamma$ amplitudes\footnote{In the remainder of this article, $WZ$ background will refer to $l\nu ll$ production process with the aforesaid generator level cuts and will include contribution from off-shell $W$, $Z$ and $\gamma$ amplitudes.}. The off-shell $W$ contributions are crucial for obtaining a reliable estimate of the background remaining after a hard cut on the transverse mass $M_{T}^{l_{W}}$, which is very efficient in reducing the background where the $W$ boson is close to its mass shell.

As we have already mentioned, the ATLAS collaboration has also derived the projected exclusion and discovery contours at $95\%$ C.L. from direct wino searches in the $WZ$ mediated $3l+\met$ searches at the HL-LHC (see Fig.~9 from \cite{ATL-PHYS-PUB-2018-048}). 
In order to check the consistency of our analysis setup, we re-derive the projection contours for direct wino searches in the $WZ$ mediated $3l+\met$ final state at the HL-LHC by using the signal regions: SRA1 - SRH1, which were obtained through optimization of the signal significances for BPA1-BPH1, respectively. The corresponding projection contours derived by using our analysis setup are illustrated in Fig.~\ref{fig:wz_hllhc_excl_wino}, in the $M_{\lsptwo,\chonepm}-M_{\lspone}$ mass plane. 
The projection contours in Fig.~\ref{fig:wz_hllhc_excl_wino} display a potential to discover (exclude) a wino like $\chonepm,\lsptwo$ up to a mass of $\sim 960~(1150) ~{\rm GeV}$ for a massless LSP. This is comparable to the corresponding projections of ATLAS which are respectively $\sim 950~(1110)~{\rm GeV}$. Our discovery (exclusion) projections degrade by less than $\sim 50~(100)~{\rm GeV}$ for LSP masses up to 400~(600)~GeV. 
Next, we move on to derive the projected reach of direct doublet higgsino searches in the $WZ$ mediated $3l+\met$ final state at the HL-LHC.

The projected exclusion and discovery contours from direct higgsino searches in the $WZ$ mediated $3l+\met$ final state at HL-LHC are  illustrated in Fig.~\ref{fig:wz_hllhc_excl} with the same color code as Fig.~\ref{fig:wz_hllhc_excl_wino}. While deriving the HL-LHC projections, we compute the value of signal significance at a grid point for all $8$ signal regions and the maximum among them is ascribed to $S_{\sigma}$. 
We conclude that HL-LHC  will be  capable of discovering (excluding) pure doublet higgsinos up to $ 820$~($1000$)~${\rm GeV}$ for a massless bino-like $\lspone$ through direct higgsino searches in the $WZ$ mediated $3l+\met$ final state.

\subsubsection{$WH_{125}$-mediated $3l~+~\met$ at HL-LHC} \label{sec:wh_hllhc}

The two important decay modes of $H_{125}$ which contribute to the $3l+\met$ final state produced from $WH_{125}$ mediated processes are: $H_{125} \to WW^{*} \to l^{\prime}\nu l^{\prime}\nu$ and $H_{125} \to \tau^{+}\tau^{-}$. Contributions from both have been considered in the present analysis. The signal decay chain proceeds as $pp \to \lsptwo/\lspthree + \chonepm \to \left(\lsptwo,\lspthree \to \lspone H_{125}\right)\left(\chonepm \to \lspone \left(W \to l^{\prime}\nu \right) \right)$. The signal samples have been generated for different values of $M_{\lspone}$ (varied from 0 - 800 GeV, with a step size of $30$ GeV) and $M_{\lsptwo,\lspthree,\chonepm}$ (varied from $100$ GeV to $1300$ GeV with a step size of $30$ GeV), with the condition, $M_{\lsptwo,\lspthree} - M_{\lspone} > 125~{\rm GeV}$, in order to ensure on-shell production of the $H_{125}$. Here, $Br\left(\lsptwo,\lspthree \to \lspone + H_{125} \right)$ and $Br\left(\chonepm \to W^{\pm} + \lspone \right)$ are assumed to be $100\%$, while the SM values of $Br(H_{125} \to W W^{*})$ and $Br(W \to l^{\prime}\nu)$ have been used. The important sources of background are: on-shell and off-shell $WZ$ and $WH_{125}$, $VVV (V=W,Z)$, $t\bar{t}V$ and $ZZ$. 

\begin{table}[!htb]
\begin{center}
\begin{tabular}{|| C{3.0cm} | C{3.0cm} C{3.0cm} C{3.0cm} C{3.0cm} ||}
\hline \hline
Kinematic & \multicolumn{4}{c||}{Signal regions} \\ \cline{2-5}
variables & SRA2 & SRB2 & SRC2 & SRD2\\\hline
$M^{inv}_{OS,min}$ [GeV] & \multicolumn{4}{c|}{$< 75$} \\ 
$\met$ [GeV]& \multicolumn{4}{c|}{$>100$} \\ 
$M_{T}^{l_{1}}$ [GeV]& $\geq 200$ & $\geq 200$ & $\geq 300$ & $\geq 400$ \\ 
$M_{T}^{l_{2}}$ [GeV]& $\geq 100$ & $\geq 150$ & $\geq 200$ & $\geq 150$ \\
$M_{T}^{l_{3}}$ [GeV]& $\geq 100$ & $\geq 100$ & $\geq 150$ & $\geq 100$ \\ \hline
\end{tabular}
\caption{List of selection cuts corresponding to the signal regions for $WH_{125}$-mediated $3l+\met$ final state. The choice of signal regions is motivated by a similar analysis in \cite{ATL-PHYS-PUB-2014-010}.}
\label{tab:wh_hllhc_sr}
\end{center}
\end{table}

\begin{figure}[!htb]
\begin{center}
\includegraphics[scale=0.15]{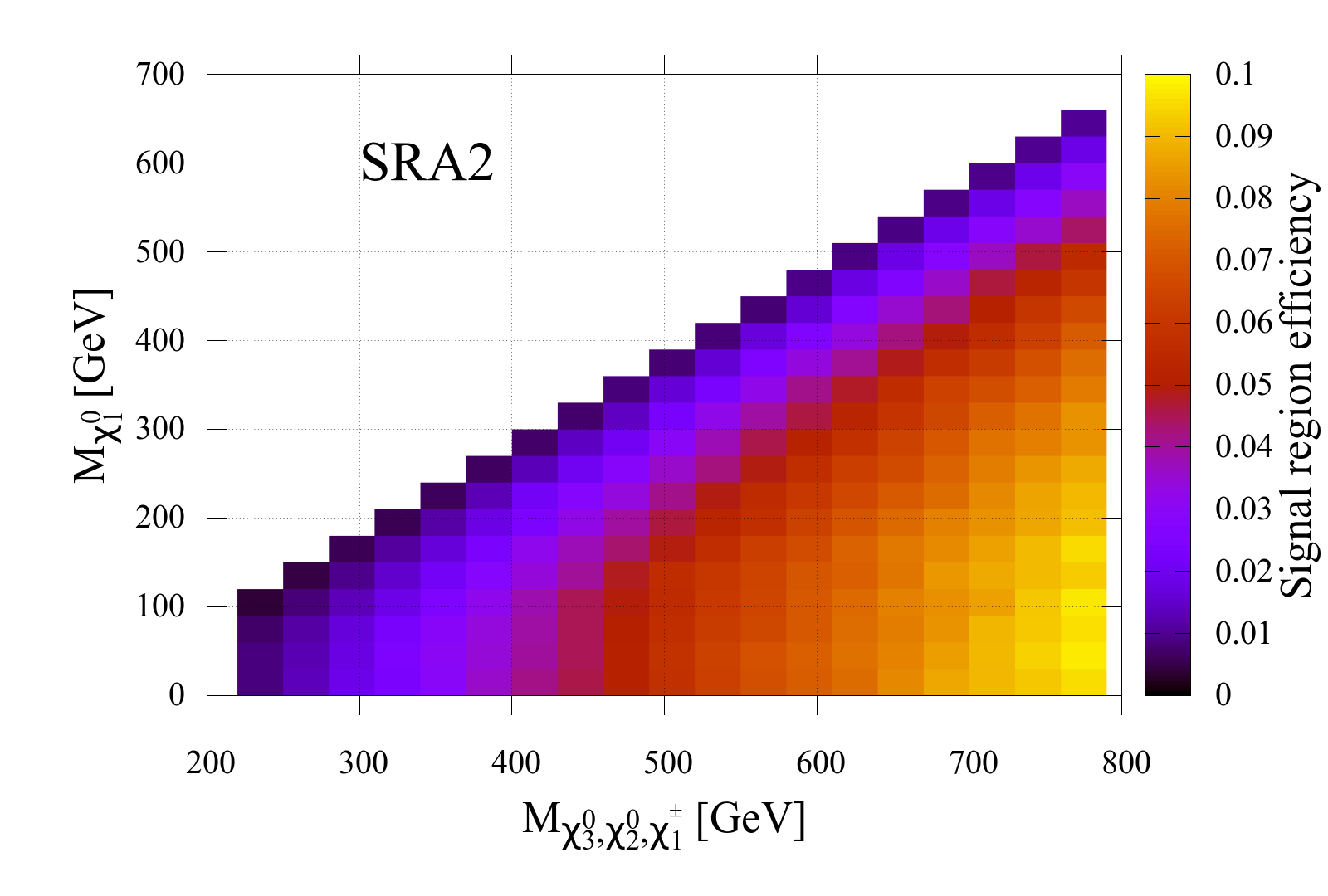}\includegraphics[scale=0.15]{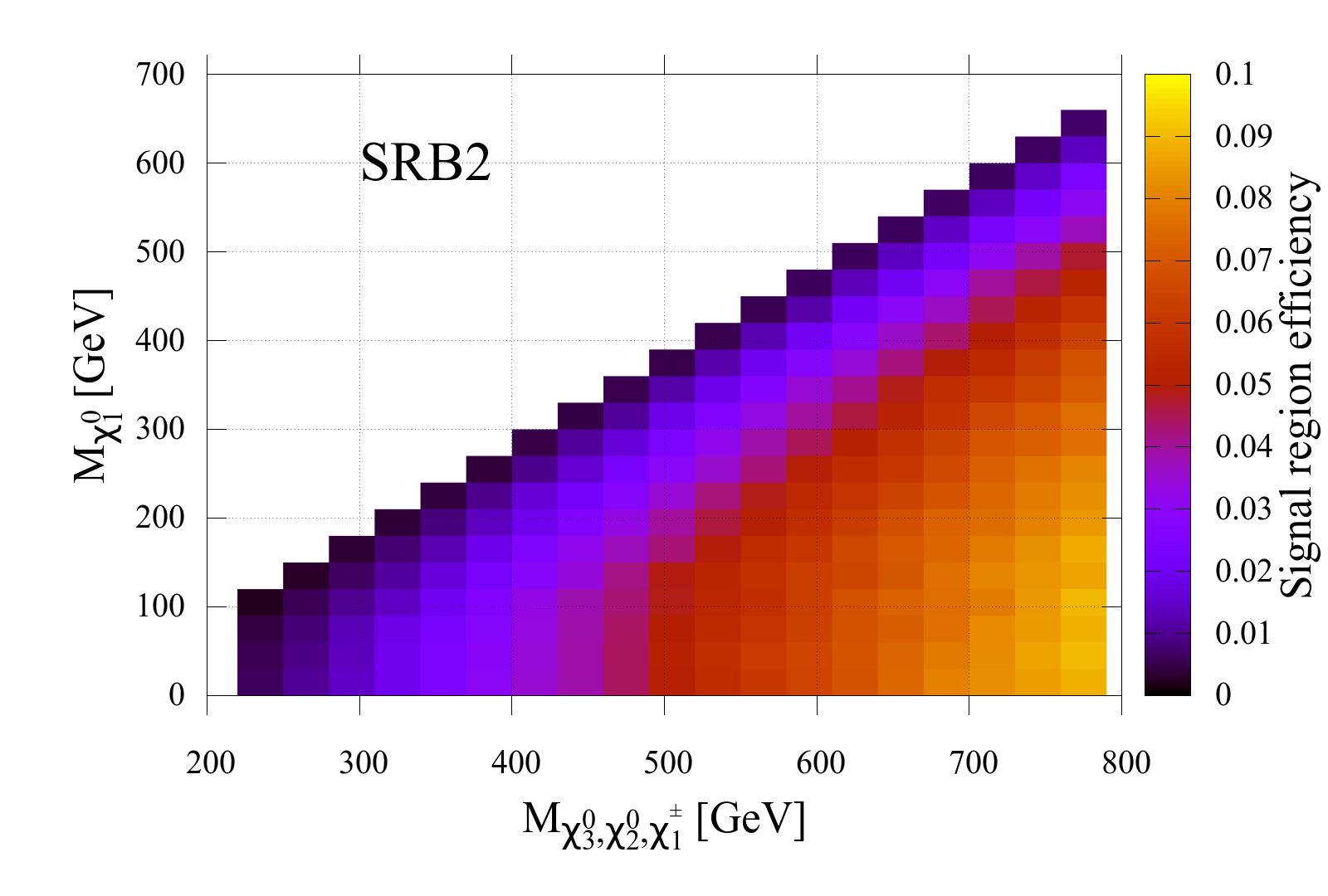}
\includegraphics[scale=0.15]{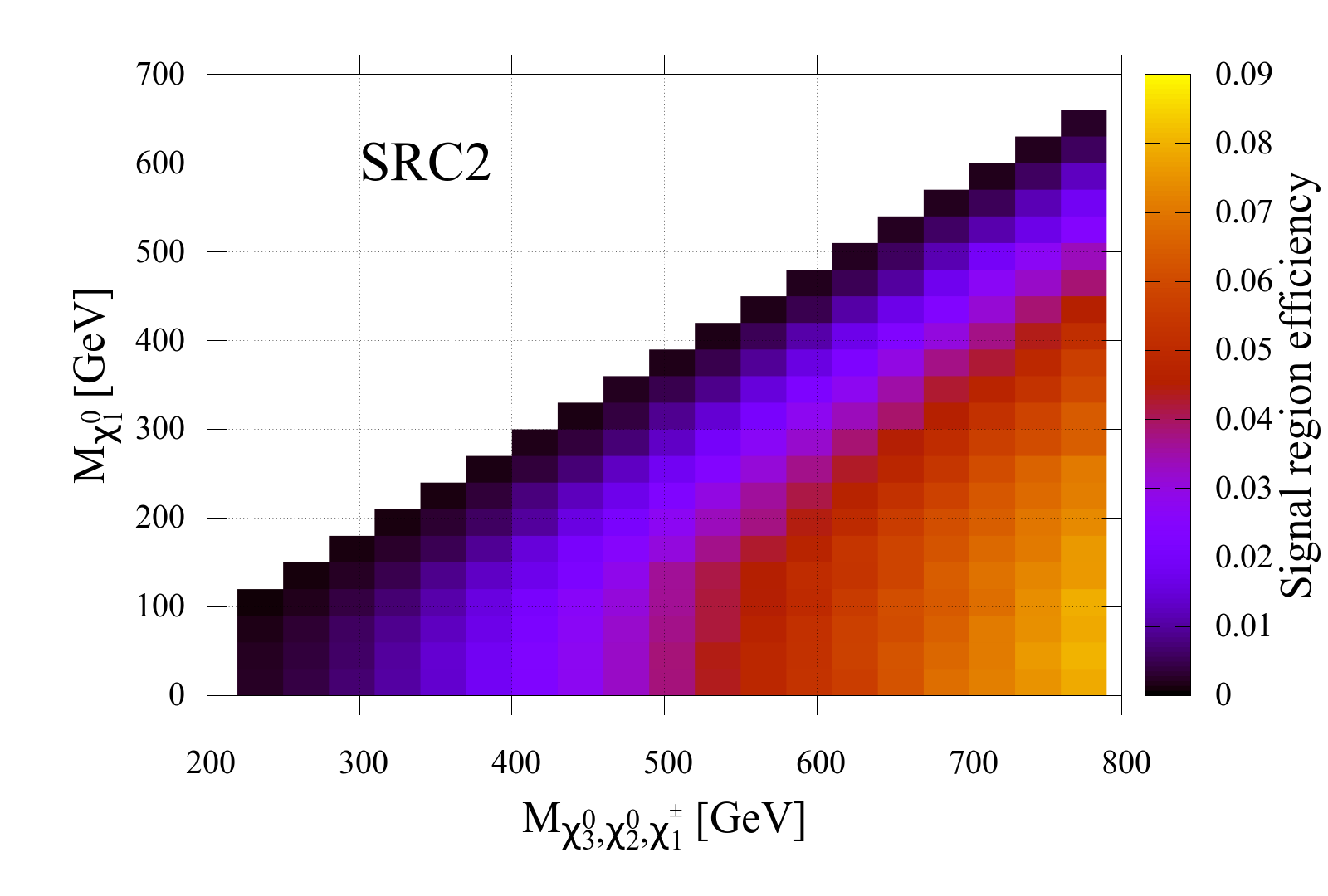}\includegraphics[scale=0.15]{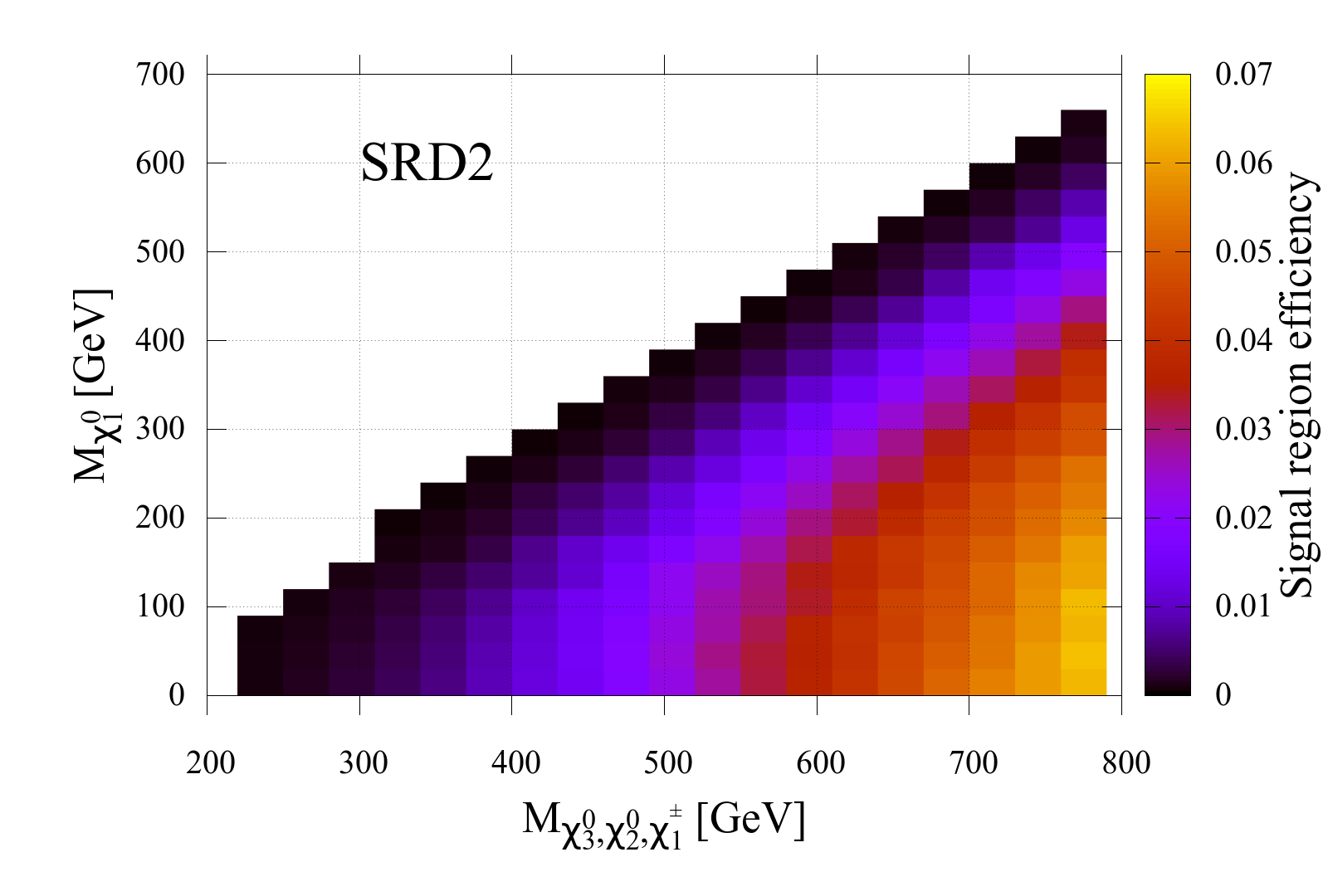}
\caption{Efficiency map of SRA2~(top left), SRB2~(top right), SRC2~(bottom left) and SRD2~(bottom right) signal regions, from searches in the $W$($H_{125} \to WW^{*} \to l^{\prime}\nu l^{\prime}\nu$) mediated $3l+\met$ final state.}
\label{fig:wh_14_ino_effmap}
\end{center}
\end{figure}

An event is required to have exactly three isolated leptons ($l = e,~\mu$) with $p_{T} > 10~{\rm GeV}$ and must lie within a pseudorapidity range of $|\eta| < 2.47$ (for electrons) and $|\eta| < 2.4$ (for muons). The lepton isolation criteria requires the ratio $p_{T}^{charged}/p_{T}^{l}$ to be $\leq 0.15$ for both, $e$ and $\mu$, with $p_{T}^{charged}$ being the sum of transverse momenta of all charged particles with $p_{T} > 1~{\rm GeV}$ within a cone of radius $\Delta R = 0.3$\footnote{Here, we have followed the lepton isolation criteria considered in \cite{ATL-PHYS-PUB-2014-010}.}. Events containing a SFOS pair of leptons with invariant mass in the range $M_{Z} \pm 12~{\rm GeV}$ are rejected in order to suppress the on-shell $WZ$ background. Events containing a $b$ tagged jet with $p_{T} > 20~{\rm GeV}$ and $|\eta| < 2.5$ are also rejected in order to suppress the $t\bar{t}V$ background. We choose four optimized signal regions: SRA2, SRB2, SRC2 and SRD2, with different set of cuts on $\met$, $M_{T}^{l_{1},l_{2},l_{3}}$ and $M^{inv.}_{OS,min}$. Here, $M_{T}^{l_{i}}$ ($i=1,2,3$) corresponds to the transverse mass of the $i^{th}$ lepton and $\met$ system, while $M^{inv.}_{OS,min}$ is the invariant mass of the opposite sign (OS) lepton pair with minimum $\Delta R$ separation. The choice of signal regions is motivated by a similar analysis in \cite{ATL-PHYS-PUB-2014-010} which probes the future prospect of direct wino searches at the HL-LHC. The list of selection cuts for the four optimized signal regions have been listed in Table~\ref{tab:wh_hllhc_sr}. 

\begin{table}[!htb]
\begin{center}
\begin{tabular}{|| C{5.0cm} || C{2.0cm} | C{2.0cm} C{2.0cm} C{2.0cm} C{2.0cm} ||}
\hline \hline
Background & Cross & \multicolumn{4}{c||}{Background yield ($14~{\rm TeV}$, $3~ab^{-1}$)} \\ \cline{3-6}
process & section [LO] & SRA2 & SRB2 & SRC2 & SRD2 \\ \hline\hline 
\textbf{$WZ$} & 686 fb & 130 & 97.4 & 57.6 & 34.3 \\ \hline 
\textbf{$t\bar{t}V$} & \multirow{3}{*}{343 fb} & \multirow{3}{*}{12.3} & \multirow{3}{*}{4.73} & \multirow{3}{*}{1.64} & \multirow{3}{*}{0.82}  \\ 
($V=W,Z$) & & & & & \\ 
($W \to l\nu$, $Z \to ll$) & & & & & \\ \hline 
\textbf{$VVV$} ($V = W,Z$) & 261 fb & 7.7 & 3.00 & 1.43 & 0.52 \\ \hline 
\textbf{$ZZ~ {\rm (leptonic)}$} & 834 fb & 6.95 & 6.95 & 2.78 & 2.78 \\ \hline 
\textbf{$WH_{125}$~$(H_{125} \to \tau^{+}\tau^{-})$} & 14.2 fb & 0.11 & 0.11 & $10^{-2}$ & 0.0 \\ \hline 
\textbf{$WH_{125}$} & \multirow{2}{*}{0.14 fb} & \multirow{2}{*}{$10^{-2}$} & \multirow{2}{*}{$10^{-2}$} & \multirow{2}{*}{$10^{-3}$} & \multirow{2}{*}{$10^{-3}$} \\
$H_{125} \to (W \to l\nu)(W^{*} \to l \nu)$ & & & & & \\ \hline
\multicolumn{2}{|c|}{\textbf{Total Background}} & 157 & 112 & 63.5 & 38.4  \\\hline\hline 
\end{tabular}
\caption{ The background yields for $\sqrt{s}=14~{\rm TeV}$ LHC corresponding to $3~ab^{-1}$ of integrated luminosity, for the $4$ different signal regions considered for the cut-based analysis, are tabulated. The leading order (LO) cross sections generated by \texttt{MadGraph5$\_$aMC@NLO} are also shown. The $t\bar{t}V$ background has been generated by decaying $V$ through leptonic decay modes while the $VVV$ sample is allowed to decay through all possible decay modes.}
\label{tab:bkg_hllhc_wh}
\end{center}
\end{table}

The background yields corresponding to the four signal regions  are shown in Table~\ref{tab:bkg_hllhc_wh} along with the LO \texttt{MadGraph5$\_$aMC@NLO} cross-section values. The background estimates for $WZ$, $WH_{125}$, $t\bar{t}V$, $VVV$ and $ZZ$ have been obtained by passing the simulated background samples through the signal region cuts. The tree-level cross-section for $WH_{125}$ has been computed using \texttt{MadGraph5$\_$aMC@NLO} with generator level cuts: $p_{T} > 5~{\rm GeV}$ and $|\eta| < 3.5$ on the leptons and includes contribution from off-shell $W$ amplitudes\footnote{Henceforth, the $WH_{125}$ background process will refer to the off-shell and on-shell $WH_{125}$ production process generated with the aforesaid generator level cuts.}. In order to correctly estimate the background yield, it is essential to consider the contribution from off-shell $W$ due to the hard cuts on $M_{T}^{l_{i}}~(i=1,2,3)$. The two final states of $H_{125}$ from the $WH_{125}$ background process which dominantly contributes to the $3l+\met$ signal are: $H_{125} \to WW^{*} \to l^{\prime} \nu_{l^{\prime}}l^{\prime} \nu_{l^{\prime}}$ and $H_{125} \to \tau^{+}\tau^{-}$. We generate these two final states separately.

We should mention that the ATLAS study \cite{ATL-PHYS-PUB-2014-010} reports a substantial
background from inclusive $t\bar{t}$ production tripling (doubling) the
background in the signal region SRA2 (SRB2) but with not very
significant contributions in the SRC2 and SRD2 signal regions. We
anticipate that $t\bar{t}W, W\rightarrow l\nu$ and both tops decaying
leptonically is the dominant physics source of {\em isolated} trileptons
from $t\bar{t}$ production. The trilepton topology can also result if a
lepton from the semi-leptonic decay of one of the bottom quarks is
accidentally isolated, or if one of jets is misidentified as an lepton.
Evaluation of these detector-dependent backgrounds is beyond the scope
of our study, but we mention it for completeness.  We have been unable
to ascertain the origin of the large background reported in
Ref.~\citep{ATL-PHYS-PUB-2014-010} and do not include it in the rest of our analysis (except where we mention it in our discussion of Fig.~\ref{fig:wh_excl_hllhc} below). The efficiency maps of SRA2, SRB2, SRC2 and SRD2 for the $W(H_{125} \to WW^{*} \to l^{\prime}\nu l^{\prime}\nu)$ mediated $3l+\met$ final state are shown in Fig.~\ref{fig:wh_14_ino_effmap}. 

\begin{figure}[!htb]
\begin{center}
\includegraphics[scale=0.3]{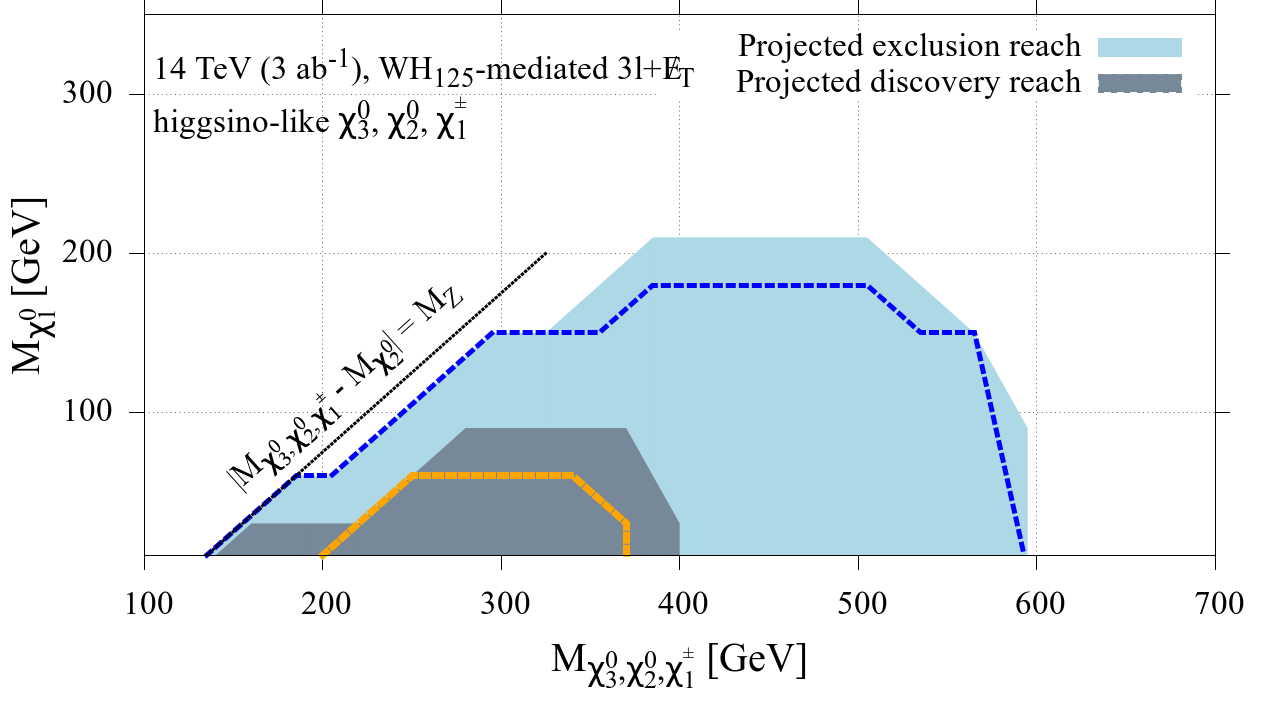}
\caption{The projected exclusion ($2\sigma$, blue color) contour and discovery ($5\sigma$, grey color) in the $M_{\lspthree,\lsptwo,\chonepm}$ - $M_{\lspone}$ plane, derived from direct search of mass degenerate higgsinos ($pp \to \lsptwo\chonepm + \lspthree\chonepm$) in the $WH_{125}$ mediated $3l+\met$ search channel at the HL-LHC. The blue dashed and orange dashed lines illustrate the $2\sigma$ and $5\sigma$ exclusion regions upon including the ATLAS $t\bar{t}$ background mentioned in the text.}
\label{fig:wh_excl_hllhc}
\end{center}
\end{figure}

The projected exclusion contour for direct higgsino searches in the $3l+\met$ final state is illustrated in light blue color in Fig.~\ref{fig:wh_excl_hllhc}. The contribution from $W(H_{125} \to WW^{*} \to l\nu l\nu)$ mediated and $W(H_{125} \to \tau^{+}\tau^{-})$ mediated $3l+\met$ signal processes have been added together in order to compute the signal yield. The signal significance for all four signal regions is computed, and the maximum among them is considered in the derivation of the projection contours shown in Fig.~\ref{fig:wh_excl_hllhc}. Furthermore, we do not consider any systematic uncertainty in the current analysis since we do not perform the signal-background optimization of our own and use the signal region cuts from \cite{ATL-PHYS-PUB-2014-010}. Our results indicate that the searches in the $WH_{125}$ mediated $3l+\met$ final state has the potential to discover (exclude) higgsinos up to $400~(590)$~${\rm GeV}$ for a massless $\lspone$. We mentioned previously that the signal regions SRA2, SRB2, SRC2 and SRD2 were motivated from Ref.~\cite{ATL-PHYS-PUB-2014-010} which probes the future reach of direct wino searches at the HL-LHC. We remark that if we include the $t\bar{t}$ background yields corresponding to the afore-mentioned signal regions from Ref.~\cite{ATL-PHYS-PUB-2014-010} (this double counts the $t\bar{t}W$ background), the projected discovery (exclusion) reach shown by the solid orange (blue-dashed) line for direct higgsino searches via the $WH_{125}$ mediated trilepton channel is altered to 375~(600)~GeV. Note that the final projections are not very sensitive to our treatment of the top background.

The goal of the next section (Sec.~\ref{sec:case1_ewino_hllhc}) is to translate the projected exclusion and discovery projections for HL-LHC derived from direct higgsino searches in the $WZ$ mediated and $WH_{125}$ mediated $3l+\met$ final state (derived in Sec.~\ref{subsec:ewino_hllhc}) on to the parameter space under study.

\subsection{Impact of projected electroweakino search limits from HL-LHC}   
\label{sec:case1_ewino_hllhc}

Because we have set the bino mass parameter to be $ 2~{\rm TeV}$, $\lsptwo,~\lspthree,~\lspfour$ as well as $\chonepm$ and $\chtwopm$ are either doublet higgsino-like, wino-like or a wino-higgsino admixture. In the present scenario, the direct chargino-neutralino pair production modes which could potentially lead to $WZ$ or $WH_{125}$ mediated $3l+\met$ final state are: $\lsptwo\chonepm$, $\lsptwo\chtwopm$, $\lspthree \chonepm$, $\lspthree \chtwopm$, $\lspfour \chonepm$ and $\lspfour \chtwopm$. The direct production cross-section for each of these chargino-neutralino pairs is computed by scaling the pure higgsino production cross-section (computed at NLO using \texttt{Prospino}) with the respective reduced squared $W\lspi\chjpm$ couplings. The reduced squared couplings have the following form: 
\begin{equation}
C_{W\lspi\chjpm}^{2} = \left\lbrace \left(N_{i3}~ V_{j2} - N_{i2}~ V_{j1} \sqrt{2} \right)^{2} + \left(N_{i4}~ U_{j2} + N_{i2}~ U_{j1} \sqrt{2} \right)^{2} \right\rbrace
\label{Eqn:Wchineut_red_sq_coup}
\end{equation}
where $V,U$ are the chargino mixing matrices. $V_{j1}/U_{j1}$ and $V_{j2}/U_{j2}$ represents the wino and higgsino component, respectively, in the $j^{th}$ chargino. $N_{i2}$ represents the wino admixture while $N_{i3/i4}$ corresponds to the higgsino admixture in the $i^{th}$ neutralino. In the case of direct production of a $ \lspi \chjpm$ pair where $\lspi$ and the two chiral components of $\chjpm$ have dominant wino composition, $N_{i3}^{2} + N_{i4}^{2},V_{j2}^{2}$ and $U_{j2}^{2}$ will be $<< 1$, while $N_{i2}^{2},~V_{j1}^{2}$ and $U_{j1}^{2}$ will be $\sim 1$. Similarly, in the case of direct production of higgsino-like chargino-neutralino pair, $N_{i3}^{2} + N_{i4}^{2},V_{j2}^{2}$ and $U_{j2}^{2}$ will be $\sim 1$, while the other three components will be negligible. From Eq.~(\ref{Eqn:Wchineut_red_sq_coup}), it can also be observed that the direct wino production cross-section is roughly twice the direct higgsino production cross-section at the LO. However, the arguments of the simplified scenario will not hold for our allowed parameter space points since the neutralinos and charginos are an admixture of both winos and higgsinos. Thereby, all terms in Eq.~\ref{Eqn:Wchineut_red_sq_coup} could be non-negligible (depending upon the electroweakino mixing structure) and contribute towards the computation of the scaled production cross-section. The production cross-sections obtained are multiplied by the relevant branching fractions needed to obtain the $3l+\met$ final state. The next step involves the computation of signal yield for all the optimized signal regions.
For the  $WZ$ mediated $3l+\met$ final state  we  compute the signal yields for all $8$ signal regions, SRA1, SRB1, SRC1, SRD1, SRE1, SRF1, SRG1 and SRH1  prescribed in Sec.~\ref{sec:wz_hllhc} while for  the $WH_{125}$ mediated search we use the four regions SRA2, SRB2, SRC2 and SRD2. The only ingredient left in the computation of the signal yield is the efficiency of the particular signal region, where efficiency is defined as the ratio of the number of events which are allowed by the selection cuts of a particular signal region to the total number of generated events. The efficiency is a function of the masses of $\lspfour/\lspthree/\lsptwo/\chtwopm/\chonepm$ and $\lspone$, and, is extracted from the signal region efficiency maps, some of which are shown in Fig.~\ref{fig:wz_hllhc_effmap} (for $WZ$ mediated $3l+\met$ searches) and Fig.~\ref{fig:wh_14_ino_effmap} (for $WH_{125}$ mediated $3l+\met$ searches). The signal significance is then computed using Eq.~(\ref{Eqn:significance}) by adopting the signal region that yields the highest value of $S_\sigma$. Parameter space points which generate a signal significance $> 2$ ($> 5$) are considered to be within the projected exclusion (discovery) reach of HL-LHC. 

\begin{figure}[!htb]
\begin{center}
\includegraphics[scale=0.20]{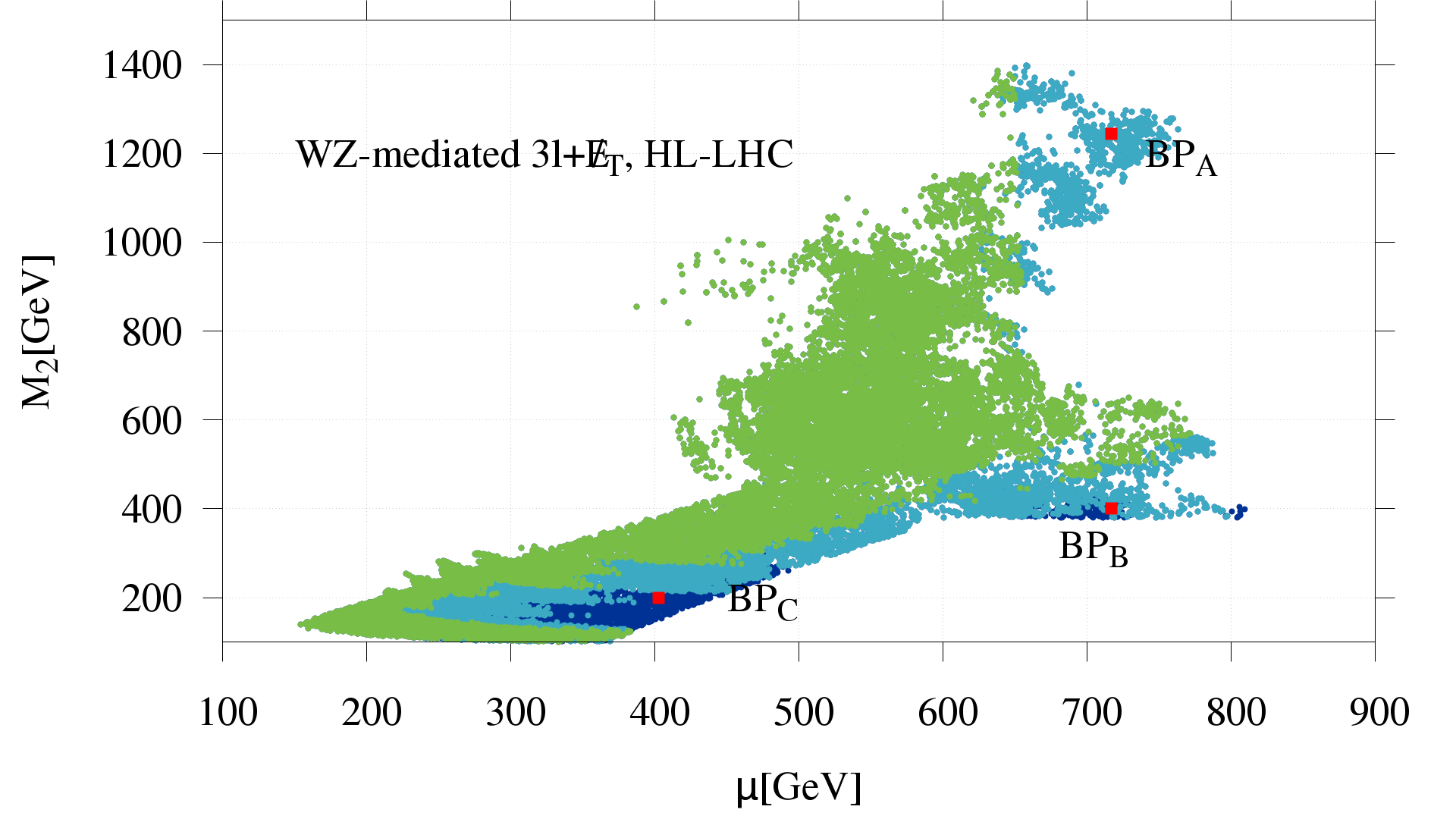}
\caption{The currently allowed parameter space points are shown in the $\mu - M_{2}$ plane. The pale blue colored (green colored) points fall within the projected exclusion reach (discovery reach) of direct higgsino searches in the $WZ$ mediated $3l+\met$ channel at the HL-LHC. The dark blue colored points fall outside HL-LHC's projected reach. The representative benchmark points: $BP_{A}$ ($\mu \sim 717~{\rm GeV} $, $M_{2} \sim 1244~{\rm GeV}$), $BP_{B}$ ($\mu \sim 717~{\rm GeV} $, $M_{2} \sim 400~{\rm GeV}$) and $BP_{C}$ ($\mu \sim 403~{\rm GeV} $, $M_{2} \sim 200~{\rm GeV}$), are also shown in this figure.}
\label{fig:m2_mu_hllhc}
\end{center}
\end{figure}

The points shown in Fig.~\ref{fig:m2_mu_hllhc} correspond to the allowed parameter space points obtained after imposing all constraints specified in Sec.~\ref{sec:constraints}, see also Fig.~\ref{fig:m2_mu_current}. The parameter space points which fall within (outside) the projected exclusion reach of direct higgsino searches in the $WZ$ mediated $3l+\met$ final state at HL-LHC are shown in pale blue (dark blue) color in Fig.~\ref{fig:m2_mu_hllhc}, while the green colored points fall within the projected discovery reach. 

Whether or not the signal is observable is the result of an intricate interplay between the production cross sections, determined by the doublet higgsino and wino masses, and the signal efficiencies shown in Fig.~\ref{fig:wz_hllhc_effmap}. In the green region with $\mu \sim M_{2} \sim 500~{\rm GeV}$, the efficiencies in Fig.~\ref{fig:wz_hllhc_effmap} are relatively large (the red/orange region) for small values of $M_{\lspone}$, and doublet higgsinos and gauginos are both in the kinematically accessible range, contributing to the $WZ$ mediated trilepton signal. For larger values of $\mu$ and $M_{2}$, though the efficiency increases, the production cross-section becomes smaller and the signal is kinematically suppressed. For smaller values of $M_{2}$ and $\mu$, though the production cross-section remains large, the efficiency reduces except for SRA1/SRB1 type cuts, for which efficiencies are small\footnote{Note that the scale is different in the efficiency maps of SRB1, SRC1, SRE1 and SRG1 in Fig.~\ref{fig:wz_hllhc_effmap}.}. We observe a dark blue colored protrusion region in the $M_{2} \sim 150~{\rm GeV}$ region from $\mu \sim 300~{\rm GeV}$ to $\mu \sim 400~{\rm GeV}$. The signal significance of the majority of the points in this region is marginally less than $2\sigma$ on account of the relatively smaller efficiency from SRB1.

In order to understand the underlying features, we  chose two benchmark points, $BP_{A}$ ($\mu \sim 717~{\rm GeV}$, $M_{2} \sim 1244~{\rm GeV}$) and $BP_B$ ($\mu \sim 717~{\rm GeV}$, $M_{2} \sim 400~{\rm GeV}$), from the allowed parameter space. The masses and the composition of $\lspone,~\lsptwo,~\lspthree,~\lspfour,~\chonepm$ and $\chtwopm$\footnote{Note that the neutralinos in our parameter space have a negligible bino admixture.} for $BP_{A}$ and $BP_{B}$ are shown in Table~\ref{tab:translation_bp1a} and Table~\ref{tab:translation_bp1b}, respectively. Furthermore, the effective production cross-section at $\sqrt{s} = 14~{\rm TeV}$ and $27~{\rm TeV}$\footnote{The projected discovery range and exclusion regions for direct higgsino searches in the $WZ$ and $WH_{125}$ mediated $3l+\met$ final state at the HE-LHC ($\sqrt{s}=27$~TeV, $\mathcal{L}=15~{\rm ab^{-1}}$) and its translation on to the allowed parameter space has been studied in the next section~(Sec.~\ref{subsec:ewino_helhc}).} of all viable chargino neutralino pairs which can eventually contribute to the signal yield of $WZ$ and $WH_{125}$ mediated $3l+\met$ final state are tabulated along with the relevant branching fractions of the charginos and neutralinos. 

\begin{table}
\begin{center}
\begin{tabular}{| C{3cm} | C{2cm} | C{2cm} | C{2cm} | C{2cm} | C{2cm} | C{2cm} |} \hline \hline 
 $BP_{A}$& $\lspone$ & $\lsptwo$ & $\lspthree$ & $\lspfour$ & $\chonepm$ & $\chtwopm$ \\ \hline
Mass~[GeV] & 58.3  & 723  & 732 & 1300  & 723 & 1300 \\ \hline
wino $\%$ & $10^{-4}$ & 0.01 & $10^{-4}$ & 0.99 & 0.01 & 0.99 \\ \hline
higgsino $\%$ & $10^{-4}$ & 0.98 & 0.99 & 0.01 & 0.99 & 0.01 \\ \hline \hline
\multicolumn{3}{|c|}{Singlino fraction in $\lspone$: 0.99} & \multicolumn{4}{c|}{$M_{H_{1}}$ = 83.7~GeV, $M_{A_{1}}$ = 34.5~GeV} \\ \hline 
Cross-section (fb) & $\lsptwo\chonepm$ & $\lsptwo\chtwopm$ & $\lspthree\chonepm$ & $\lspthree \chtwopm$ & $\lspfour \chonepm$ & $\lspfour \chtwopm $ \\ 
$\sqrt{s}=14~{\rm TeV}$ & 2.4 & $5 \times 10^{-3}$ & 2.3 & $4 \times 10^{-3}$ & $5 \times 10^{-3}$ & 0.3 \\ 
$\sqrt{s}=27~{\rm TeV}$ & 11.8 & 0.03 & 11.1 & 0.03 & 0.03 & 3.7 \\ \hline

\multirow{3}{*}{Branching ratio} & \multicolumn{6}{c|}{$\lsptwo \to \lspone Z$ (0.50), $\lspone H_{125}$ (0.43), $\lspone H_{1}$ (0.07)} \\ \cline{2-7}
 & \multicolumn{6}{c|}{$\lspthree \to \lspone Z$ (0.49), $\lspone H_{125}$ (0.44),  $\lspone H_{1}$ (0.06)} \\ \cline{2-7}
 & \multicolumn{6}{c|}{$\lspfour \to \lsptwo H_{125}$ (0.22), $\lsptwo Z$ (0.01), $\chonepm W^{\mp}$ (0.51), $\lspthree Z$ (0.24)}\\ \hline
\multicolumn{7}{|c|}{Significance at HL-LHC: $WZ$ mediated $3l+\met$: 3.8, $WH_{125}$ mediated $3l+\met$: 0.4 } \\ 
\multicolumn{7}{|c|}{Significance at HE-LHC: $WZ$ mediated $3l+\met$: 14, $WH_{125}$ mediated $3l+\met$: 6.6 } \\ \hline 
\end{tabular}
\caption{The masses, wino and higgsino fractions of neutralinos and charginos, neutralino-chargino production cross-sections, branching ratios for neutralinos and signal significance for $BP_{A}$. The input parameters of $BP_{A}$ are tabulated in Appendix~\ref{Appendix:BP}. The HE-LHC projections and their translation on to the allowed parameter space has been studied in Sec.~\ref{subsec:ewino_helhc}.}
\label{tab:translation_bp1a}
\end{center}
\end{table}

\begin{table}
\begin{center}
\begin{tabular}{| C{3cm} | C{2cm} | C{2cm} | C{2cm} | C{2cm} | C{2cm} | C{2cm} |} \hline \hline 
 & $\lspone$ & $\lsptwo$ & $\lspthree$ & $\lspfour$ & $\chonepm$ & $\chtwopm$ \\ \hline
Mass~[GeV] & 60.4  & 421 & 734 & 742 & 421 & 741 \\ \hline
wino $\%$ & $10^{-5}$ & 0.96 & $2\times 10^{-3}$ & 0.04 & 0.94 & 0.06 \\ \hline
higgsino $\%$ & $10^{-4}$ & 0.04 & 0.99 & 0.96 & 0.06 & 0.94 \\ \hline \hline
\multicolumn{3}{|c|}{Singlino fraction in $\lspone$: 0.99} & \multicolumn{4}{c|}{$M_{H_{1}}$ = 97.2~GeV, $M_{A_{1}}$ = 99~GeV} \\ \hline 
Cross-section (fb) & $\lsptwo\chonepm$ & $\lsptwo\chtwopm$ & $\lspthree\chonepm$ & $\lspthree \chtwopm$ & $\lspfour \chonepm$ & $\lspfour \chtwopm $ \\ 
$\sqrt{s}=14~{\rm TeV}$ &  104 & 0.27 & 0.28 & 2.1 & 0.25 & 2.3 \\
$\sqrt{s}=27~{\rm TeV}$ &  363 & 1.1 & 1.1 & 10.2 & 1.0 & 11.2 \\ \hline
\multirow{4}{*}{Branching ratio} & \multicolumn{6}{c|}{$\lsptwo \to \lspone Z$ (0.04), $\lspone H_{125}$ (0.82), $\lspone H_{1}$ (0.14)} \\ \cline{2-7}
 & \multicolumn{6}{c|}{$\lspthree \to \lspone Z$ (0.13), $\lspone H_{125}$ (0.10), $\lspone H_{1}$ (0.01), $\chonepm W^{\mp}$ (0.51), $\lsptwo Z$ (0.23), $\lsptwo H_{125}$ (0.01)} \\  \cline{2-7}
 & \multicolumn{6}{c|}{$\lspfour \to \lspone Z$ (0.12), $\lspone H_{125}$ (0.11), $\chonepm W^{\mp}$ (0.53)} \\
 & \multicolumn{6}{c|}{$\lspfour \to \lsptwo Z$ (0.02), $\lsptwo H_{125}$ (0.21)} \\\hline 
\multicolumn{7}{|c|}{Significance at HL-LHC: $WZ$ mediated $3l+\met$: 1.5, $WH_{125}$ mediated $3l+\met$: 5.3} \\
\multicolumn{7}{|c|}{Significance at HE-LHC: $WZ$ mediated $3l+\met$: 4.4, $WH_{125}$ mediated $3l+\met$: 34} \\ \hline 
\end{tabular}
\caption{The masses, wino and higgsino fractions of neutralinos and charginos, neutralino-chargino production cross-sections, branching ratios for neutralinos and signal significance for $BP_{B}$.  The input parameters of $BP_{B}$ are tabulated in Appendix~\ref{Appendix:BP}.}
\label{tab:translation_bp1b}
\end{center}
\end{table}

\begin{table}
\begin{center}
\begin{tabular}{| C{3cm} | C{2cm} | C{2cm} | C{2cm} | C{2cm} | C{2cm} | C{2cm} |} \hline \hline 
 & $\lspone$ & $\lsptwo$ & $\lspthree$ & $\lspfour$ & $\chonepm$ & $\chtwopm$ \\ \hline
Mass~[GeV] & 3.0 & 205 & 415 & 427 & 206 & 432 \\ \hline
wino $\%$ & $10^{-4}$ & 0.91 & $7\times 10^{-3}$ & 0.08 & 0.87 & 0.13 \\ \hline
higgsino $\%$ & $10^{-4}$ & 0.08 & 0.99 & 0.92 & 0.13  & 0.87 \\ \hline \hline
\multicolumn{3}{|c|}{Singlino fraction in $\lspone$: 0.99} & \multicolumn{4}{c|}{$M_{H_{1}}$ = 6.6~GeV, $M_{A_{1}}$ = 6.3~GeV} \\ \hline 
Cross-section (fb) & $\lsptwo\chonepm$ & $\lsptwo\chtwopm$ & $\lspthree\chonepm$ & $\lspthree \chtwopm$ & $\lspfour \chonepm$ & $\lspfour \chtwopm $ \\ 
$\sqrt{s}=14~{\rm TeV}$ & 1593 & 6.1 & 7.7  & 24.8 & 7.0 & 28.8 \\
$\sqrt{s}=27~{\rm TeV}$ & 4399 & 19.1 & 23.8 & 86.5 & 21.9 & 100.9 \\ \hline
\multirow{3}{*}{Branching ratio} & \multicolumn{6}{c|}{$\lsptwo \to \lspone Z$ (0.04), $\lspone H_{125}$ (0.92), $\lspone H_{1}$ (0.03)} \\ \cline{2-7}
 & \multicolumn{6}{c|}{$\lspthree \to \chonepm W^{\mp}$ (0.68), $\lsptwo Z$ (0.29), $\lsptwo H_{125}$ (0.02)} \\ \cline{2-7}
 & \multicolumn{6}{c|}{$\lspfour \to \chonepm W^{\mp}$ (0.72), $\lsptwo Z$ (0.04), $\lsptwo H_{125}$ (0.22))} \\\hline 
\multicolumn{7}{|c|}{Significance at HL-LHC: $WZ$ mediated $3l+\met$: 1.3, $WH_{125}$ mediated $3l+\met$: 8.8} \\
\multicolumn{7}{|c|}{Significance at HE-LHC: $WZ$ mediated $3l+\met$: 2.1, $WH_{125}$ mediated $3l+\met$: 48} \\ \hline 
\end{tabular}
\caption{The masses, wino and higgsino fractions of neutralinos and charginos, neutralino-chargino production cross-sections, branching ratios for neutralinos and signal significance for $BP_{C}$. The input parameters of $BP_{C}$ are tabulated in Appendix~\ref{Appendix:BP}.}
\label{tab:translation_bp1c}
\end{center}
\end{table}

In the case of $BP_{A}$, $\chonepm,~\lsptwo,~\lspthree$ have a dominant higgsino composition while $\lspfour,~\chtwopm$ are wino dominated and much heavier. The $\lspone$, on the other hand, is singlino dominated.The $\lsptwo$ and $\lspthree$ decay into $Z+\lspone$ and $H_{125}+\lspone$ with a branching fraction of $\sim 50\%$ and $\sim 43\%$, respectively, in each mode (shown in Table~\ref{tab:translation_bp1a}). $BP_{A}$ yields a signal significance of $3.8$ in direct higgsino searches in the $WZ$ mediated $3l+\met$ searches at the HL-LHC, and falls short of the discovery reach of the HL-LHC. In case of $BP_{A}$, although $Br(\lspthree/\lsptwo \to \lspone H_{125})$ is comparable with $Br(\lspthree/\lsptwo \to \lspone Z)$, the signal significance of $BP_{A}$ in the $WH_{125}$ mediated $3l+\met$ final state is $\lesssim 2$ due to the larger cascade decay chain in the latter case which reduces the signal significance to $0.4$. 
  
\begin{figure}
\includegraphics[scale=0.2]{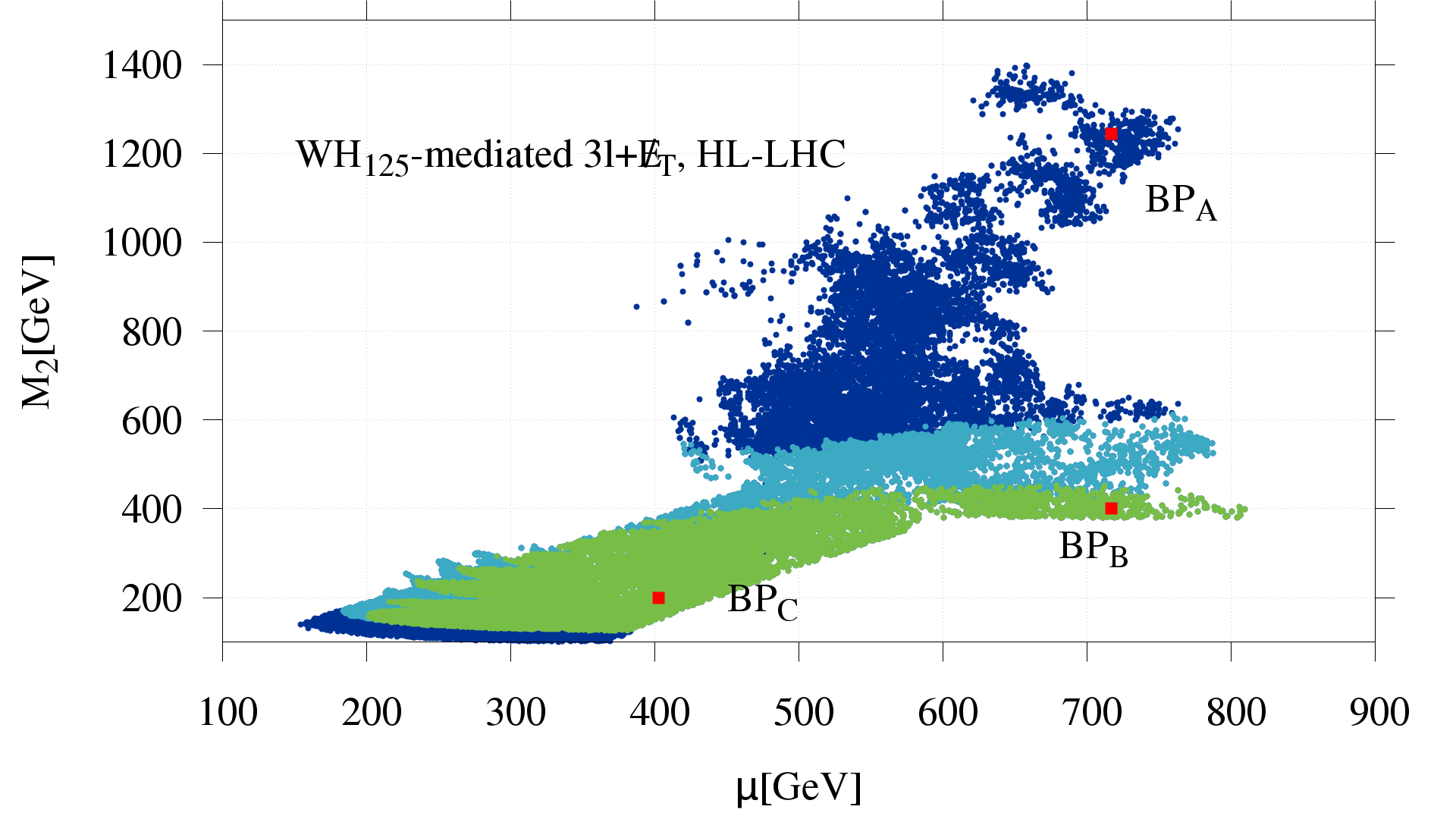}
\caption{The currently allowed parameter space points have been shown in the $\mu - M_{2}$ plane. The pale blue and green colored points fall within the projected exclusion and discovery reach, respectively, of HL-LHC through direct higgsino searches in the $WH_{125}$ mediated $3l+\met$ final state. The dark blue colored points remain outside HL-LHC's projected reach. The representative benchmark points: $BP_{A}$ ($\mu \sim 717~{\rm GeV} $, $M_{2} \sim 1244~{\rm GeV}$), $BP_{B}$ ($\mu \sim 717~{\rm GeV} $, $M_{2} \sim 400~{\rm GeV}$) and $BP_{C}$ ($\mu \sim 403~{\rm GeV} $, $M_{2} \sim 200~{\rm GeV}$), are also shown in this figure.}
\label{fig:wh_hllhc}
\end{figure}

In the case of $BP_{B}$, $\lsptwo$ is  wino-like and decays dominantly into $\lspone H_{125}$~($Br(\lsptwo \to \lspone H_{125}) \sim 82\%$)  while the branching ratio into $\lspone Z$ is only $4 \%$. The $\chonepm$ which is also wino-like decays into $W^{\pm} \lspone$ with $100\%$ branching fraction. Among the six possible combinations of chargino neutralino pairs, the $\lsptwo\chonepm$ pair has by far the highest cross-section $\sim 104~{\rm fb}$ and thereby, plays the most significant role in signal yield computation. The signal significance of $BP_{B}$ in the $WZ$ mediated $3l+\met$ search channel is $\sim 1.5$, and therefore, falls marginally outside the projected exclusion reach of HL-LHC. However, considering the large branching fraction of $\lsptwo \to H_{2} \lspone$, it is expected that direct searches in the $WH_{125}$  channel will be more effective in probing $BP_{B}$. The reach of the HL-LHC for NMSSM doublet higgsinos via the $WH_{125}$ mediated $3l+\met$ search channel is illustrated in Fig.~\ref{fig:wh_hllhc}, where we plot the allowed points from Fig.~\ref{fig:m2_mu_current}, with the same color convention as in Fig.~\ref{fig:m2_mu_hllhc}. Indeed $BP_{B}$ falls within the projected discovery reach of this search  with a signal significance of $5.3$.

Another important aspect to be noted for the $BP_{B}$ benchmark point is the presence of other cascade decay modes with considerable branching ratios. For instance, $\lspthree$ decays to $Z\lsptwo$ while $\lsptwo$ in addition to its decay to $\lspone H_{125}$, also decays to $\lspone H_1$ with a branching ratio $\sim 14\%$. We see from Table~\ref{tab:translation_bp1b} that $\lspthree$ is dominantly produced in addition to $\chtwopm$ which decays to $Z$ or $H_{125}$ plus $\chonepm$, or via $W$ and lighter neutralinos. The final states from $\lspthree \chtwopm$ production can, therefore, be very rich and include $VV+\met$~\cite{Baer:2017gzf} and $V/Z/H_{1}+\met$ events in the final state. Unfortunately, at least for this point $\sigma(\lspthree\chtwopm)$ is just 2.1~fb at $\sqrt{s}=14~{\rm TeV}$ and 10.2~fb at $\sqrt{s}=27~{\rm TeV}$, but see another case in Table~\ref{tab:translation_bp1c} where $\sigma(\lspthree\chtwopm)$ is much larger (24.8~fb at $\sqrt{s}=14~{\rm TeV}$ and 86.5~fb at $\sqrt{s}=27~{\rm TeV}$). The search strategies devised to optimize the $3l+\met$ searches will not be particularly efficient in the presence of these cascade decay channels and Fig.~\ref{fig:m2_mu_hllhc} and Fig.~\ref{fig:wh_hllhc} may not provide the complete picture. Dedicated searches beyond the scope of this paper will be needed to explore these novel signals.

It can also be observed from Fig.~\ref{fig:wh_hllhc} that future direct searches in the $WH_{125}$ mediated $3l+\met$ channel are more effective in probing the parameter space region with $M_{2} \lesssim \mu$. However, below $M_{2} \lesssim 125~{\rm GeV}$, an on-shell $H_{125}$ cannot be produced resulting in zero signal efficiency, and therefore this channel cannot be used. It is very striking to see the complementarity in the search power via the $WZ$ and the $WH_{125}$-mediated trilepton search channels exhibited in Fig.~\ref{fig:m2_mu_hllhc} and Fig.~\ref{fig:wh_hllhc}, respectively. For instance, in the dark blue coloured band near $M_2\sim 180$~GeV and $\mu\sim 300$-400~GeV in Fig.~\ref{fig:m2_mu_hllhc}, the signal significance is less than 2, while in the same region in Fig.~\ref{fig:wh_hllhc}, the signal is above the discovery limit! Likewise, in much of the green region at large values of $M_2$ where the signal can be discovered via the $WZ$-mediated channel (Fig.~\ref{fig:m2_mu_hllhc}), the signal is unobservable via the $WH_{125}$ channel in Fig.~\ref{fig:wh_hllhc}. To obtain a better understanding of what is happening in the low $M_2$ region, we examine a representative benchmark point $BP_{C}$~(see Table~\ref{tab:translation_bp1c}). $BP_{C}$ features a wino-like $\lsptwo,~\chonepm$ and thus a large production cross-section of $pp \to \lsptwo\chonepm$ (1593~fb), a large branching fraction of $\lsptwo \to \lspone H_{2}$ ($92 \%$) and $\chonepm \to \lspone W^{\pm}$ ($100 \%$). However, due to the small doublet higgsino component in $\lsptwo$, the branching ratio of $\lsptwo \to \lspone Z$ is only $4\%$. Additionally, the signal regions in Table~\ref{tab:wz_hllhc_sr} generate relatively lower efficiencies in the $M_{NLSP} \sim 180~{\rm GeV}$ region compared to the smaller and larger values of $M_{\lsptwo}$, from SR1A and SR1B. Consequently, the signal significance of $BP_{C}$ in the $WZ$ channel is only $\sim 1.3$ and thereby falls outside the projected reach of HL-LHC. $BP_{C}$ is, however, within the projected discovery reach of HL-LHC in the $WH_{125}$ channel where the signal significance is $ 8.8$.

An examination of Fig.~\ref{fig:m2_mu_hllhc} and Fig.~\ref{fig:wh_hllhc} shows that it should be possible to discover SUSY with $> 5\sigma$ confidence at the HL-LHC over almost the entire allowed parameter space (with the exception of the island at large values of $M_2$ and $\mu$ near the benchmark point $BP_A$) if it is realized as in the NMSSM scenario with $m_{\lspone} < 62.5$~GeV.

\subsection{Electroweakino searches at the HE-LHC}
\label{subsec:ewino_helhc}

We have just seen that the HL-LHC will be able to probe most of the NMSSM parameter space via $3l+\met$ searches via at least one or the other of the $WZ$ or $WH_{125}$ mediated channels. Here, we turn to the exploration of the capabilities of the proposed HE-LHC for the corresponding search. Our motivation for this is two-fold. First, it is clear that the signal might be just above the discovery limit at the HL-LHC so that a larger signal may be obtained at the proposed energy upgrade, allowing for a detailed study of the new physics. Second, we want to map out the NMSSM region where the signal would be discoverable via both the $WZ$ and the $WH_{125}$ mediated channels, since observations in multiple channels would clearly help to elucidate its origin. To this end, we devise new strategies to isolate the SUSY signal from the background, and delineate the corresponding $5\sigma$ discovery reach and $2\sigma$ exclusion regions for the cases of both the $WZ$ and the $WH_{125}$ mediated $3l+\met$ channels. We recognize that other superpartners would also be accessible, but for definiteness focus only on the golden trilepton signal from the lightest ino states.

\subsubsection{$WZ$ mediated $3l+\met$ searches at the HE-LHC}
\label{subsec:wz_helhc}
We select events with exactly three isolated leptons, where lepton isolation as described in Sec.~\ref{sec:wz_hllhc} is used.
Two isolated leptons are required to form a SFOS lepton pair with invariant mass in the range $M_{Z}\pm10~{\rm GeV}$. In the presence of two such SFOS lepton pairs, the one which minimizes the transverse mass of the non-SFOS lepton and $\met$ is chosen as the correct SFOS pair. In addition, a veto on $b$ jet with $p_{T} > 30~{\rm GeV}$ and $|\eta| < 2.5$ is also applied. On-shell and off-shell $WZ$, $VVV$, $t\bar{t}V$ and $ZZ$ are the background sources. Signal events have been generated by varying $M_{\lspone}$ over the range of $0-1000~{\rm GeV}$ with a step size of $30~{\rm GeV}$ while $M_{\lspthree,\lsptwo,\chonepm}$ (assuming, $M_{\lspthree} = M_{\lsptwo} = M_{\chonepm}$) has been varied from $100~{\rm GeV}$ to $1780~{\rm GeV}$ with a step size of $30~{\rm GeV}
$.

As before, we have examined various distributions for several benchmark points with a diverse range of parent ino massses and mass gaps: [$M_{\lspthree,\lsptwo,\chonepm}$, $M_{\lspone}$] (in GeV): BPA3 [130,30], BPB3 [250,30], BPC3 [340,30], BPD3 [520,0], BPE3 [520,390], BPF3 [1000,0], BPG3 [1000,480], BPH3 [1000,780], BPI3 [1420,0], BPJ3 [1420,990]. Based on this analysis, we choose the ten signal regions that optimize the significance of the signal for a variety of ino and LSP masses (though, of course, for the NMSSM analysis of interest, the LSP is light). The selection cuts are listed in Table~\ref{tab:wz_helhc_sr}.

\begin{table}[!htb]
\begin{center}\scalebox{0.8}{
\begin{tabular}{ C{2.2cm} | C{1.6cm} C{1.6cm} C{1.6cm} C{1.6cm} C{1.6cm} C{1.9cm} C{1.6cm} C{1.6cm}  C{1.9cm} C{1.6cm} }
\hline \hline
Cuts & SRA3 & SRB3 & SRC3 & SRD3 & SRE3 & SRF3 & SRG3 & SRH3 & SRI3 & SRJ3 \\\hline 
$M_{\lsptwo,\lspthree,\chonepm}$ [GeV] & 130 & 250 & 340 & 520 & 520 & 1000 & 1000 & 1000 & 1420 & 1420 \\
$M_{\lspone}$ [GeV] & 30 & 30 & 30 & 0 & 390 & 0 & 480 & 780 & 0 & 990 \\ \hline 
$\Delta \phi_{SFOS-\met}$ & - & $[2.6:\pi]$ & $[2.6:\pi]$ & $[1.4:\pi]$ & $[1.4:\pi]$ & $[1.6:\pi]$ & $[1.5:\pi]$ & $[2.1:\pi]$ & $[1.7:\pi]$ & $[1.4:\pi]$ \\
$\Delta R_{SFOS}$ & $[1.3:3.8]$ & $[0.2:1.4]$ & $[0.1:1.3]$ & - & $[0.3:2.1]$ & $[0.0:1.2]$ & $[0.0:1.2]$ & $[0.3:1.5]$ & $\leq 1.0$ & $[0.2:0.8]$ \\ 
$\met$ [GeV]& $[80:160]$ & $\geq 150$ & $\geq 200$ & $\geq 250$ & $\geq 400$ & $\geq 220$  & $\geq 300$ & $\geq 250$ & $\geq 220$ & $\geq 250$ \\ 
$M_{T}^{l_{W}}$ [GeV]& $[100:130]$ & $\geq 130$ & $\geq 140$ & $\geq 200$ & $\geq 100$ & $\geq 230$ & $\geq 160$ & $\geq 140$ & $\geq 250$ & $\geq 210$\\ 
$M_{CT}^{l_{W}}$ [GeV]& $[70:230]$ & $\geq 200$ & $\geq 200$ & $\geq 150$ & $\geq 100$ & $\geq 200$ & $\geq 200$ & $\geq 140$ & $\geq 250$ & $\geq 180$\\ 
$p_{T}^{l_{1}}$ [GeV]& $[80:160]$ & $\geq 90$ & $\geq 110$ & $\geq 170$ & $\geq 100$ & $\geq 220$ & $\geq 200$  & $\geq 130$ & $\geq 270$ & $\geq 200$\\ 
$p_{T}^{l_{2}}$ [GeV]& $\geq 50$ & $\geq 50$ & $\geq 70$ & $\geq 90$ & $\geq 50$ & $\geq 120$ & $\geq 120$  & $\geq 90$ & $\geq 130$ & $\geq 150$\\
$p_{T}^{l_{3}}$ [GeV]& $\geq 40$ & $\geq 40$ & $\geq 40$ & $\geq 50$ & $\geq 40$ & $\geq 70$ & $\geq 60$  & $\geq 50$ & $\geq 80$ & $\geq 50$\\ \hline
\end{tabular}}
\caption{List of selection cuts corresponding to the signal regions for searches in $WZ$ mediated $3l+\met$ final state at the HE-LHC. The signal regions have been optimized through a cut-based analysis to yield maximum significance for the signal samples corresponding to the respective values of $M_{\lsptwo,\lspthree,\chonepm}$ and $M_{\lspone}$.}
\label{tab:wz_helhc_sr}
\end{center}
\end{table}

\begin{figure}[htb!]
\begin{center}
\includegraphics[scale=0.18]{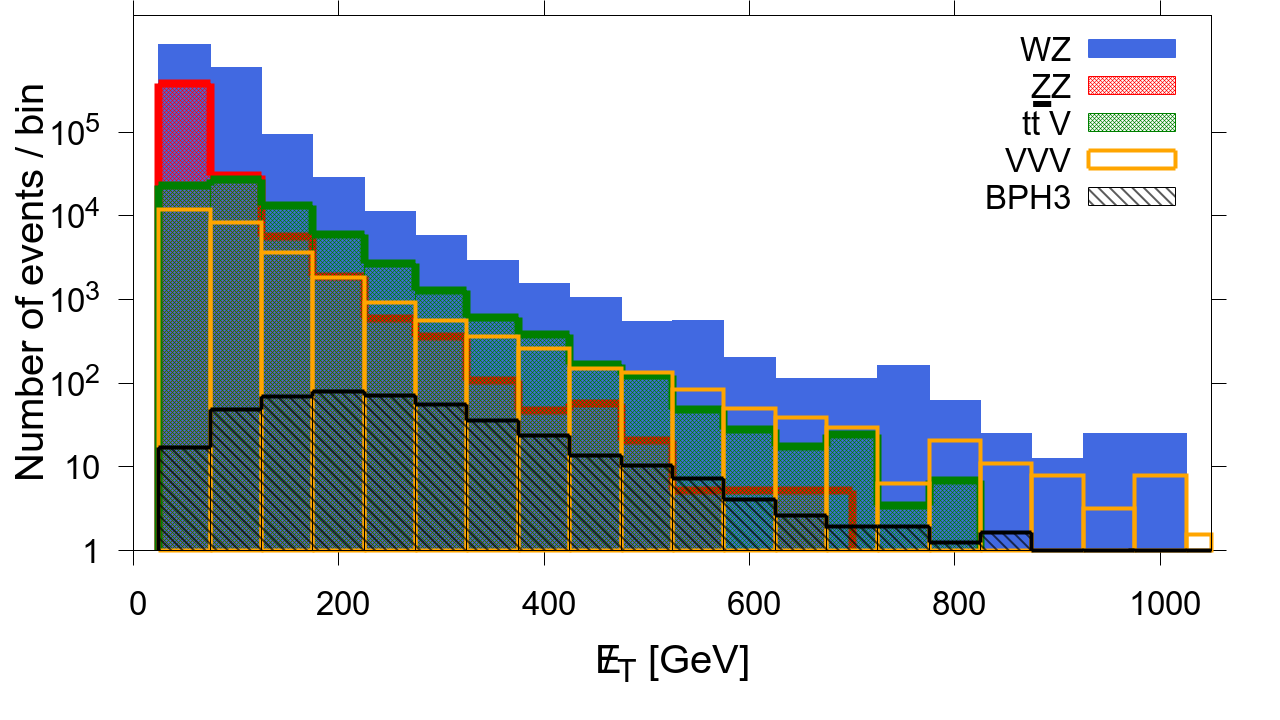}\includegraphics[scale=0.18]{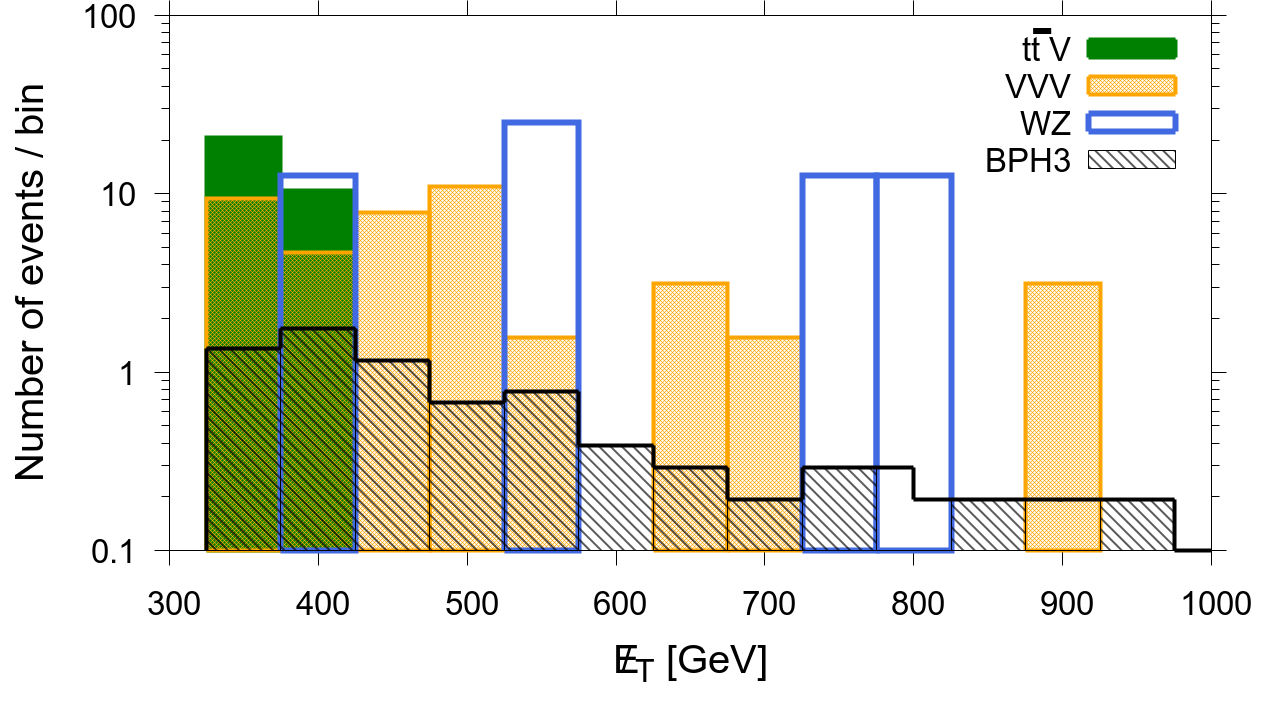}\\
\includegraphics[scale=0.18]{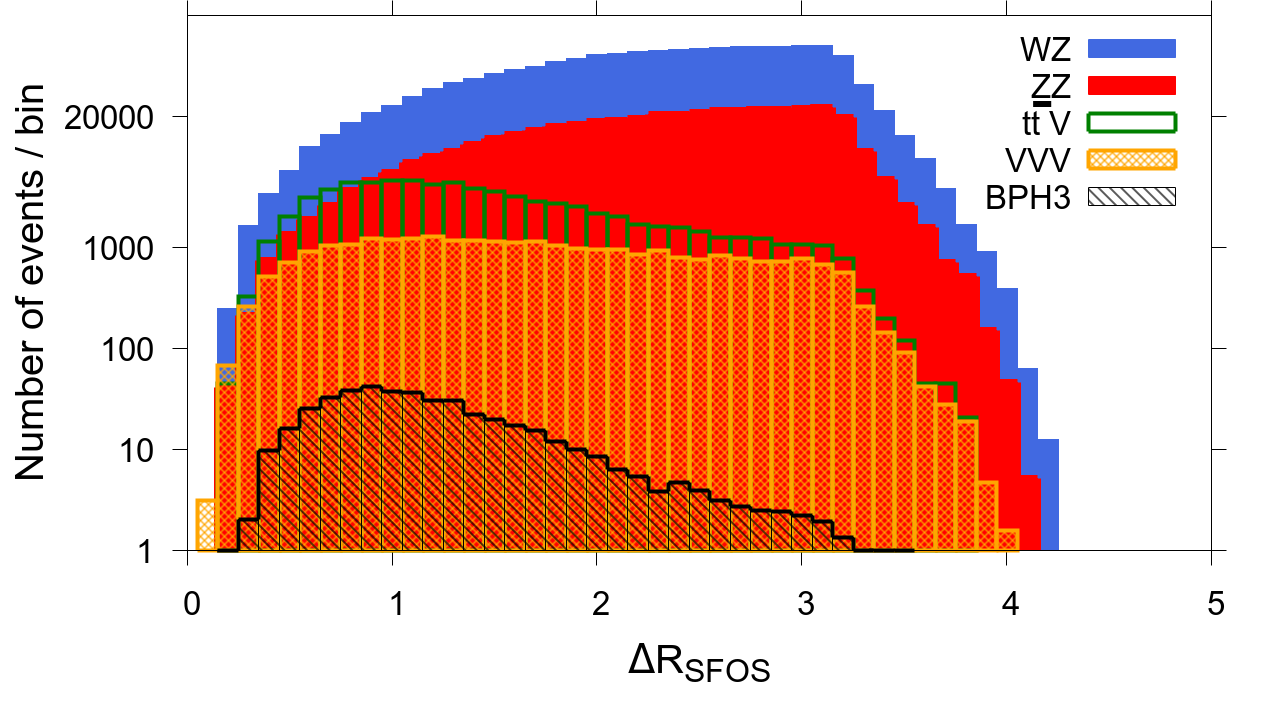}\includegraphics[scale=0.18]{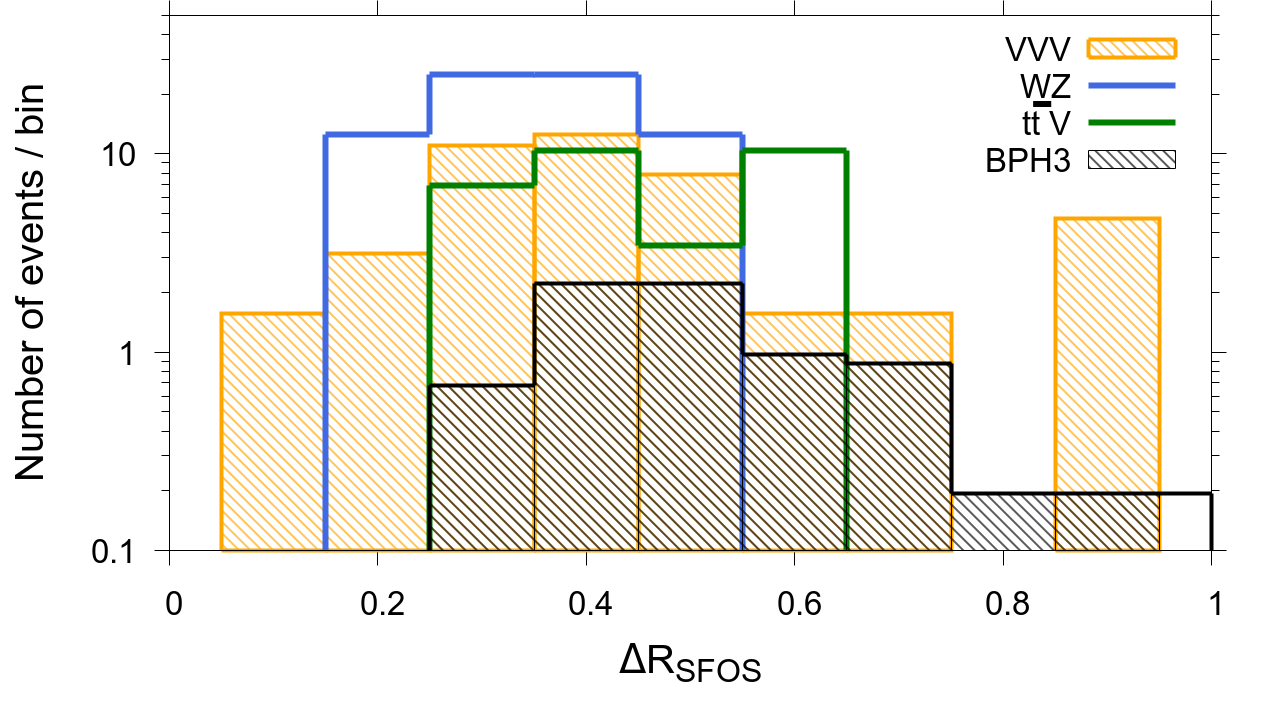}\\
\caption{The event distribution of $\met$ (top) and $\Delta R_{SFOS}$ (bottom) at the HE-LHC is shown for the signal benchmark point BPH3( $M_{\lspthree,\lsptwo,\chonepm}=1000~{\rm GeV}$, $M_{\lspone}=780~{\rm GeV}$) in black color. The corresponding event distributions for on-shell and off-shell $WZ$, $VVV$, $ZZ$ and $t\bar{t}V$ have been shown in blue, orange, red and green colors, respectively. The left (right) panel shows the corresponding distributions before (after) the application of SRH3 cuts.}
\label{fig:WZ_kin_dist_B3}
\end{center}
\end{figure}

\begin{table}[!htb]
\begin{center}\scalebox{0.85}{
\begin{tabular}{||C{3.2cm} || C{1.5cm} | C{1.2cm} C{1.2cm} C{1.2cm} C{1.2cm} C{1.2cm} C{1.2cm} C{1.2cm} C{1.2cm} C{1.2cm} C{1.2cm} ||}
\hline \hline
Background & Cross & \multicolumn{10}{c||}{Background yield ($27~{\rm TeV}$, $15~ab^{-1}$)} \\ \cline{3-12}
process & section [LO] & SRA3 & SRB3 & SRC3 & SRD3 & SRE3 & SRF3 & SRG3 & SRH3 & SRI3 & SRJ3 \\ \hline\hline 
\textbf{$t\bar{t}V$} & \multirow{3}{*}{1385 fb} & \multirow{3}{*}{79.6} & \multirow{3}{*}{166} & \multirow{3}{*}{107} & \multirow{3}{*}{69.2} & \multirow{3}{*}{38.1} & \multirow{3}{*}{41.5} & \multirow{3}{*}{31.2} & \multirow{3}{*}{96.9} & \multirow{3}{*}{20.8} & \multirow{3}{*}{27.7} \\ 
($V=W,Z$) & & & & & & & & & & & \\ 
($W \to l\nu$, $Z \to ll$) & & & & & & & & & & & \\ \hline 
\textbf{$WZ$} & 1263 fb & 341  & 164 & 126 & 101 & 88.4 & 75.8 & 75.8 & 75.8 & 63.1 & 63.1 \\ \hline 
\textbf{$VVV$} ($V = W,Z$) & 681 fb & 31.4 & 132 & 99 & 70.7 & 66.0 & 33.0 & 44.0 & 70.7 & 25.1 & 36.1 \\ \hline
$ZZ$&  \multirow{2}{*}{2092} & \multirow{2}{*}{0.0} & \multirow{2}{*}{0.0} & \multirow{2}{*}{0.0} & \multirow{2}{*}{10.4} & \multirow{2}{*}{5.18} & \multirow{2}{*}{0.0} & \multirow{2}{*}{0.0} & \multirow{2}{*}{5.18} & \multirow{2}{*}{0.0} & \multirow{2}{*}{0.0} \\ 
(leptonic) &  &  &  &  &  &  &  &  &  &  &  \\ \hline 
\multicolumn{2}{|c|}{\textbf{Total Background}} & 452 & 462 & 333 & 251 & 198 & 150 & 151 & 249 & 109 & 127 \\\hline\hline 
\end{tabular}}
\caption{The background yields for $\sqrt{s}=27~{\rm TeV}$ LHC corresponding to $15~ab^{-1}$ of integrated luminosity, for the $8$ different signal regions considered for the cut-based analysis, are presented. The leading order (LO) cross sections generated by \texttt{MadGraph5$\_$aMC@NLO} have also been shown. The $t\bar{t}V$ background has been specifically decayed into leptonic final states.}
\label{tab:bkg_helhc}
\end{center}
\end{table}

As an illustration, we show the $\met$ and $\Delta R_{SFOS}$ distributions for the benchmark case BPH3 together with those for the main SM backgrounds (for HE-LHC) in Fig.~\ref{fig:WZ_kin_dist_B3}. The signal event distributions are represented in a black colored shade while the distributions for $WZ$, $VVV$, $t\bar{t}V$ and $ZZ$ backgrounds are shown in blue, orange, green and red colors, respectively. Note that BPH3 is the signal benchmark point with intermediate mass gap between the NLSP and the LSP, and was used to derive the optimized set of selection cuts corresponding to SRH3. The plots in the left panel of Fig.~\ref{fig:WZ_kin_dist_B3} have been obtained by imposing the requirement of three isolated leptons in the final state, presence of at least one SFOS pair of leptons with invariant mass in the range $ M_{Z}\pm 10~{\rm GeV}$, and the absence of any $b$ $jet$ with $p_{T} \geq 30~{\rm GeV}$ and $|\eta| < 2.5$.  
The distribution in the right hand panels of Fig.~\ref{fig:WZ_kin_dist_B3} are obtained after all the SRH3 cuts in Table~\ref{tab:wz_helhc_sr}.

\begin{figure}[!htb]
\begin{center}
\includegraphics[scale=0.15]{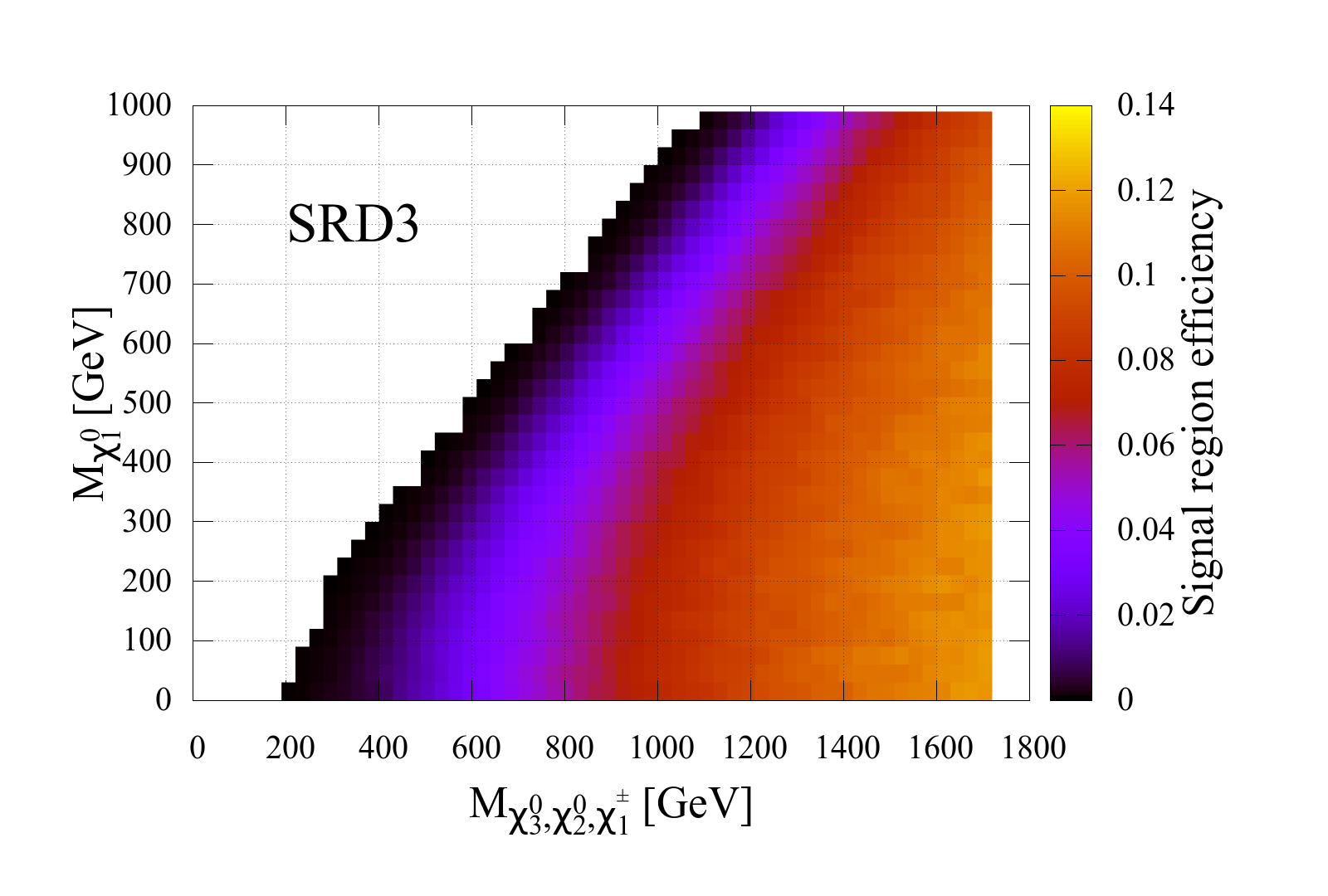}\includegraphics[scale=0.15]{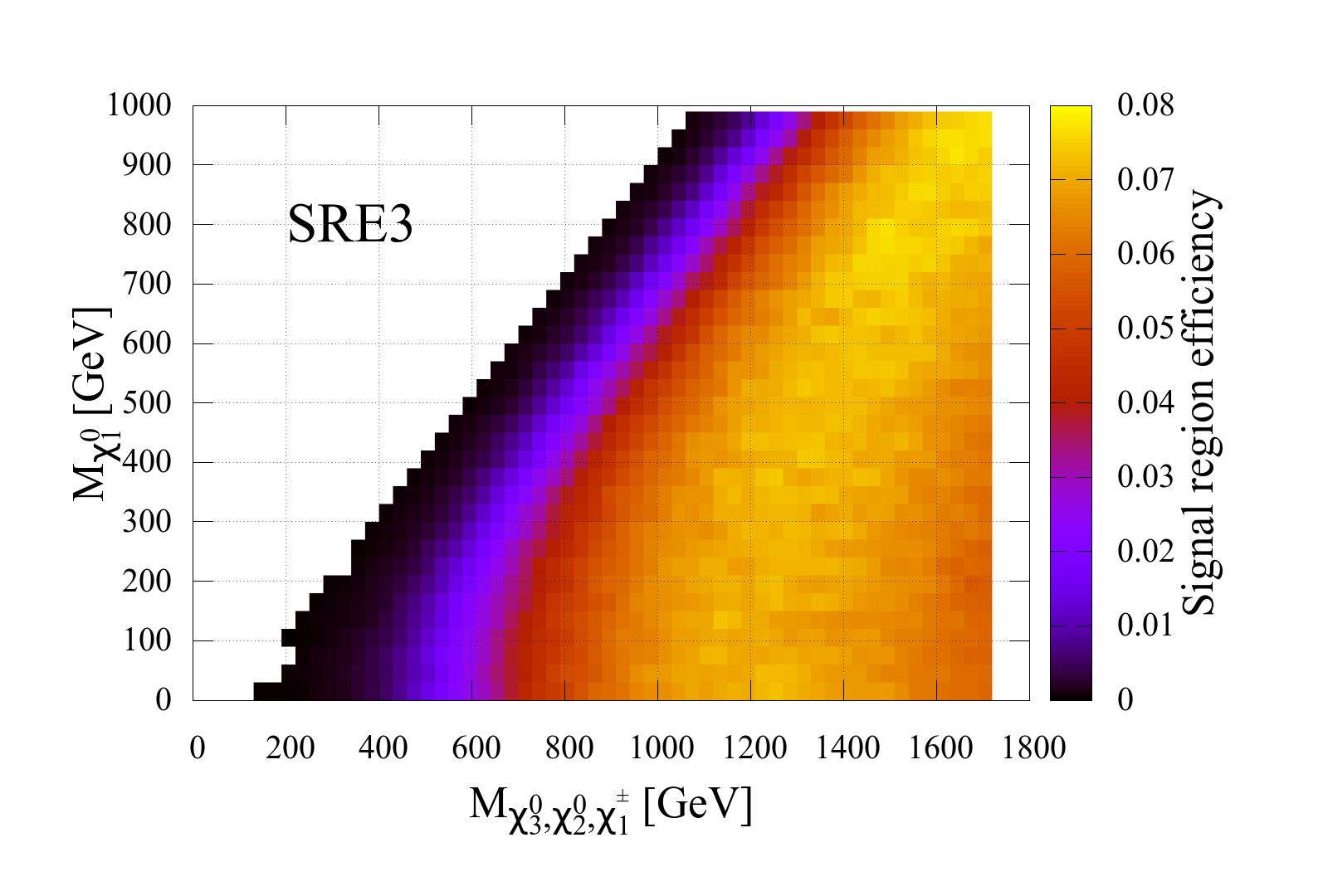}\\
\includegraphics[scale=0.15]{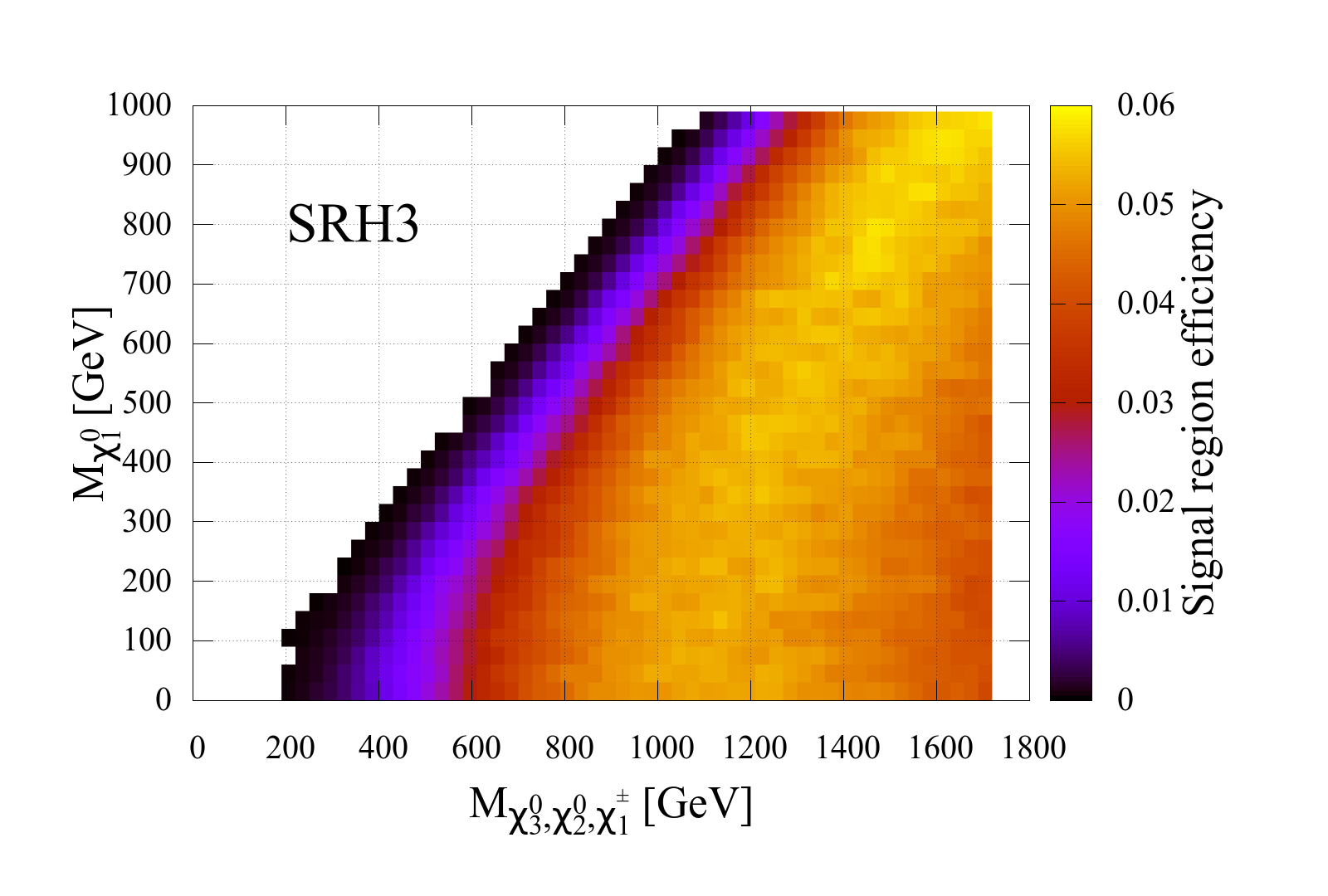}\includegraphics[scale=0.15]{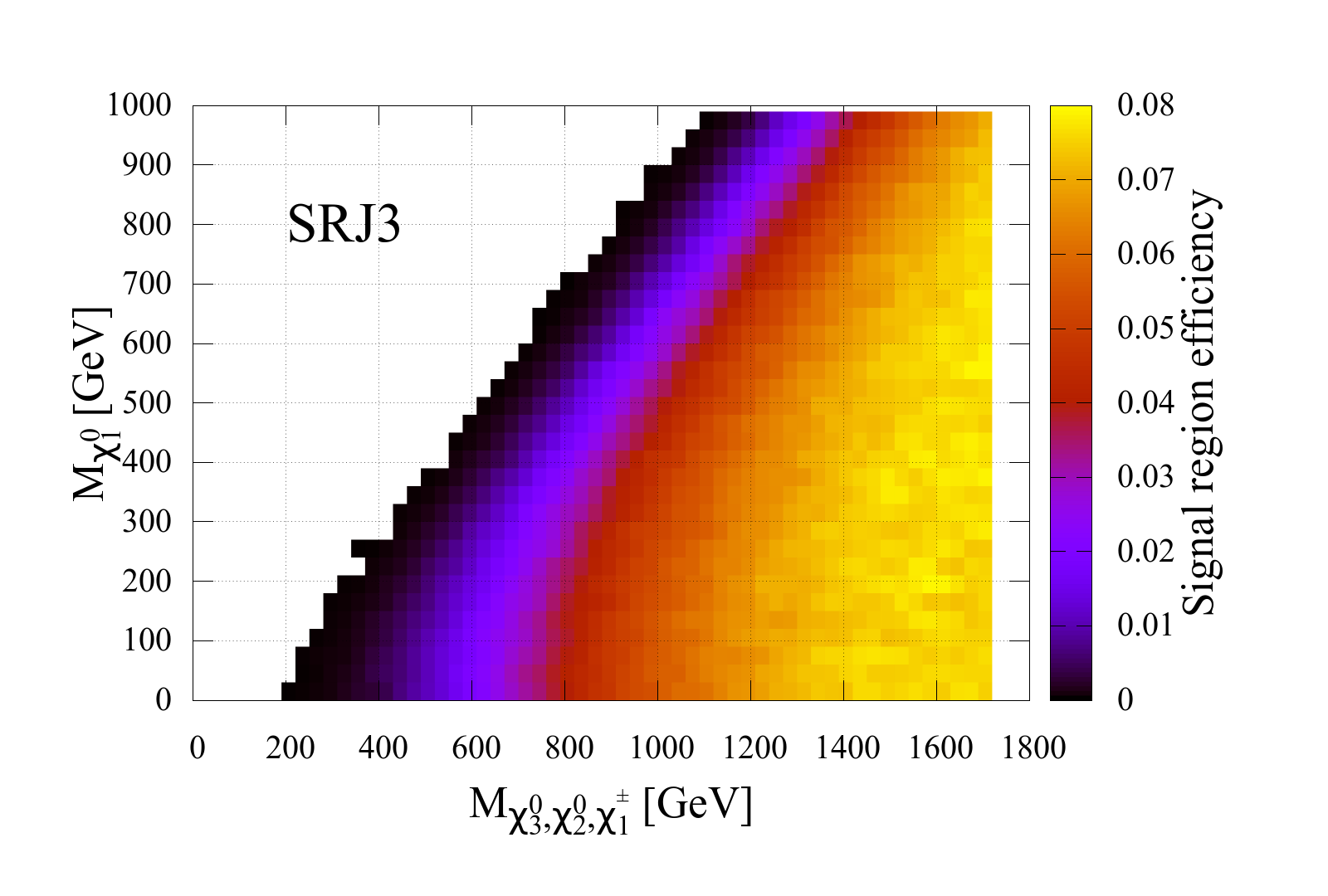}\\
\caption{ The efficiency map for SRD3, SRE3, SRH3 and SRJ3 has been shown as a color palette in the $M_{\lspone}-M_{\lsptwo}$ plane. The top-left and top-right figures represent the efficiency map of SRD3 and SRE3, respectively. The bottom grid represents the efficiency map for SRH3 (left) and SRJ3 (right), respectively. Note the difference in the vertical scales in the four frames.}
\label{fig:27_eff_grid}
\end{center}
\end{figure}

As before, the dominant physics SM backgrounds to the hadronically quiet $3l+\met$ signal is expected to come from $WZ$ (which includes $W^{(*)}Z$ and $W^{(*)}\gamma^*$) production, $t\bar{t}V$ production, $VVV$ production and $ZZ$ production. The corresponding background levels obtained using \texttt{MadGraph5$\_$aMC@NLO} are shown in Table~\ref{tab:bkg_helhc} after all our analysis cuts. The efficiency map of SRD3, SRE3, SRH3 and SRJ3 are illustrated in Fig.~\ref{fig:27_eff_grid} in the $M_{\lspthree,\lsptwo,\chonepm}-M_{\lspone}$ plane with the color palette representing the signal efficiency. Again, we have assumed a systematic uncertainty of $5\%$. For this analysis, we took  $\mathcal{L}= 15~ab^{-1}$. The total background yield, $B$, has been computed by summing over the yields from all relevant backgrounds.

\begin{figure}[!htb]
\begin{center}
\includegraphics[scale=0.26]{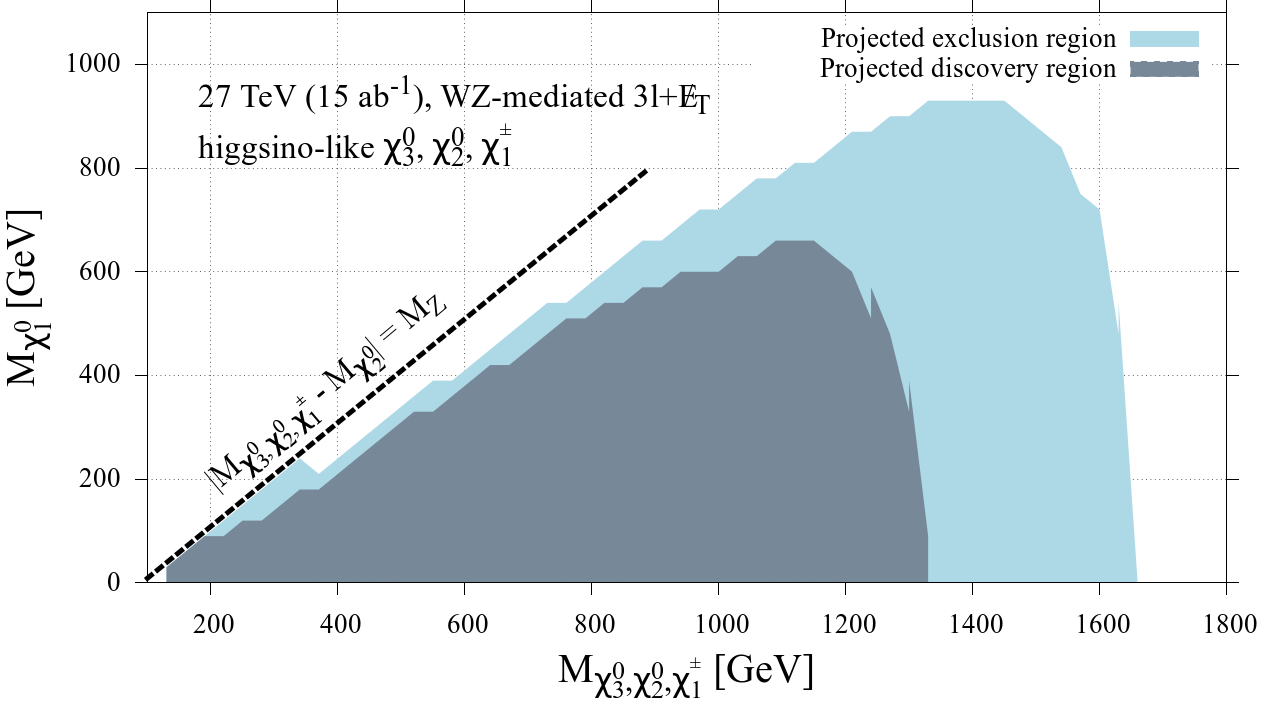}
\caption{ The projected exclusion (blue color) and discovery regions (grey color) from searches in the $WZ$ mediated $3l+\met$ final state produced from the cascade decay of directly produced mass degenerate higgsino-like $\lsptwo\chonepm+\lspthree\chonepm$ at $\sqrt{s}=27~{\rm TeV}$ LHC corresponding to an integrated luminosity of $15~ab^{-1}$, are shown.}
\label{fig:27_higgsino_contour}  
\end{center}
\end{figure}

We show the projected exclusion region ($S_{\sigma} >2\sigma$, blue color) and the projected discovery region ($S_{\sigma}>5\sigma$, grey color) obtained from doublet higgsino searches in the $WZ$ mediated $3l+\met$ final state at the HE-LHC in Fig.~\ref{fig:27_higgsino_contour}. It can be seen from Fig.~\ref{fig:27_higgsino_contour} that HE-LHC will be able to discover (exclude) higgsino-like $\lspthree,~\lsptwo$ and $\chonepm$ with masses up to $\sim 1330~(1660)~{\rm GeV}$ if the LSP is light. 

\subsubsection{$WH_{125}$-mediated $3l+\met$ searches at the HE-LHC}

As in Sec.~\ref{sec:wh_hllhc}, we consider the signal from chargino-neutralino pair production where the neutralinos decay to $H_{125}$ and the chargino decays to $W$. Dominant contributions to the $3l+\met$ signal arise when $W$ decays leptonically, and $H_{125} \rightarrow WW^* \rightarrow l^{\prime}\nu l^{\prime}\nu$ or $H_{125} \rightarrow \tau^+\tau^-$ and the taus decay leptonically.
Signal events have been generated for $M_{\lspthree,\lsptwo,\chonepm}$ varying from $100~{\rm GeV}$ to $1700~{\rm GeV}$, while $M_{\lspone}$ has been varied from $0~{\rm GeV}$ to $1000~{\rm GeV}$ with a step size of $30~{\rm GeV}$. The event selection criteria prescribed in Sec.~\ref{sec:wh_hllhc} is followed here. Background contributions arise from $WH_{125}$ and $WZ$, $VVV$, $t\bar{t}V$ and $ZZ$. 

\begin{table}[!htb]
\begin{center}
\begin{tabular}{|| C{2.5cm} || C{1.55cm} C{1.55cm} C{1.55cm} C{1.55cm} C{1.55cm} C{1.55cm} C{1.55cm}||}
\hline \hline
Cuts & SRA4 & SRB4 & SRC4 & SRD4 & SRE4 & SRF4 & SRG4 \\\hline 
$M_{\lsptwo,\lspthree,\chonepm}$ [GeV] & 160 & 550 & 550 & 910 & 910 & 1240 & 1240 \\
$M_{\lspone}$ [GeV] & 30 & 0 & 420 & 0 & 750 & 0 & 900  \\ \hline \hline
$p_{T}^{l_{1}}$ [GeV] & $ \geq 80$ & $\geq 120$ & $[120:170]$ & $\geq 200$ & $[100:200]$ & $\geq 200$ & $\geq 150$  \\ 
$p_{T}^{l_{2}}$ [GeV] & $ \geq 50$ & $\geq 80$ & $\geq 40$ & $\geq 100$ & $\geq 40$ & $\geq 120$ & $\geq 50$ \\  
$p_{T}^{l_{3}}$ [GeV] & $ \geq 40$  & $\geq 40$ & $\geq 30$ & $\geq 50$ & $\geq 30$ & $\geq 50$ & $\geq 30$ \\
$M_{T}^{l_{1}}$ [GeV] & $[125:225]$ & $\geq 235$ & $\geq 150$ & $\geq 250$ & $\geq 150$ & $\geq 350$ & $\geq 220$ \\ 
$M_{T}^{l_{2}}$ [GeV] & $\geq 100$ & $\geq 150$ & $\geq 100$ & $\geq 170$ & $\geq 100$ & $\geq 150$ & $\geq 150$ \\ 
$M_{T}^{l_{3}}$ [GeV] & $\geq 100$ & $\geq 120$ & $\geq 100$ & $\geq 120$ & $\geq 100$ & $\geq 120$ & $\geq 120$ \\
$\Delta R_{OS}^{min}$ & $[0.2:0.8]$ & $\leq 0.9$ & $[0.2:1.2]$ & $\leq 0.6$ & $[0.2:0.8]$ & $ \leq 0.9 $ & $\leq 1.2$  \\      
$\Delta R_{OS}^{max}$ & $[0.7-2.5]$ & $[0.4:2.8]$ & $[0.3:3.0]$ & $[0.3:3.3]$ & $[0.5:3.5]$ & $[0.2:2.6]$ & $[0.4:2.4]$  \\ 
$M_{OS,min}^{inv.}$ [GeV] & $ \leq 60$ & $\leq 60$ & $\leq 60$ & $\leq 120$ & $\leq 150$ & $\leq 150$ & $\leq 50$  \\ 
$\met$ [GeV] & $[80:180]$ & $\geq 150$ & $\geq 150$ & $\geq 150$ & $\geq 150$ & $\geq 200$ & $\geq 200$ \\ \hline
\end{tabular}
\caption{ List of selection cuts corresponding to the signal regions for $WH_{125}$-mediated $3l+\met$ final state. The signal regions have been optimized through a cut-based analysis to yield maximum significance for the signal samples corresponding to the respective values of $M_{\lsptwo,\lspthree,\chonepm}$ and $M_{\lspone}$.}
\label{tab:wh_helhc_sr}
\end{center}
\end{table}

\begin{figure}[h]
\begin{center}
\includegraphics[scale=0.18]{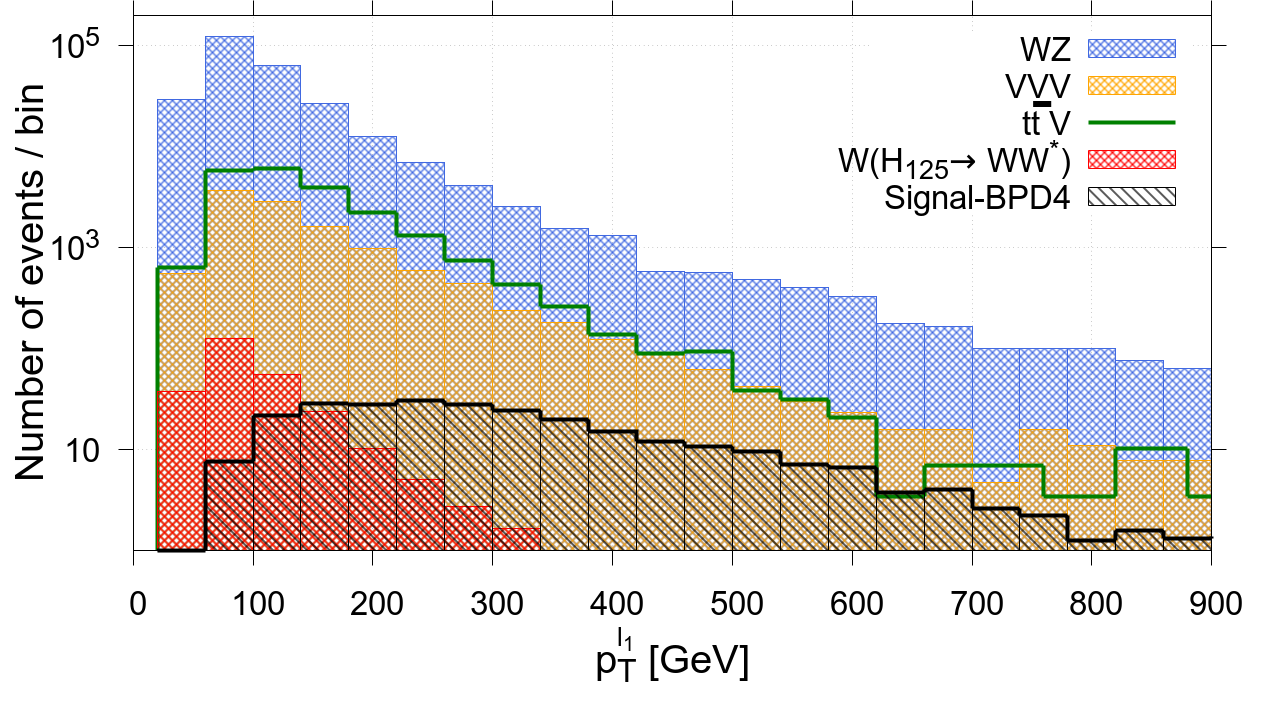}\includegraphics[scale=0.18]{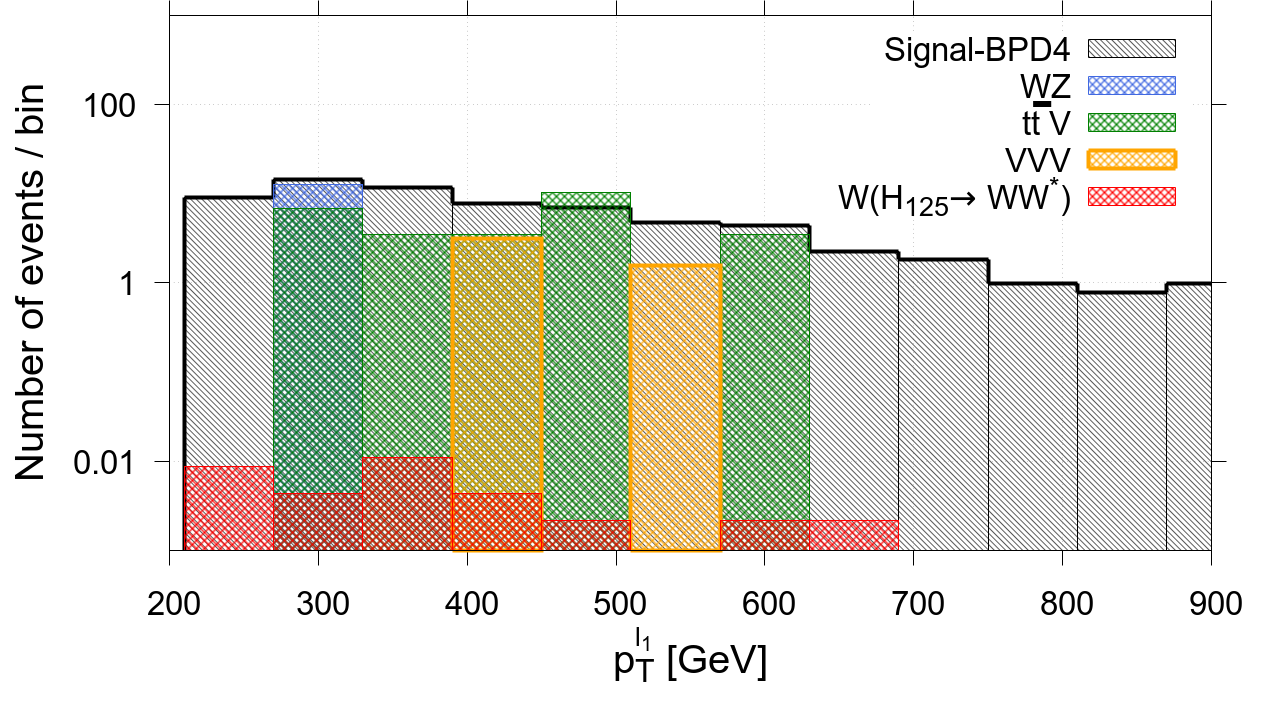}\\
\includegraphics[scale=0.18]{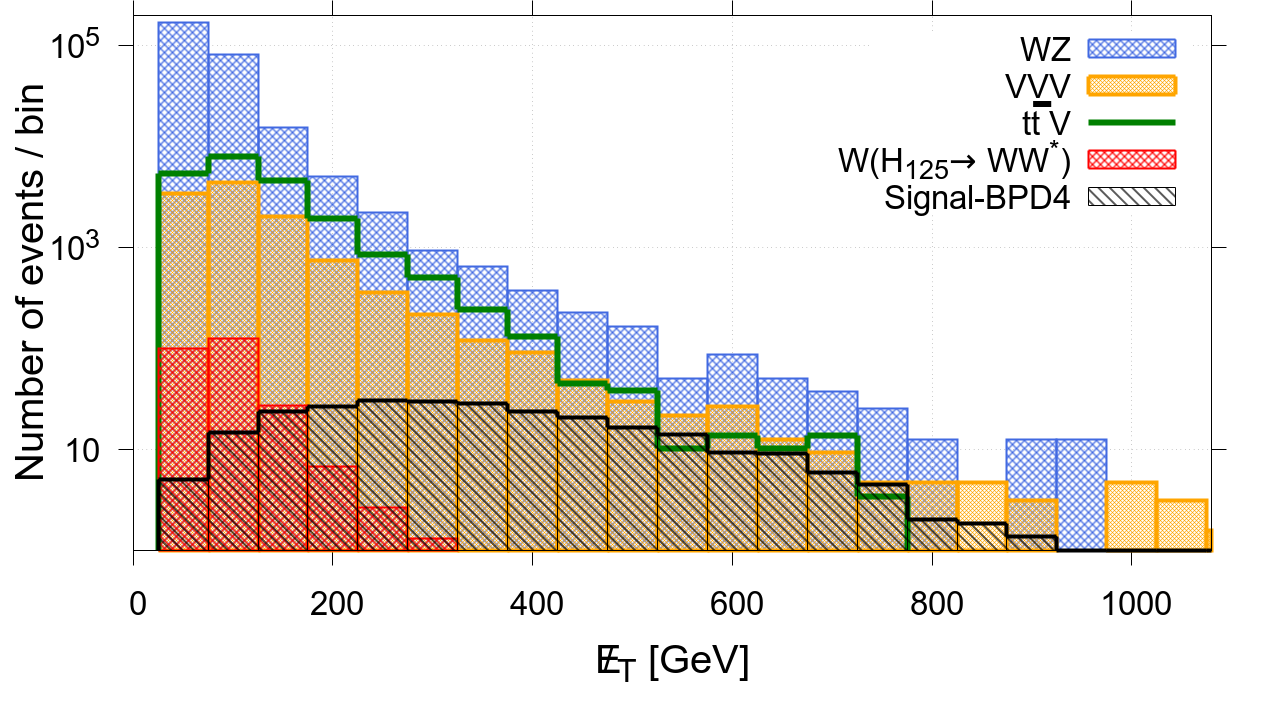}\includegraphics[scale=0.18]{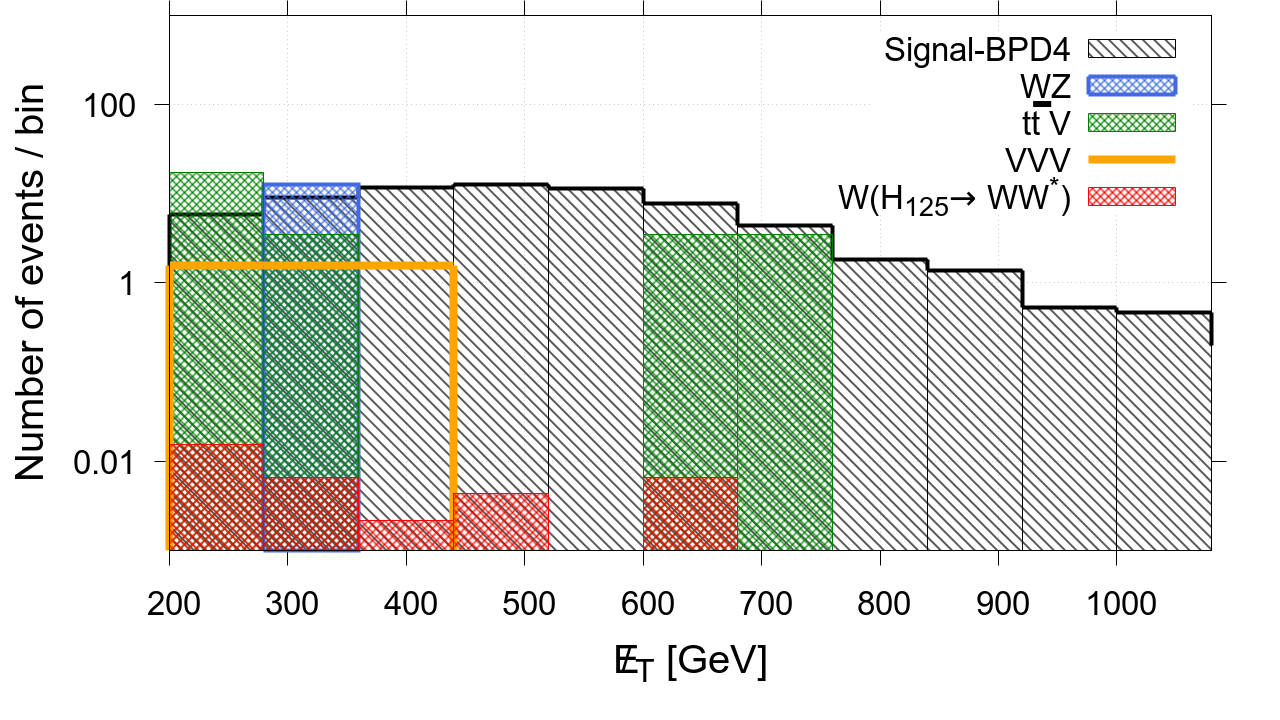}\\
\caption{Event distribution of $p_{T}^{l_{1}}$ (top) and $\met$ (bottom) at the HE-LHC for BPD4~[$M_{\lsptwo,\lspthree,\chonepm} = 910~{\rm GeV}$, $M_{\lspone} = 0~{\rm GeV}$] used to obtain the optimized signal region SRD4 (shown in black shade), and the background processes: $t\bar{t}V$ (shown in green color), $WZ$ (shown in blue color), $WH_{125}$ (shown in red color) and $VVV$ (shown in orange color). The figures on the left have been obtained after demanding the presence of three isolated leptons in the final state, $b$ $jet$ veto and a veto on SFOS pair with invariant mass in the range $M_{Z} \pm 12~{\rm GeV}$. The figures on the right have been obtained upon a further application of SRD4 cuts.}
\label{fig:wh_27_550_420}
\end{center}
\end{figure}

We choose $7$ optimized signal regions: SRA4, SRB4, SRC4, SRD4, SRE4, SRF4 and SRG4 in order to perform the current analysis. SRA4, SRC4, SRE4 and SRG4 (SRB4, SRD4 and SRF4) have been optimized to maximize the signal significance of signal events with small (large) mass difference between the higgsino NLSPs and the LSP. The signal regions have been obtained by optimizing the signal significances of signal processes corresponding to the following values of $\left[M_{\lspthree,\lsptwo,\lspone},M_{\lspone}\right]$ (in $\rm GeV$): BPA4~$\left[160,30\right]$, BPB4~$\left[550,0\right]$, BPC4~$\left[550,420\right]$, BPD4~$\left[910,0\right]$, BPE4~$\left[910,750\right]$, BPF4~$\left[1240,0\right]$ and BPG4~$\left[1240,900\right]$. The kinematic variables used to design the optimized signal regions are: $p_{T}^{l_{1}}$, $p_{T}^{l_{2}}$, $M_{T}^{l_{1}}$, $M_{T}^{l_{2}}$, $M_{T}^{l_{3}}$, $\Delta R_{OS}^{min}$, $\Delta R_{OS}^{max}$, $M_{OS,min}^{inv.}$ and $\met$. Here, $\Delta R_{OS}^{min}$ and $\Delta R_{OS}^{max}$ represents the $\Delta R$ between OS leptons with minimum and maximum $\Delta R$ separation, respectively, while, $M_{OS,min}^{inv.}$ represents the invariant mass of the OS lepton pair with minimum $\Delta R$. $M_{T}^{l_{i}}$ ($i=1,2,3$) corresponds to the transverse mass of the $\met$ and $p_{T}$ ordered $i^{th}$ final state lepton. The set of optimized cuts corresponding to the signal regions are presented in Table~\ref{tab:wh_helhc_sr}.

\begin{figure}[!htb]
\begin{center}
\includegraphics[scale=0.18]{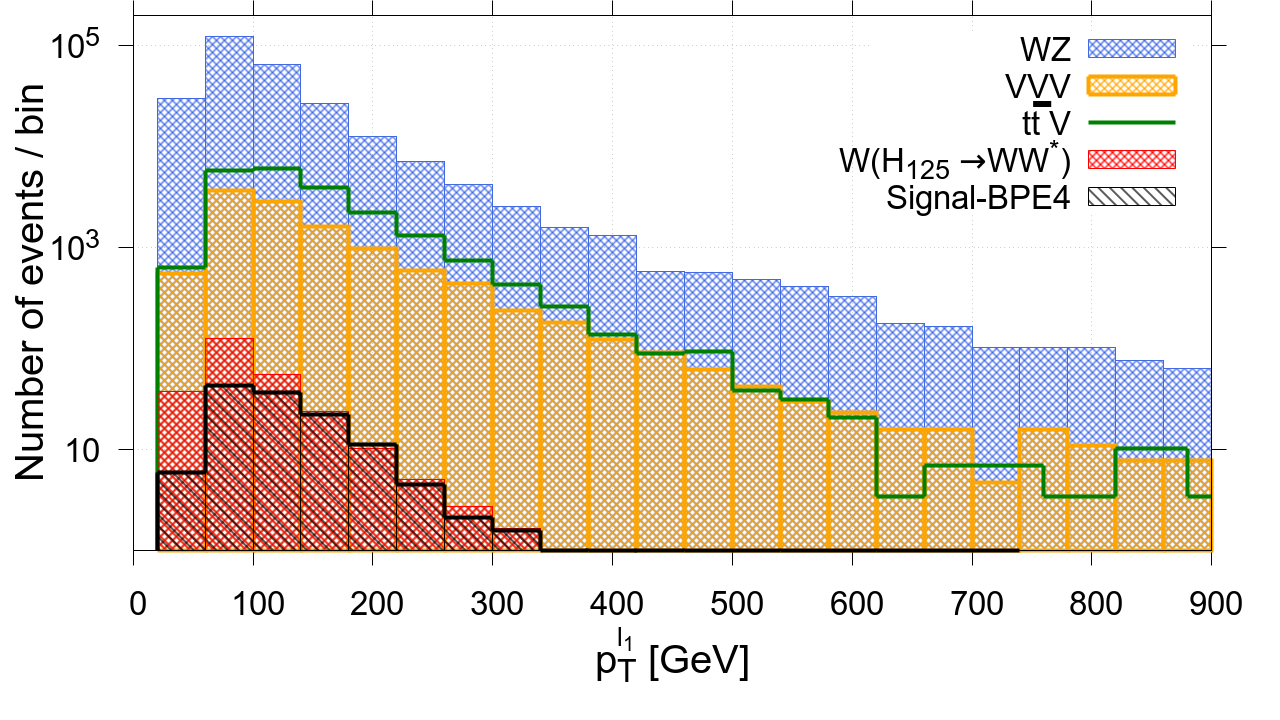}\includegraphics[scale=0.18]{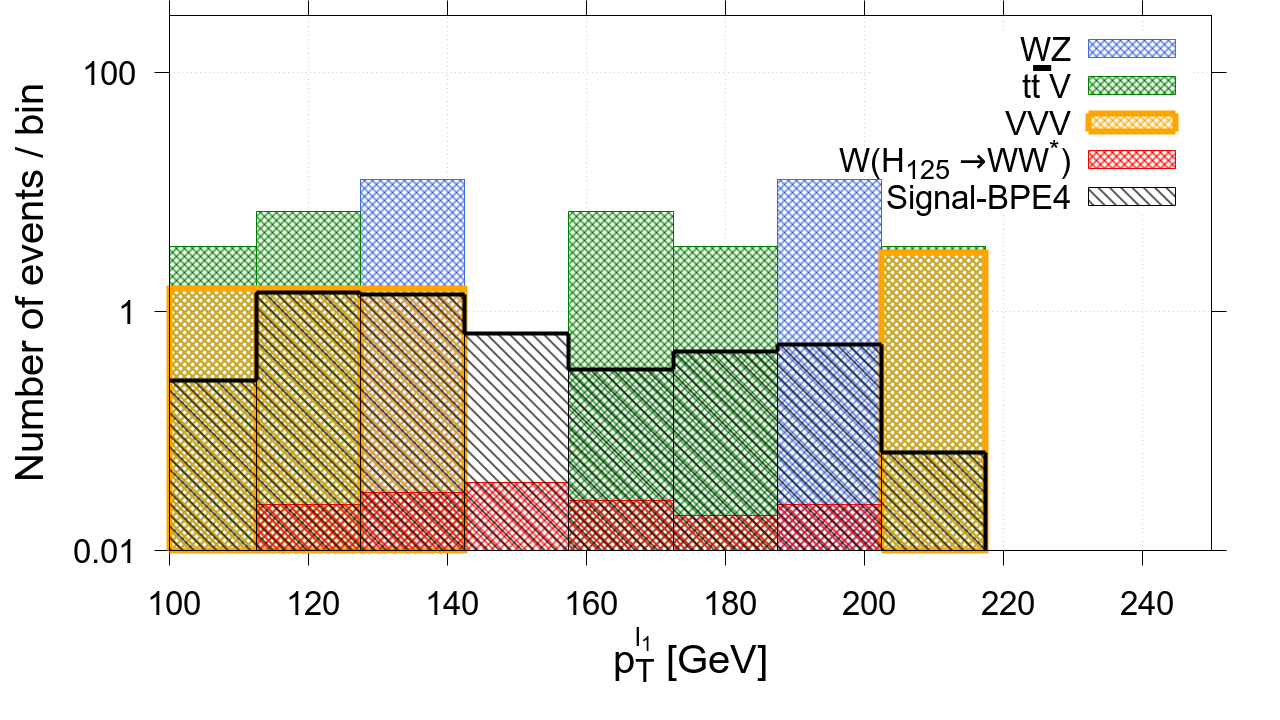}\\
\includegraphics[scale=0.18]{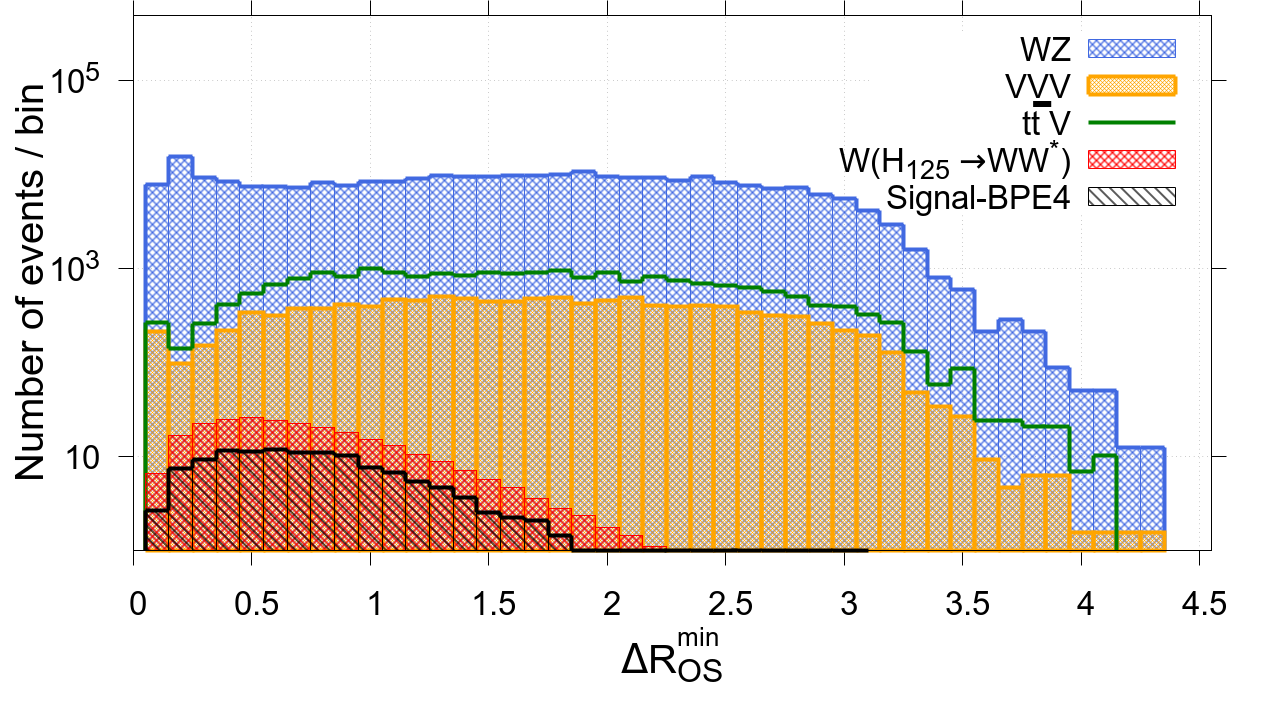}\includegraphics[scale=0.18]{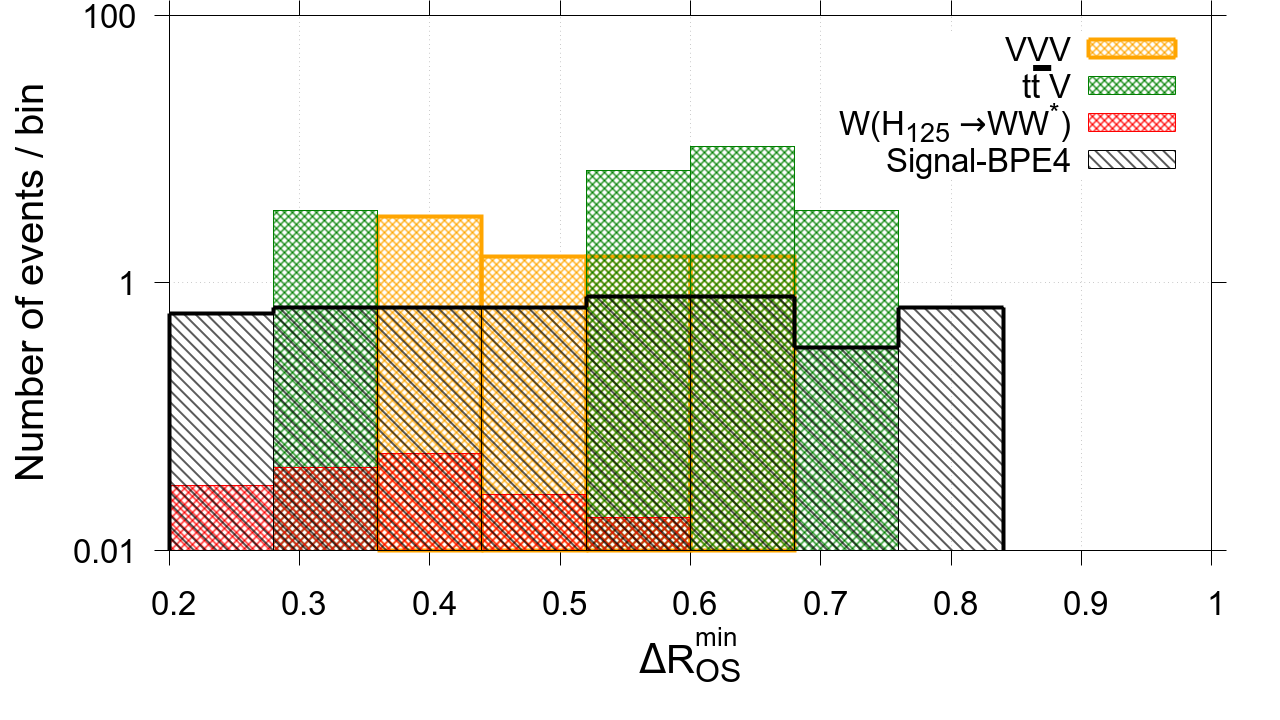}\\
\caption{Event distribution of $p_{T}^{l_{1}}$ (top) and $\Delta R_{OS}^{min}$ (bottom) at the HE-LHC for BPE4~[$M_{\lsptwo,\lspthree,\chonepm} = 910~{\rm GeV}$, $M_{\lspone} = 750~{\rm GeV}$] used to obtain the optimized signal region SRE4 (shown in black shade), and the background processes: $t\bar{t}V$ (shown in green color), $WZ$ (shown in blue color), $WH_{125}$ (shown in red color) and $VVV$ (shown in orange color). The figures on the left have been obtained after demanding the presence of three isolated leptons in the final state, $b$ $jet$ veto and a veto on SFOS pair with invariant mass in the range $M_{Z} \pm 12~{\rm GeV}$. The figures on the right have been obtained upon a further application of SRE4 cuts.}
\label{fig:wh_27_850_0}
\end{center}
\end{figure}

Although not directly germane to the NMSSM signal with a light LSP, we find it interesting to show the event distribution of $p_T^{l_1}$ and $\met$  for BPD4 at the HE-LHC. The signal distribution is shown by the black-hatched
region in Fig.~\ref{fig:wh_27_550_420} together with the distributions of the relevant background processes ($t\bar{t}V$: green color, $W H_{125}$: red color, $VVV$: orange color, and, $WZ$: blue color). The distributions on the left panels of Fig.~\ref{fig:wh_27_550_420} have been obtained by requiring the presence of exactly three isolated leptons in the final state, the absence of any SFOS lepton pair with invariant mass in the range $M_{Z} \pm 12~{\rm GeV}$ and $b$-$jet$ veto. In the case of BPD4, the large mass difference between $\lspthree,\lsptwo,\chonepm$ and $\lspone$ results in the leading $p_{T}$ lepton getting produced with a larger boost and therefore peaks at a higher value (around $ 200~{\rm GeV}$) and has a flatter distribution as compared to the background processes. The larger mass difference between the NLSPs and the LSP also results in a $\met$ distribution which extends to much larger values ($ > 900~{\rm GeV}$). The $\met$ distribution for the background processes, on the other hand, falls down more steeply. The figures in the right panel of Fig.~\ref{fig:wh_27_850_0} have been obtained after passing through the SRD4 cuts.

The event distribution of $p_{T}^{l_{1}}$ and $\Delta R_{OS}^{min}$ at the HE-LHC for BPE4 and the relevant background processes have been shown in Fig.~\ref{fig:wh_27_850_0} following the color code of Fig.~\ref{fig:wh_27_550_420}. The left panels of Fig.~\ref{fig:wh_27_850_0} have been obtained by imposing the following criteria: presence of three isolated leptons, veto on SFOS pair with invariant mass $M_{Z} \pm 12~{\rm GeV}$ and $b$-jet veto. Due to a smaller mass difference between the $\lspthree,\lsptwo,\chonepm$ and $\lspone$ in BPE4, the leading $p_{T}$ lepton is produced with a smaller boost and peaks at a slightly lower $p_{T}$ (around $ 50~{\rm GeV}$) which is roughly similar to the background processes (all of which peak around $50-100~{\rm GeV}$). Similarly, the $\Delta R_{min}^{OS}$ distribution for BPE4 falls off before $ \lesssim 3.0$ unlike the $WZ$, $WH_{125}$, $t\bar{t}V$ and $VVV$ backgrounds which extend up to $\gtrsim 4.0$. The plots on the right panels of Fig.~\ref{fig:wh_27_550_420} show the event distribution obtained upon the further imposition of SRE4 cuts.

\begin{figure}[!htb]
\begin{center}
\includegraphics[scale=0.15]{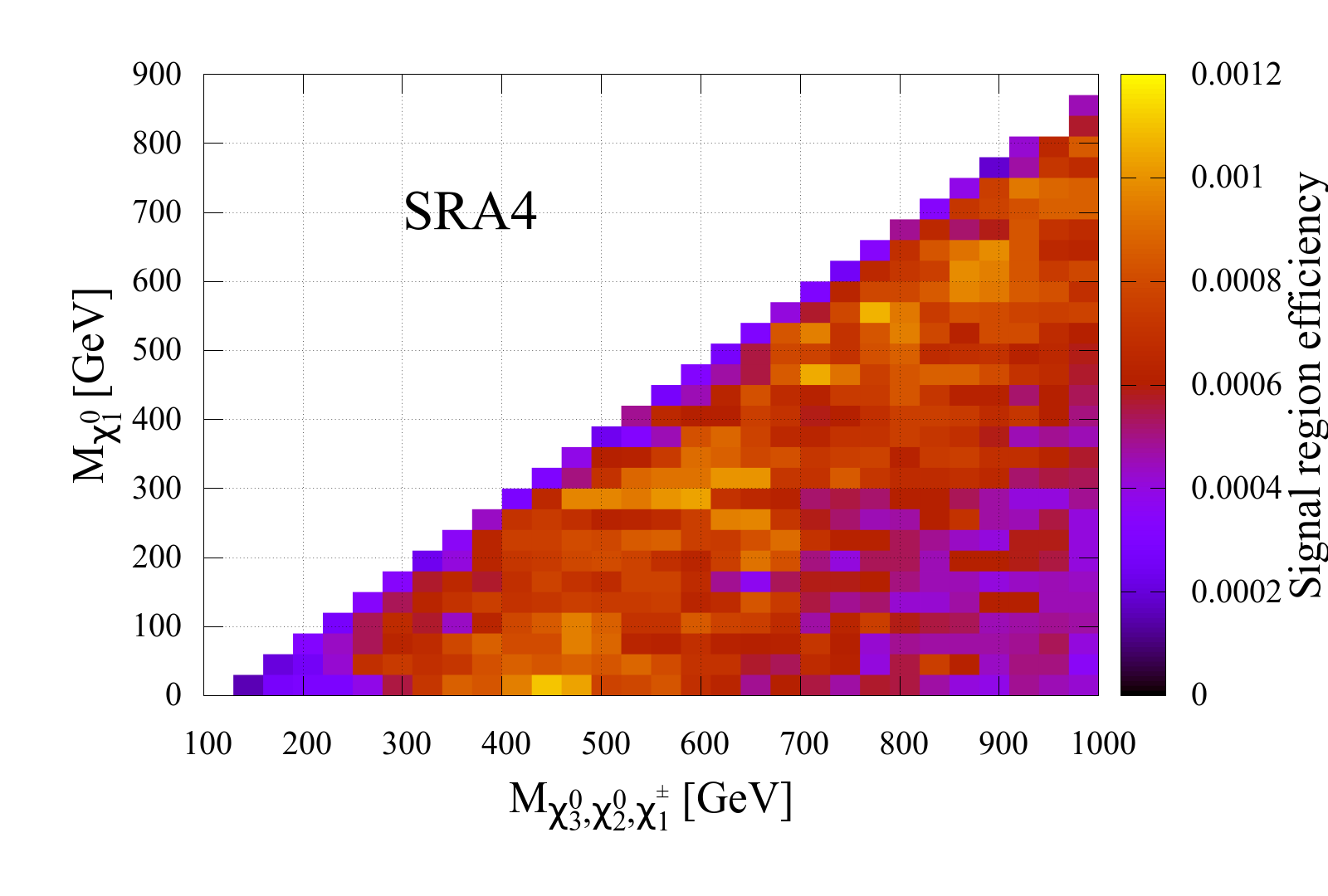}\includegraphics[scale=0.15]{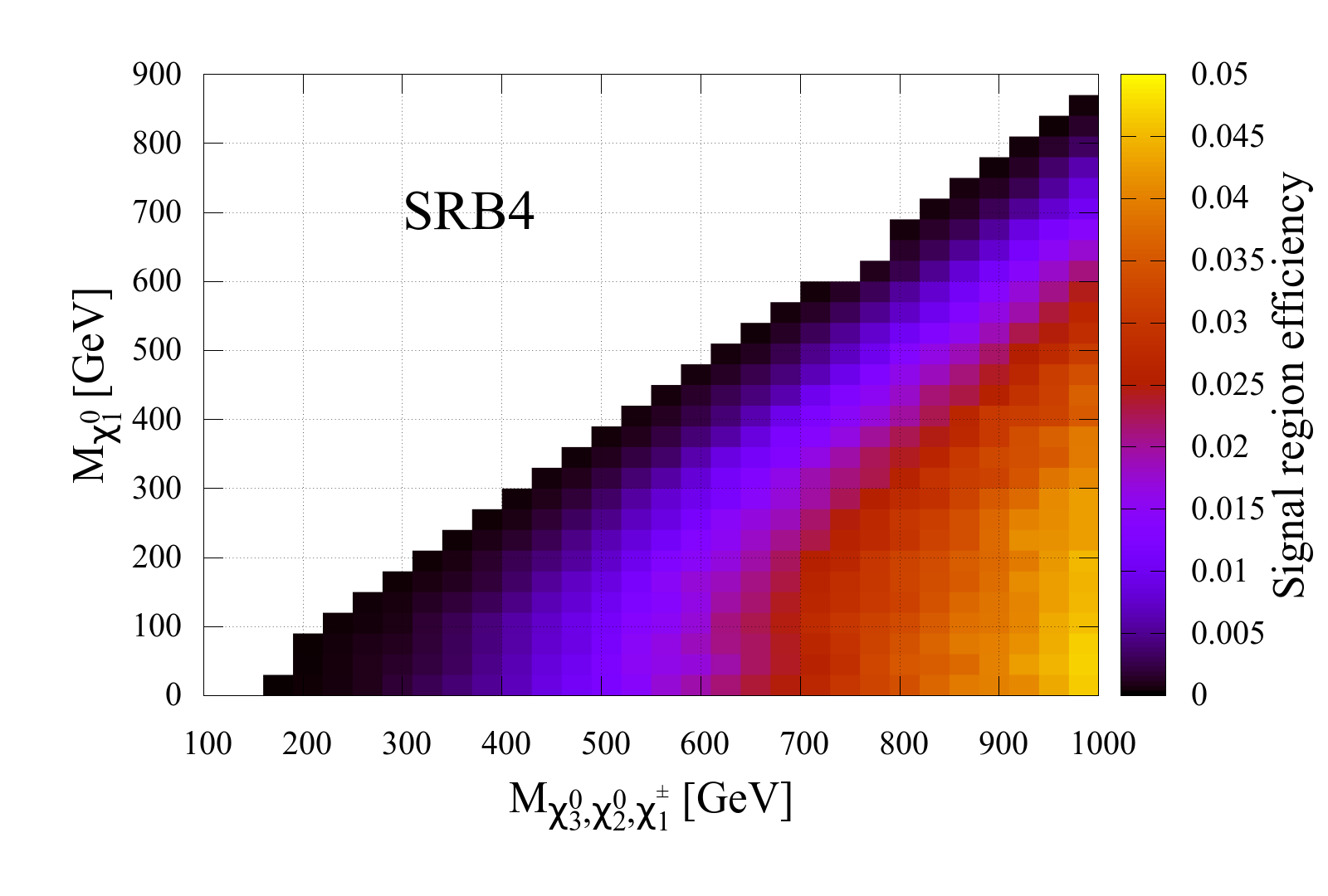}\\
\includegraphics[scale=0.15]{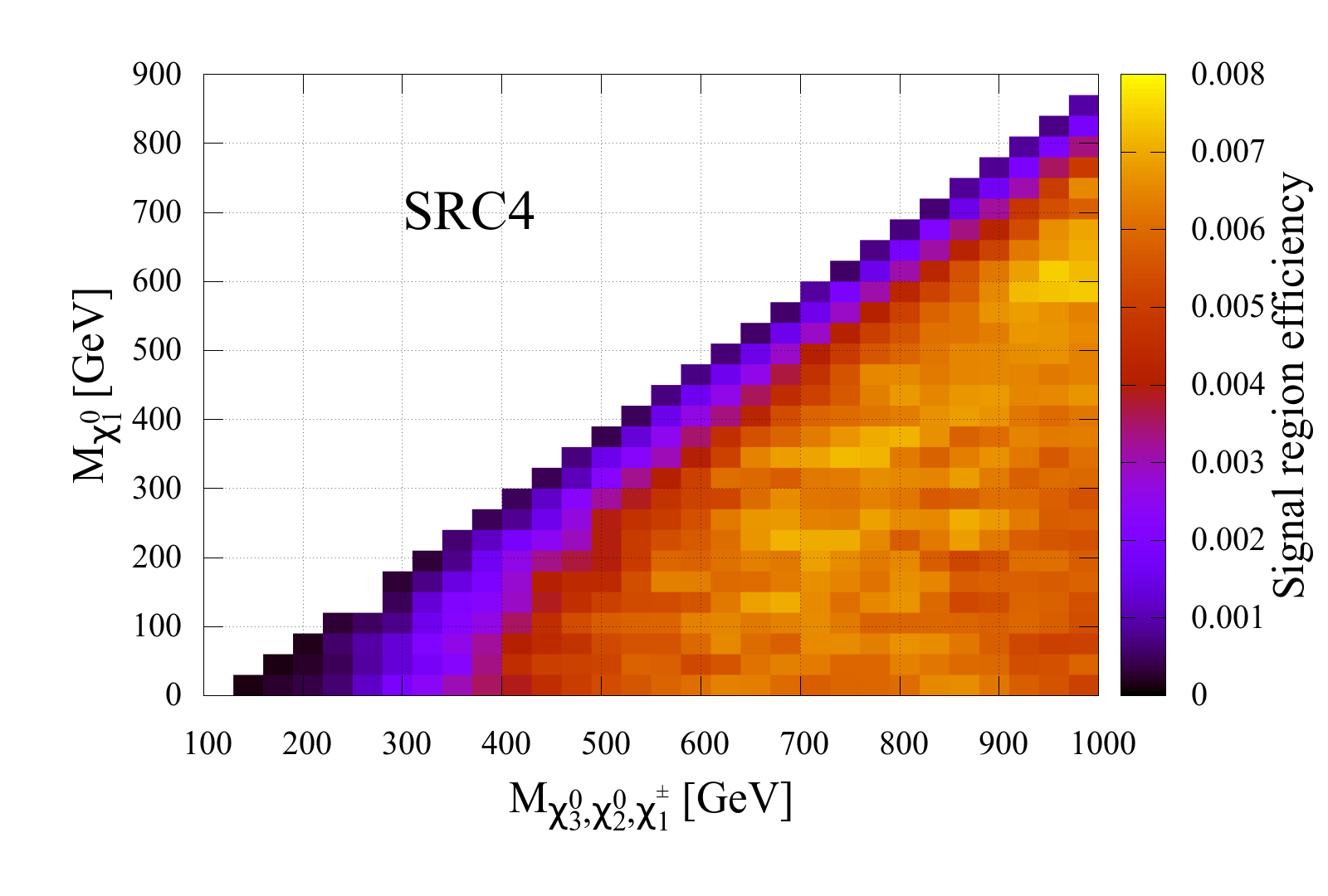}\includegraphics[scale=0.15]{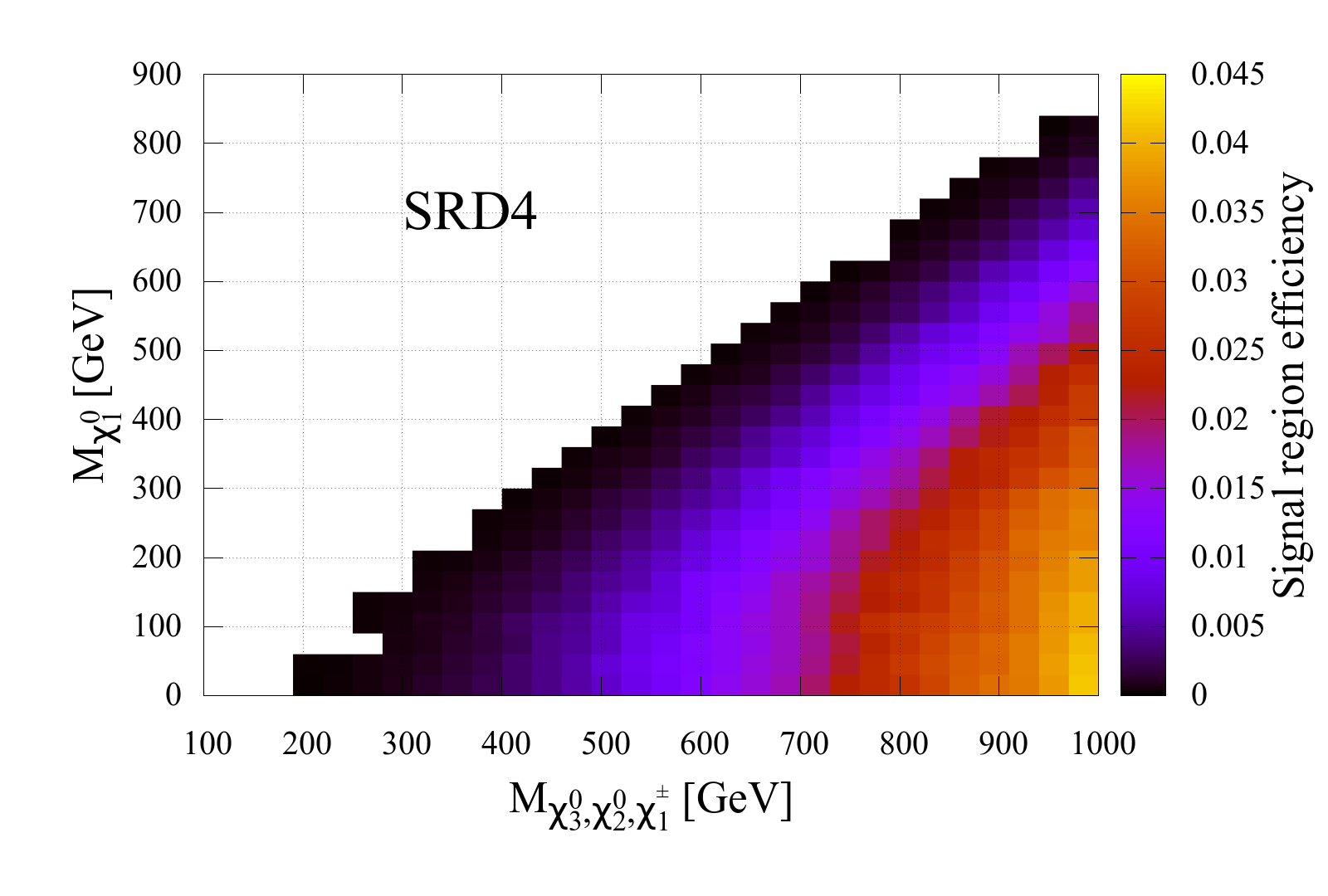}
\caption{Efficiency map of SRA4 (top-left), SRB4 (top-right), SRC4 (bottom-left) and SRD4 (bottom-right) is shown as a color palette in the $M_{\lspthree,\lsptwo,\chonepm}$-$M_{\lspone}$ plane.}
\label{fig:eff_wh_27}
\end{center}
\end{figure}

\begin{table}[!htb]
\begin{center}\scalebox{0.83}{
\begin{tabular}{|| C{3.0cm} | C{2.0cm} | C{1.5cm} C{1.5cm} C{1.5cm} C{1.5cm} C{1.5cm} C{1.5cm} C{1.5cm} ||}
\hline \hline
Background & Cross & \multicolumn{7}{c||}{Background yield ($27~{\rm TeV}$, $15~ab^{-1}$)} \\ \cline{3-9}
process & section [LO] & SRA4 & SRB4 & SRC4 & SRD4 & SRE4 & SRF4 & SRG\\ \hline\hline 
\textbf{$ZZ$} & 2092 fb & 0.0 & 10.4 & 0.0 & 0.0 & 0.0 & 0.0 & 5.18 \\ \hline 
\textbf{$t\bar{t}V$} & \multirow{3}{*}{1385 fb} & \multirow{3}{*}{10.4} & \multirow{3}{*}{10.4} & \multirow{3}{*}{6.92} & \multirow{3}{*}{27.7} & \multirow{3}{*}{24.2} & \multirow{3}{*}{17.3} & \multirow{3}{*}{6.93} \\ 
($V=W,Z$) & & & & & & & & \\ 
($W \to l\nu$, $Z \to ll$) & & & & & & & &\\ \hline
\textbf{$WZ$} & 1263 fb & 37.9 & 37.9 & 12.6 & 12.6 & 25.3 & 12.6 & 0.00 \\ \hline   
\textbf{$VVV$} ($V = W,Z$) & 681 fb & 1.57 & 0.0 & 1.57 & 4.72 & 7.86 & 1.57 & 0.0 \\ \hline
\textbf{$WH_{125}$ ($H_{125} \to \tau^{+}\tau^{-}$)} & 30.2~fb & 0.0 & 0.0 & 0.45 & 0.0 & 0.70 & 0.01 & 0.0 \\ \hline 
\textbf{$WH_{125}$ $(H_{125} \to WW^{*})$} & \multirow{3}{*}{0.29 fb} & \multirow{3}{*}{0.40} & \multirow{3}{*}{$0.05$} & \multirow{3}{*}{$0.08$} & \multirow{3}{*}{$0.04$} & \multirow{3}{*}{$0.18$} & \multirow{3}{*}{$0.01$} & \multirow{3}{*}{$0.02$} \\ 
$(W \to l\nu)$ & & & & & & & & \\ \hline
\multicolumn{2}{||c|}{\textbf{Total Background}} & 50.7 & 58.7 & 21.3 & 45.1 & 57.7 & 31.5 & 12.2 \\ \hline\hline 
\end{tabular}}
\caption{The background yields for $\sqrt{s}=27~{\rm TeV}$ LHC corresponding to $15~ab^{-1}$ of integrated luminosity, for the $5$ different signal regions considered for the cut-based analysis, are presented. The leading order (LO) cross sections generated by \texttt{MadGraph5$\_$aMC@NLO} have been considered.}
\label{tab:bkg_helhc_wh}
\end{center}
\end{table}

The LO background cross-sections and the background yields corresponding to the $7$ signal regions are shown in Table~\ref{tab:bkg_helhc_wh}. We also show the efficiency grids for SRA4, SRB4, SRC4 and SRD4 for the $H_{125} \to WW^{*}$ scenario in Fig.~\ref{fig:eff_wh_27} in the  $M_{\lspthree,\lsptwo,\chonepm}$ - $M_{\lspone}$ plane. 

\begin{figure}[!htb]
\begin{center}
\includegraphics[scale=0.3]{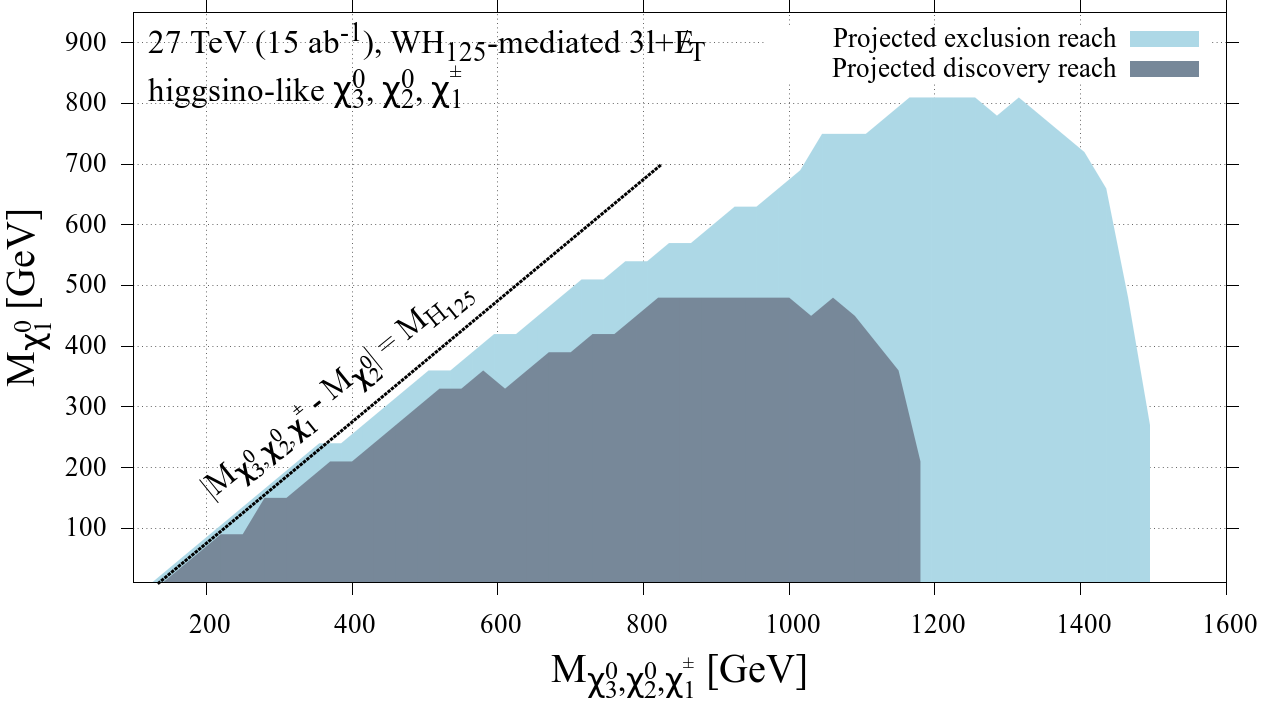}
\caption{ The projected exclusion (blue color) and discovery regions (grey color) from searches in the $WH_{125}$-mediated $3l+\met$ final state produced from direct higgsino searches ($pp \to \lsptwo\chonepm+\lspthree\chonepm$) at the HE-LHC, is shown.}
\label{fig:27_higgsino_contour_wh}  
\end{center}
\end{figure}

The strategy prescribed in Sec.~\ref{sec:wh_hllhc} has been used for the computation of signal significance. The signal significance is computed for the seven signal regions, and the one which yields the highest value for a particular point is considered in deriving the projection contours. We have assumed a systematic uncertainty of $5\%$ in the present analysis. The projected exclusion and discovery contours for direct higgsino searches in the $WH_{125}$-mediated $3l+\met$ channel at the HE-LHC is  shown in Fig.~\ref{fig:27_higgsino_contour_wh}. The projected discovery (exclusion) region reaches up to $M_{\lspthree,\lsptwo,\chonepm} \sim 1180~(1500) ~{\rm GeV}$ for a massless LSP.

\subsection{Projected limits on NMSSM-inos at the HE-LHC}
\label{sec:ewino_helhc}

\begin{figure}[!htb]
\begin{center}
\includegraphics[scale=0.18]{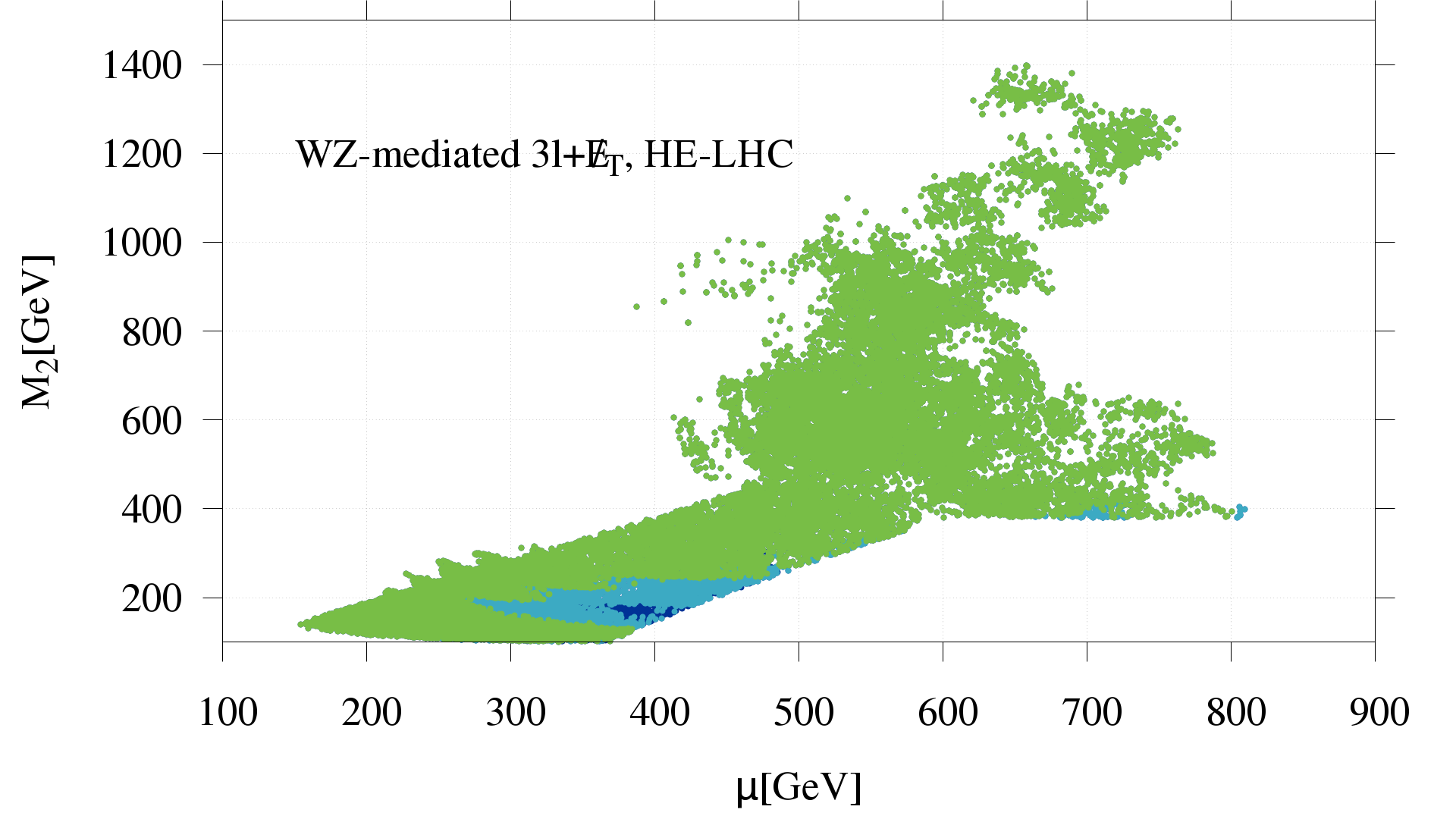}
\caption{ The currently allowed parameter space points are shown in the $M_{2}-\mu$ plane. The pale blue and green colored points fall within the projected exclusion and discovery reach, respectively, of direct higgsino searches in the $WZ$ mediated $3l+\met$ channel at the HE-LHC.
}
\label{fig:mu_m2_helhc}
\end{center}
\end{figure}

\begin{figure}[!htb]
\begin{center}
\includegraphics[scale=0.18]{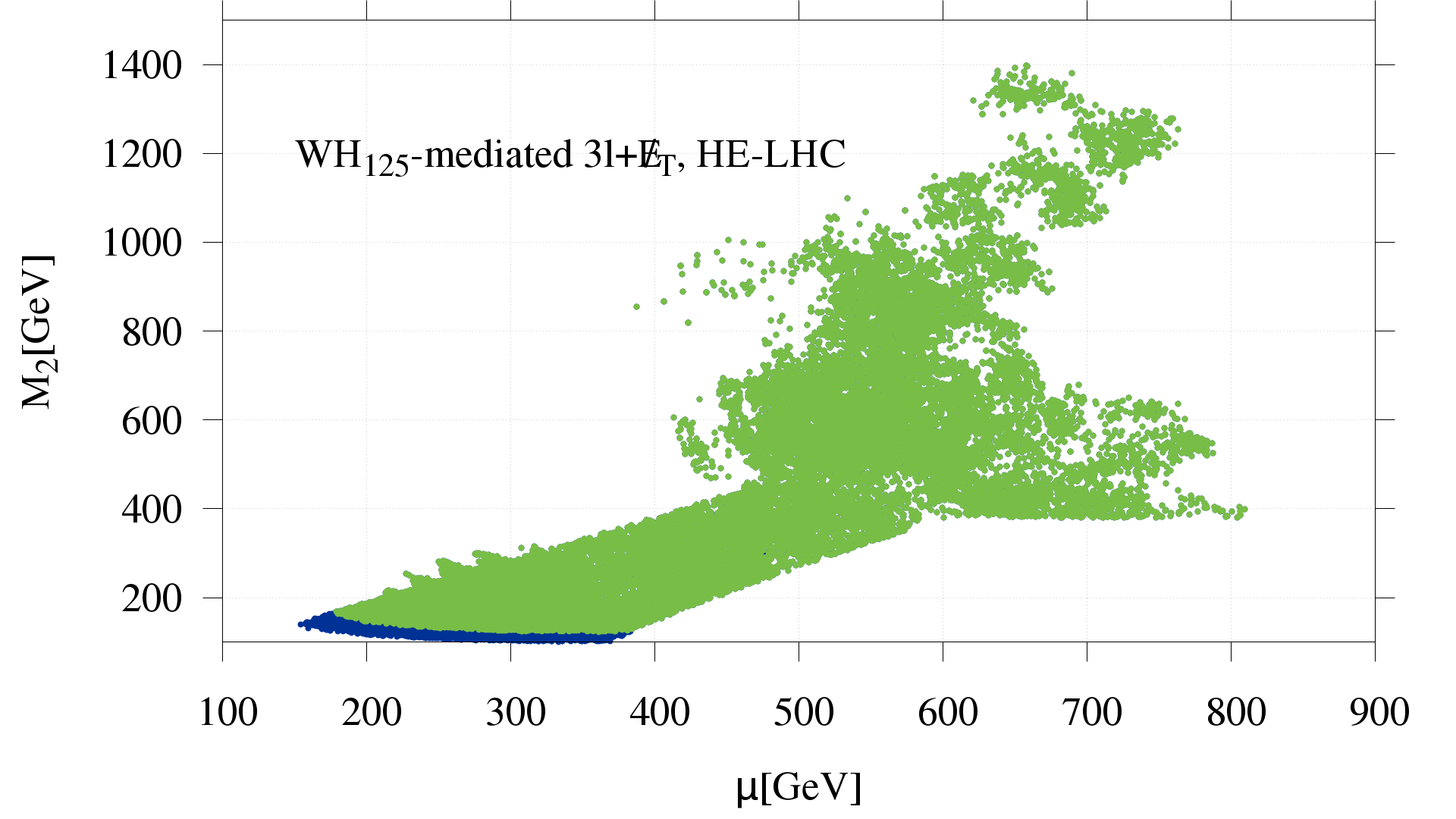}
\caption{The allowed parameter space points have been shown in the $M_{2}-\mu$ plane. The pale blue and green colored points fall within the projected exclusion and discovery reach, respectively, of direct electroweakino searches in the $WH_{125}$ mediated $3l+\met$ channel at the HE-LHC. The dark blue colored points fall outside the HE-LHC's projected reach.}
\label{fig:mu_m2_helhc_wh}
\end{center}
\end{figure}

Extending the discussion and analysis prescribed in Sec.~\ref{sec:case1_ewino_hllhc}, we evaluate the projected reach of doublet higgsino searches at the HE-LHC in probing the NMSSM parameter space with light neutralinos. The direct higgsino production cross-sections of the relevant chargino neutralino pairs is computed at $\sqrt{s}=27~{\rm TeV}$ and  is then scaled with the relevant reduced couplings for each  parameter space point. The signal significance values are then computed using the strategy described in Sec.~\ref{sec:case1_ewino_hllhc}. The projected impact of direct higgsino searches in the $WZ$ mediated $3l+\met$ final state at the HE-LHC is  shown in Fig.~\ref{fig:mu_m2_helhc}.
Comparing with Fig.~\ref{fig:m2_mu_hllhc}, we see that the HE-LHC provides a larger discovery opportunity than the HL-LHC for the detection of NMSSM-inos with a light LSP. Nearly all the NMSSM allowed points are within the discovery reach of the HE-LHC.
The parameter space points shown in Fig.~\ref{fig:mu_m2_helhc} correspond to the allowed points obtained after imposing all the current constraints discussed in Sec.~\ref{sec:constraints}. The green colored points in Fig.~\ref{fig:mu_m2_helhc} are within the projected discovery reach (signal significance $> 5$), while the pale blue colored points are within the projected exclusion reach (signal significance $>2$) of direct higgsino searches in the $WZ$ mediated $3l+\met$ final state at HE-LHC. The dark blue colored points, however, will be undetectable to HE-LHC, these correspond to a 
small band of points at $M_{2} \sim 200~{\rm GeV}$. 
We note that the signal cases, and hence the cuts, that we have chosen in
Table~\ref{tab:wz_helhc_sr} are not designed to probe the very light wino region where the lightest NLSP mass is around $M_{NLSP} \sim 180-200$~GeV 
resulting in reduced signal efficiency in the dark blue region. Furthermore, the lightest NLSP in the dark blue region is dominantly wino in nature resulting in small $\lsptwo \to \lspone Z$ branching rates. We have
checked that even so, these points only narrowly escape the projected
exclusion. We fully expect that choosing additional signal regions to
optimize the signal significance around $M_{NLSP} \sim 190$~GeV will prove
effective in extending the exclusion/discovery region of the HE-LHC to
this region of parameter space.

\begin{table}[!htb]
\begin{center}
\begin{tabular}{C{4cm}|C{2cm}|C{2cm}|C{2cm}|C{2cm}}
Benchmark point & \multicolumn{2}{c|}{$WZ$ mediated} & \multicolumn{2}{|c}{$WH_{125}$ mediated} \\ \cline{2-5}
($M_{2}$,~$\mu$) [in GeV] & HL-LHC & HE-LHC & HL-LHC & HE-LHC \\ \hline 
$BP_{A}$ & 13 & 180 & 4 & 23\\
(1244,~717) & (3.8) & (14) & (0.4) &  (6.6) \\ \hline
$BP_{B}$ & 7 & 86 & 63 & 272\\
(400,~717) & (1.5) & (4.4) & (5.3) & (34) \\ \hline
$BP_{C}$ & 7 & 65 & 131 &  388\\
(200,~403) & (1.3) & (2.1) & (8.8) & (48) \\ \hline
$BP_{D}$ & 20 & 231 & 8 &  35\\
(952,~585) & (6.1) & (18) & (1.0) & (9.8) \\ \hline 
$BP_{E}$ & 23 & 408 & 12 & 36\\
(696,~518) & (7.0) & (20) & (1.2) & (10) \\ \hline 
$BP_{F}$ & 28 & 418 & 18 & 79 \\
(555,~571) & (8.6) & (21) & (2.1) & (22) \\ \hline 
$BP_{G}$ & 23 & 233 & 78 &  206\\
(396,~515) & (5.2)  & (12) & (5.3) & (27) \\ \hline 
$BP_{H}$ & 17 & 167 & 125 &  368 \\
(204,~302) & (3.4) & (5.3) &  (8.4) & (45) \\ \hline 
$BP_{I}$ & 27 & 257 & 110 & 321\\
(210,~262) & (5.3) & (8.1) & (7.4) & (40) \\ \hline 
\end{tabular}
\caption{Comparison of the expected number of signal events in the $WZ$ and $WH_{125}$ mediated $3l+\met$ channels after all cuts in the accumulated data set at the HL-LHC and the HE-LHC for the benchmark points, $BP_A$, $BP_B$, $BP_C$, $BP_D$, $BP_E$, $BP_F$, $BP_G$, $BP_H$ and $BP_I$, chosen from the allowed parameter space. $BP_A$, $BP_B$ and $BP_C$ are also shown in Figs.~\ref{fig:m2_mu_hllhc} and \ref{fig:wh_hllhc}. For each case, the signal significance is shown in parenthesis. The input parameters of these benchmark points are tabulated in Appendix~\ref{Appendix:BP}.}
\label{tab:signal_yield}
\end{center}
\end{table}

Before concluding this section, we also study the future reach of direct higgsino searches in the $WH_{125}$ mediated $3l+\met$ final state at the HE-LHC. The results are illustrated in Fig.~\ref{fig:mu_m2_helhc_wh} using the color code from Fig.~\ref{fig:mu_m2_helhc}. It can be seen from Fig.~\ref{fig:mu_m2_helhc_wh} that the projected reach of HE-LHC extends much beyond that of the HL-LHC (shown in Fig.~\ref{fig:wh_hllhc}) for the detection of NMSSM-inos. Here, the dark blue colored points are concentrated in the $M_{2} \lesssim 125~{\rm GeV}$ (where the decay to $H_{125}$ is kinematically forbidden). The projected discovery reach covers the rest of the NMSSM parameter space. We see that a combination of doublet higgsino searches at the HE-LHC
in the $WH_{125}$ and $WZ$ mediated $3l+\met$ channels will probe
the entire region of the currently allowed NMSSM parameter space with
light neutralinos at the $>5\sigma$ level.

Even though discovery may be possible also at the HL-LHC, the reader may find it interesting to compare the signal size after the final cuts at these facilities. We show such a comparison in Table~\ref{tab:signal_yield} for the benchmark points $BP_A$, $BP_B$ and $BP_C$ in Fig.~\ref{fig:m2_mu_hllhc} and Fig.~\ref{fig:wh_hllhc}, and $6$ other benchmark points ($BP_D$, $BP_E$, $BP_F$, $BP_G$, $BP_H$ and $BP_I$), for both the $WZ$ and the $WH_{125}$ mediated decays. The corresponding signal significance is shown in parenthesis. We see that the accumulated signal is appreciably larger at the HE-LHC. This larger data sample should enable detailed studies of event shapes and distributions, and help zero in on the origin of the new physics.

\section{Conclusion}
\label{Sec:NMSSM:Conclusion}

In this paper, our goal was to analyze the current status and future
prospects of a light neutralino DM ($M_{\lspone} <  M_{H_{125}} /2$) within the NMSSM
framework. Previous studies of light neutralinos
in the MSSM  have shown that such scenarios could  be entirely probed 
at future direct detection experiments such as Xenon-nT, with potential
complementary signatures in invisible Higgs decays. We therefore
concentrated only on regions of the  NMSSM with features clearly
distinct from those of the MSSM regions explored before. In particular we focused our study on
the case where a new light singlet Higgs was present (with $M_{A_{1}},M_{H_{1}}<122$
GeV). Such scenarios can provide a new mechanism for efficient
neutralino annihilation in the early Universe via the exchange of a
light scalar and/or pseudoscalar Higgs at roughly twice the mass of the
neutralino DM. Thus in the NMSSM framework neutralino DM at the GeV
scale can satisfy the relic density constraint.

After imposing relevant current constraints from collider and
astrophysical experiments,  the allowed parameter space of the model was
identified. This parameter space features both a singlino- dominated
LSP  and a singlet-like light Higgs. We showed that direct detection
experiments and Higgs invisible width measurements at future colliders
such at the ILC, CEPC and FCC have the potential to probe some of the model's
parameter space, however two regions in particular remain out of
reach. The first correspond to a neutralino with a mass below about 10
GeV where not only direct detection experiments  lose their sensitivity
but also where the Higgs invisible width can often be very small due to
the singlino nature of the LSP.  The second  had a LSP mass near ~62
GeV. For such a neutralino with a mass close to $M_{H_{125}}/2$, the relic
density constraint can be satisfied
even when the coupling of the  neutralino LSP to the Higgs is strongly
suppressed,  thus reducing the direct detection signal as well as  the
Higgs invisible width. A significant fraction of the first region is
also out of reach of searches for light scalars/pseudoscalars in SM-like
Higgs decays at the future HL-LHC since those searches which typically
involve a decay channel of a light Higgs into b-quarks lose their
sensitivity when  $M_{A_{1},H_{1}} \lesssim 11~ {\rm GeV}$, which in our framework is linked to $M_{DM}< 5.5~ {\rm GeV}$.
Thus a large fraction of the NMSSM parameter space with light
neutralinos will remain out of reach of these future searches.

We found that these difficult-to-probe regions came with light electroweakinos,  in particular  with  NLSP winos barely above
100 GeV or a NLSP higgsino as light as 400 GeV, hence have the potential
to be discovered at future runs of the LHC. After recasting and
validating the current searches for light electroweakinos in the
$3l+\met$ final state from neutralino-chargino pairs in
the $WZ$ and $WH_{125}$ channel, we investigated how these searches
could probe the allowed parameter space of the model at the future
high-luminosity and high-energy extension of the LHC. We showed that over a wide range of parameters, the doublet higgsino could be probed at the HL-LHC via the $WZ$ mediated channel, at least at the $2\sigma$ level, but for the most part even at the $5\sigma$ level. Moreover, we found that the $WH_{125}$ mediated channel was truly complementary particularly in the wino NLSP region ($M_2< \mu$) where the $WZ$ channel had the least power. We found that over almost all the allowed NMSSM parameter space with a light LSP, experiments at the HL-LHC would be able to discover ($5\sigma$) SUSY via at least one of the two channels. We have limited our study to the golden $3l+\met$ signal from electroweakino production. It may be worth examining other signal channels which occur at larger rates, but also have larger backgrounds.

The higher energy and luminosity available at the HE-LHC guarantees
discovery via both the $WZ$ and $WH_{125}$ mediated trilepton channels over essentially the entire allowed parameter space.
We also highlighted some regions where additional search channels,
involving the decays of neutralinos into light scalar/pseudoscalar
Higgses could offer distinctive probes of the NMSSM electroweakino sector.
The tools developed for this analysis are not restricted to NMSSM
scenarios under consideration and the recasting presented here can be
applied to other prospective studies of electroweakino searches at the
HL-LHC and HE-LHC where the LSP neutralinos can be heavier than $60~{\rm GeV}$.

\section*{Acknowledgements}

We thank Jason Kumar for reminding us about the CMB constraints on light dark matter. We thank Amit Adhikary, Rhitaja Sengupta and Prabhat Solanki for the helpful discussions. This work was supported in part by the CNRS LIA-THEP (Theoretical High EnergyPhysics) and the INFRE-HEPNET (IndoFrench Network on High Energy Physics) of CEFIPRA/IFCPAR (Indo-French Centre for the Promotion of Advanced Research). The work of BB is supported by the Department of Science and Technology,  Government of India, under the Grant Agreement number IFA13-PH-75 (INSPIRE Faculty Award). The work of DS is supported by the National Science Foundation under Grant No. PHY- 1915147. The work of RMG is supported by the Department of Science and Technology, India under Grant No. SR/S2/JCB-64/2007. BB, RMG and RKB acknowledges the hospitality at LAPTh where a part of this work was carried out. 
XT thanks the Centre for High Energy Physics, Indian Institute of Science
where this work was begun for its hospitality, and also the Infosys
Foundation for making this visit possible.

\endpage 
\pagebreak


\newpage

\appendix

\section{Benchmark points}
\label{Appendix:BP}

\begin{table}[!htb]
\begin{center}
\begin{tabular}{C{3cm}| C{12cm}}
Benchmark points & Input parameters \\ \hline
\multirow{2}{*}{$BP_{A}$} & $\lambda$~=~0.3, $\kappa$~=~0.01, $\tan\beta$~=~9.5, $A_{\lambda}$~=~6687~GeV, $A_{\kappa}$~=~5.2~GeV,\\
 & $\mu$~=~717~GeV, $M_{2}$~=~1244~GeV, $M_{3}$~=~2301~GeV \\ \hline
 \multirow{2}{*}{$BP_{B}$} & $\lambda$~=~0.44, $\kappa$~=~0.02, $\tan\beta$~=~11.8, $A_{\lambda}$~=~8894~GeV, $A_{\kappa}$~=~-57~GeV,\\
 & $\mu$~=~717~GeV, $M_{2}$~=~400~GeV, $M_{3}$~=~4323~GeV \\ \hline
 \multirow{2}{*}{$BP_{C}$} & $\lambda$~=~0.08, $\kappa$~=~$3\times 10^{-4}$, $\tan\beta$~=~18, $A_{\lambda}$~=~6563~GeV, $A_{\kappa}$~=~-7.9~GeV,\\
 & $\mu$~=~403~GeV, $M_{2}$~=~,200~GeV $M_{3}$~=~3080~GeV \\ \hline
 \multirow{2}{*}{$BP_{D}$} & $\lambda$~=~0.44, $\kappa$~=~0.02, $\tan\beta$~=~15.6, $A_{\lambda}$~=~585~GeV, $A_{\kappa}$~=~9501~GeV,\\
 & $\mu$~=~585~GeV, $M_{2}$~=~952~GeV, $M_{3}$~=~4457~GeV \\ \hline
 \multirow{2}{*}{$BP_{E}$} & $\lambda$~=~0.27, $\kappa$~=~0.02, $\tan\beta$~=~11.6, $A_{\lambda}$~=~5875~GeV, $A_{\kappa}$~=~12~GeV,\\
 & $\mu$~=~518~GeV, $M_{2}$~=~696~GeV, $M_{3}$~=~3634~GeV \\ \hline
 \multirow{2}{*}{$BP_{F}$} & $\lambda$~=~0.30, $\kappa$~=~0.01, $\tan\beta$~=~11.2, $A_{\lambda}$~=~6319~GeV, $A_{\kappa}$~=~17~GeV,\\
 & $\mu$~=~571~GeV, $M_{2}$~=~555~GeV, $M_{3}$~=~2687~GeV \\ \hline
 \multirow{2}{*}{$BP_{G}$} & $\lambda$~=~0.42, $\kappa$~=~0.02, $\tan\beta$~=~15.9, $A_{\lambda}$~=~8638~GeV, $A_{\kappa}$~=~43.4~GeV,\\
 & $\mu$~=~515~GeV, $M_{2}$~=~396~GeV, $M_{3}$~=~2903~GeV \\ \hline
 \multirow{2}{*}{$BP_{H}$} & $\lambda$~=~0.02, $\kappa$~=~$7\times 10^{-5}$, $\tan\beta$~=~25.5, $A_{\lambda}$~=~7348~GeV, $A_{\kappa}$~=~-7.3~GeV,\\
 & $\mu$~=~302~GeV, $M_{2}$~=~204~GeV, $M_{3}$~=~2239~GeV \\ \hline
 \multirow{2}{*}{$BP_{I}$} & $\lambda$~=~0.02, $\kappa$~=~$6\times 10^{-5}$, $\tan\beta$~=~27.6, $A_{\lambda}$~=~6924~GeV, $A_{\kappa}$~=~-5.7~GeV,\\
 & $\mu$~=~262~GeV, $M_{2}$~=~210~GeV, $M_{3}$~=~2217~GeV \\ \hline
\end{tabular}
\caption{The input parameters corresponding to the benchmark points considered in this paper.}
\end{center}
\end{table}

\end{document}